\RequirePackage{fix-cm}
\documentclass[twocolumn]{svjour3}          
\smartqed  

\usepackage{graphicx}
\usepackage{hyperref}
\hypersetup{
    colorlinks=true,
    linkcolor=blue,
    filecolor=magenta,      
    urlcolor=cyan,
    pdftitle={Overleaf Example},
    pdfpagemode=FullScreen,
    }
\usepackage{amssymb}
\usepackage{amsmath}
\usepackage{psfrag}
\usepackage{mathtools}
\usepackage{color}
\usepackage{xcolor}
\usepackage[export]{adjustbox}
\usepackage[titletoc,title]{appendix}
\usepackage[utf8]{inputenc}

\newcommand{\tr}[1]{\textcolor{black}{#1}} 
\newcommand{\fc}[1]{\textcolor{black}{#1}} 

\definecolor{mygreen1}{cmyk}{0.55, 0, 0.85,0}
\definecolor{darkblue}{rgb}{0,0,0.55}
\definecolor{cyan}{cmyk}{0.48,0,0.13,0}
\definecolor{magenta}{cmyk}{0.41, 0.78, 0, 0}
\definecolor{darkblue2}{cmyk}{0.76,0.65,0,0}
\definecolor{myred}{cmyk}{0,0.95,0.92,0}
\definecolor{lightblue}{cmyk}{1,0,0,0}
\definecolor{lightorange}{cmyk}{0,0.5,1,0}
\definecolor{darkblue2}{cmyk}{1,.9,0,0}
\definecolor{darkblue3}{cmyk}{0.76,0.65,0,0}
\definecolor{myblue}{cmyk}{0.85, 0.5, 0, 0}
\definecolor{myred2}{cmyk}{0, 1, 1,0}
\definecolor{myyellow}{cmyk}{0.12, 0, 0.91,0}
\definecolor{ablue}{cmyk}{0.93, 0.75, 0,0}
\definecolor{gramp}{cmyk}{0.65,0,1,0}

\begin{document}
\title{Steady State Statistics of Emergent Patterns in a Ring of \fc{Oscillators}\thanks{This project is funded by the Swiss National Science Foundation under Grant agreement 184617.}}


\titlerunning{Steady State Statistics of Emergent Patterns in a Ring of Oscillators}        

\author{Tiemo Pedergnana         \and
        Nicolas Noiray 
}


\institute{CAPS Laboratory, Department of Mechanical and Process Engineering, ETH Zürich, Sonneggstrasse 3, 8092 Zürich, Switzerland
}

\date{Received: date / Accepted: date}

\maketitle

\begin{abstract}
Networks of coupled nonlinear oscillators \\model a broad class of physical, chemical and biological systems. Understanding emergent patterns in such networks is an ongoing effort with profound implications for different fields. In this work, we analytically and numerically study a symmetric ring of $N$ coupled \fc{self-}oscillators of Van der Pol type under external stochastic forcing. The system is proposed as a \fc{model} of the thermo- and aeroacoustic interaction\fc{s} of sound fields in rigid enclosures \fc{with} compact source regions in a can-annular combustor.

The oscillators are \fc{connected} via linear resistive coupling with nonlinear saturation. After transforming the system to amplitude-phase coordinates, deterministic and stochastic averaging is performed to eliminate the fast oscillating terms. By projecting the potential of the slow-flow dynamics onto the phase-locked quasi-limit cycle solutions, we obtain a compact, low-order description of the (de-)synchronization transition for an arbitrary number of oscillators. The stationary probability density function of the state variables is derived from the Fokker--Planck equation, studied for varying parameter values and compared to time series simulations. We leverage our analysis to offer explanations for features of acoustic pressure spectrograms observed in real-world gas turbines.
\end{abstract}

\section{Introduction \label{Section 1}}
 \begin{figure*}[h!]
\begin{psfrags}
\psfrag{o}{\hspace{-0.3cm}\fc{$U_\mathrm{tot}$}}
\psfrag{a}{$\varphi_j(t)$}
\psfrag{b}{\hspace{-0.14cm}\fc{$A_j(t)$}}
\psfrag{P}{$0$}
\psfrag{k}{\hspace{-0.2cm}\bf \textcolor{lightorange}{Shear layer}}
\psfrag{q}{\bf \footnotesize \textcolor{darkblue2}{\begin{tabular}{@{}c@{}}
Mean \\
axial\\
velocity
\end{tabular}}}
\psfrag{r}{\hspace{0.2cm}\bf \footnotesize \textcolor{lightblue}{\begin{tabular}{@{}c@{}}
Turbulent \\
wake
\end{tabular}}}
\psfrag{c}{$W$}
\psfrag{m}{\hspace{-0.7cm}\fc{{\begin{tabular}{@{}c@{}}
Acoustic\\
\,pressure $\eta_j$
\end{tabular}}}}
\psfrag{d}{$x$}
\psfrag{e}{$L$}
\psfrag{f}{$0$}
\psfrag{p}{$q_j$}
\psfrag{g}{$C$}
\psfrag{h}{\hspace{-0.05cm}Burner outlet}
\psfrag{j}{Can outlet}
\psfrag{i}{\bf {Flame}}
\psfrag{u}{\bf
{\begin{tabular}{@{}c@{}}
\textcolor{myred2}{Resistive coupling} \\
{}\\
\textcolor{myblue}{Reactive coupling}
\end{tabular}}}
\psfrag{v}{$A_{j+1}$}
\psfrag{V}{$A_{j-1}$}
\psfrag{x}{\bf \textcolor{myblue}{Reactive coupling}}
\psfrag{s}{\textcolor{myred2}{$\boldsymbol{\lambda}$}}
\psfrag{t}{\textcolor{myblue}{$\boldsymbol{\mu}$}}
\psfrag{X}{\textbf{(b)}}
\psfrag{Y}{\textbf{(c)}}
\psfrag{Z}{\textbf{(a)}}
\psfrag{K}{{\begin{tabular}{@{}c@{}}
First row of\\
\,turbine vanes
\end{tabular}}}
\psfrag{1}{$\mathcal{C}_j$}
\psfrag{2}{\hspace{-0.35cm}\fc{{\begin{tabular}{@{}c@{}}
Can combustion\\
chamber
\end{tabular}}}}
\psfrag{3}{$\boldsymbol{\ldots}$\hspace{9.22cm}$\boldsymbol{\ldots}$}

    \centering
    \includegraphics[width=0.675\textwidth,center]{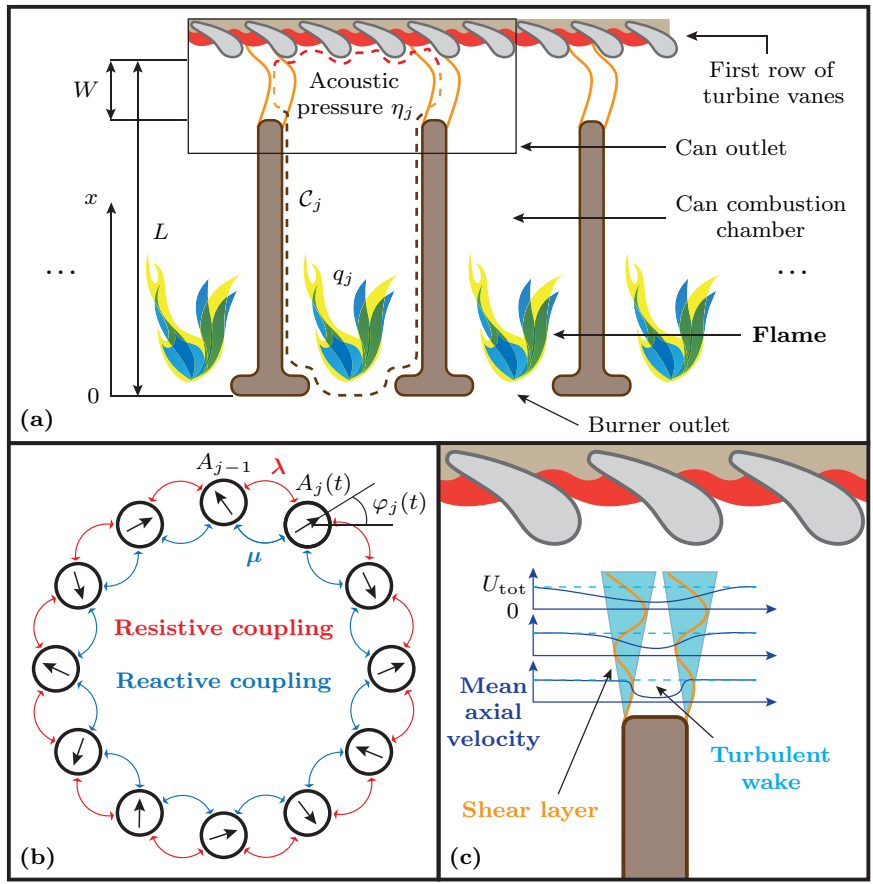}
    \end{psfrags}
    \caption{\textbf{(a)} Sketch of the modeled \fc{can-annular combustor}, from a radial view. \fc{The dimensions are not true to scale.} Shown \fc{are} the control volume $\mathcal{C}_j$, the acoustic pressure at the can outlet $\eta_j$, the aperture width $W$, the can length $L$ and the coherent heat release fluctuations of the flame $q_j$, which characterize the flame response to acoustic perturbations. \textbf{(b)} Abstraction \fc{of} the system illustrated in \textbf{(a)}, viewed from an axial perspective, for $N=12$ cans. Shown \fc{are the amplitude $A_j$ and phase $\varphi_j$ of the acoustic pressure at the can outlet $\eta_j$, respectively,} the \fc{linear} resistive coupling $\lambda$ and the reactive coupling $\mu$. \tr{For simplicity,} reactive coupling is neglected in this study. \textbf{(c)} Sketch of the turbulent wake in the aperture between neighboring cans. Shown is a typical profile of the mean axial velocity, the mean \fc{axial} bulk flow speed of the combustion products $U_\mathrm{tot}$ and the turbulent wake, which is bounded by two shear layers. The acoustic-hydrodynamic interaction in the aperture leads to the resistive and reactive coupling between the oscillators \fc{shown} in \textbf{(b)}.}
    \label{Figure 1}
\end{figure*}
\subsection{Thermoacoustic instabilities in can-annular combustors}
In a rigid enclosed volume, the \fc{heat release rate} response of an compact unsteady flame \fc{to acoustic perturbations} forms a feedback loop with the acoustics of the enclosure. If, at a given condition, this interaction between the sound field and the flame \fc{exceeds the dissipation due to radiation losses and viscous effects in the fluid, occurring mainly at the boundary of the enclosure}, it can give rise to so-called thermoacoustic instabilities. The study of this physical phenomenon goes back to the work of Rayleigh \cite{Rayleigh1878explanation}. When an instability is insufficiently damped, it can lead to high-amplitude acoustic pressure oscillations in the enclosure. In industrial machines, such as stationary gas turbines, the resulting pulsations induce high-cycle fatigue cracks in the metal parts surrounding the enclosure, which cause down-time, incurring fees for the manufacturer, \fc{and can sometimes lead to catastrophic system failures if a broken part flies through the combustor and collides at high speed with the vanes and blades of the turbine}. \fc{A} review of high-cycle fatigue in gas turbines is given in Ref. \cite{Nicholas1999}. Applied studies on Helmholtz dampers to suppress thermoacoustic instabilities in gas turbine combustors \fc{are} found in \fc{Refs. \cite{Bellucci2004271,Bothien2014}}. 

Since the turn of the \fc{millennium}, the modeling, prediction and suppression of thermoacoustic instabilities has drawn renewed interest due to internationally imposed emission limits on power generation systems and the resulting increased demand for lean-premixed combustion. In a lean-premixed combustor, the entire air flow passes through the burner to ensure a sufficiently lean mixture \fc{and to avoid rich reactant pockets, which produce significantly more pollutants. There are two consequences to this concept: First, lean premixed flames are much more sensitive to flow perturbations than non-premixed flames, which increases \tr{critically} their response to acoustic forcing. Secondly, in a premixed setting, one does not need to dilute the combustion products to temperatures that are acceptable for the available cooling of the turbine parts, and hence there are no dilution holes, which provide \tr{strong} acoustic damping in non-premixed combustors. These two factors lead to combustion instability problems in lean-premixed gas turbines similar to those encountered in rocket engines. In both applications, these instabilities remain a key problem in the design process \cite{keller95,Poinsot20171}.}  

\fc{Traditionally, most} literature on modeling thermoacoustic instabilities focused on \textit{silo-type} (one single can) or \textit{annular} combustors (see, for instance, \fc{\cite{crocco1951aspects,keller1985thermally,schuller_poinsot_candel_2020,Polifke2020}} and \cite{Noiray2013dynamicnature,Ghirardo2013,faure-beaulieu_indlekofer_dawson_noiray_2021}, respectively). \fc{However,} as energy demand continues to rise \fc{and energy networks are changing}, larger gas turbines are required. \fc{For these types of gas turbines, which produce more than $750$ MW in combined cycles at more than $62\%$ efficiency, and which can be supplied with sustainably produced H$_2$ (hydrogen) that is blended with natural gas, the silo-type and annular combustor designs become uneconomical due to several engineering trade-offs. Therefore,} modern high-efficiency H-class gas turbines exclusively feature \textit{can-annular} combustor architectures. In this type of system, combustion takes place in a number (typically 12 or 16) of cans evenly distributed along the turbine annulus. The cans are thermodynamically decoupled from each other, but the annular turbine inlet is shared by all can outlets and allows for acoustic cross-talk between neighboring cans. Theoretical, numerical and experimental studies performed at Siemens \cite{bethke2002thermoacoustic,krebs2005thermoacoustic,Kaufmann08,Farisco17}, General Electric \cite{Bethke19,MOON2020178,MOON2021295} and Ansaldo Energia Switzerland \cite{ghirardo18,ghirardo2020effect} are testament to the increased interest of industry in the thermoacoustics of can-annular combustors. In Ref. \cite{ghirardo18}, so-called Bloch modes, which describe wave-like \fc{phase} patterns of the acoustic pressure field along the \fc{turbine annulus}, are discussed. \fc{In Fig. 8 therein, the authors provide indirect experimental evidence of such Bloch modes in a real-world gas turbine by decomposing measured acoustic pressure signals from different cans into Bloch mode components using the discrete Fourier transform. Direct evidence of Bloch modes occurring in a four-can system, showing the synchronized state and a wave-like phase pattern along the annulus, is presented in Figs. 5 and 6 of Ref. \cite{MOON2020178}, respectively.} The theoretical studies in Refs. \cite{VONSALDERN20216145,von2021analysis} use periodic \fc{(Bloch-type)} boundary conditions to simplify the analysis of reduced-order models of can-annular combustors. \tr{The effect of resistive and reactive coupling on the linear stability of a ring of thermoacoustic oscillators is investigated in Ref. \cite{pedergnana2021coupling}.} We also mention that nonlinear phenomena in can-annular configurations, such as amplitude death and quenching, have drawn increased interest in recent years \cite{Biwa2015,Thomas18AmpDeathinCoup,Hyodo2018,Dange2019}.

In this work, we combine analytical and numerical tools to study a low-order model of a can-annular combustor. Our goal is to gain a qualitative understanding of \fc{some of the system's} dynamic features that are independent of the number of cans $N$. \fc{In particular, we are interested in how, in the steady state, the amplitude and phase statistics of the acoustic pressure in the cans are related to the Bloch modes observed in real-world gas turbines.} \tr{For this, we propose a symmetric model where all reactive effects of the flame response and the aeroacoustic coupling are neglected. We show that this simple model is able to describe a wide variety of possible emergent dynamics, including synchronization, rotating waves as well as quasi-steady superpositions of clockwise (CW) and counterclockwise (CCW) waves, and that it reproduces the intermittent energy transfer between different Bloch mode components of the acoustic pressure observed in real-world gas turbines.}

A sketch of the physical system we consider is shown in Fig. \ref{Figure 1}\textbf{(a)} from a radial view. Shown \fc{are} the control volume $\mathcal{C}_j$, the acoustic pressure at the can outlet $\eta_j$, the aperture width $W$, the can length $L$ and the coherent \fc{(at the acoustic frequency)} heat release fluctuations of the flame $q_j$, which characterize the flame response to acoustic perturbations. Figure \ref{Figure 1}\textbf{(b)} shows an abstraction the same system for $N=12$ cans from an axial perspective, the amplitude $A_j$ and phase $\varphi_j$ of the acoustic pressure at the can outlet $\eta_j$, the linear resistive coupling $\lambda$ and the reactive coupling $\mu$. \tr{For simplicity,} reactive coupling is neglected in this study. We use the convention that a positive increment in $j$ implies a CW rotation by \fc{an angle of} $2\pi/N$ along the ring, and that $j=1$ corresponds to the oscillator at the $12/N$ o'clock position. The turbulent wake in the aperture between neighboring cans is sketched in Fig. \ref{Figure 1}\textbf{(c)}. Shown is a typical profile of the mean axial velocity in the aperture, the mean axial bulk flow speed of the combustion products $U_\mathrm{tot}$ and the turbulent wake, which is bounded by two shear layers.

\fc{In our model, the flame drives a single natural (longitudinal) eigenmode $\psi_0$ of the can with corresponding eigenfrequency $\omega_0$. In a first approximation, we assume that the mode shape of $\psi_0$ is unperturbed by the thermo- and aeroacoustic interactions and that the acoustic pressure signal is close to harmonic. These are often reasonable assumptions in practice \cite{Culick19711,Lieuwen20033JSV}. Following the low-order modeling approach presented in the latter two references as well as Ref. \cite{Noiray16}, the thermoacoustic instability in each can is described by a Van der Pol oscillator with linear growth rate $\nu$ and nonlinear saturation $\kappa$. For simplicity, all oscillators are assumed to be identical. For $\omega\approx\omega_0$, where $\omega=2\pi f$ and $f$ is the frequency of oscillation, the acoustic pressure field $p_j$ in the $j^\mathrm{th}$ can is approximated by the unimodal projection}
\begin{equation}
    p_j(\fc{x},t)\approx \eta_j(t)\psi_0(\fc{x}),
\end{equation}
where $\eta_j$ is the dominant modal amplitude. It is \fc{the} dynamic variable of the Van der Pol oscillator. \fc{Following \cite{YOON2020115774}, we assume a pressure antinode at the turbine inlet $\psi_0(L)=1$ (see Fig. \ref{Figure 1}\textbf{(a)}). In this case, $\eta_j$ is equivalent to the acoustic pressure \fc{$p_j$} at the can outlet. }

The oscillators are driven by white noise, which represents broadband, quasi-random acoustic perturbations induced by turbulent fluctuations of the mean \fc{axial} flow in the cans. 

The thermoacoustic growth rate $\nu$ can be further decomposed into \fc{a sum of the} acoustic gains from the flame response and \fc{dissipative losses at the boundary of the enclosure}, \fc{each of which may} be measured separately \cite{Boujo2016245}. The parameter $\kappa$ describes the saturation of the coherent response of the flame to acoustic perturbations: for $\nu>0$, a small acoustic perturbation is amplified by the flame and grows, until the response saturates at \fc{a} point where the acoustic gain from the flame equals the \fc{dissipative losses}, and a limit cycle is reached.

\subsection{Aeroacoustic coupling between the cans \label{Section: aeroac. coupling}}
The cans are acoustically coupled through the apertures at the turbine inlet. The coupling is strongly influenced by the mean axial \fc{bulk} flow speed of the combustion products  $U_\mathrm{tot}$. When a sound wave with frequency $f$ from inside the can is reflected at the coupling aperture of width $W$, the acoustic (compressible) velocity field inside the can interacts with the hydrodynamic (incompressible) fluid motion in the aperture. As a consequence of this interaction, depending on the value of the nondimensional Strouhal number $fW/U_\mathrm{tot}$, either destructive or constructive interference can occur.

On the hydrodynamic side, the acoustic pressure fluctuations can lead to shear layer flapping \fc{in the turbulent wake between the cans (see Fig. \ref{Figure 1}\textbf{(c)})}, which causes vorticity fluctuations that generate sound according to Howe's acoustic energy corollary \cite{howe1980dissipation}. For a study on sound production by shear layer flapping in a lab-scale aeroacoustic system (a whistling beer bottle under grazing flow over the bottle top) the reader is referred to Ref. \cite{BOUJO20}. In extreme cases, at high pressure amplitudes, the shear layer can roll up and form discrete vortices which are shed periodically from the upstream edge of the aperture and produce sound when they meet the downstream edge. \fc{Numerical studies of sound generation by vortex shedding in side-branch apertures can be found in Refs. \cite{Bauerheim20,Ho2021}.} 
 
The main consequence of the acoustic-hydrodynamic interaction described above is the following. For fixed aperture width $W$ and frequency \fc{$f$}, depending on the mean flow speed $U_\mathrm{tot}$, the sound field may lose or gain acoustic energy from the interaction with the grazing mean flow \fc{over} the aperture. Experimental and numerical studies highlight also the role of the nonlinear saturation of the aeroacoustic response of a shear layer \cite{bourquard_faure-beaulieu_noiray_2021,boujo_bauerheim_noiray_2018}: \fc{When the acoustic forcing amplitude is increased, the response of the shear layer, in terms of its kinetic energy or its acoustic energy production at the fundamental frequency of excitation, decreases. A theoretical model of this nonlinear shear layer response, based on Howe's formulation, was derived and validated against experiments in \cite{Pedergnana2021}.} The same saturation mechanism leads to self-excited aeroacoustic instabilities, which occur, for example, when we whistle. The first modern study of this type of instability \fc{was conducted by} Sondhauss in the $19^\mathrm{th}$ century \cite{sondhauss1852ueber}. Readers interested in the role of aeroacoustic instabilities in whistling and musical instruments are referred to the classic experiments of Wilson \cite{wilson1971experiments} as well as \fc{the more recent review of Fabre et al.} \cite{fabre2012aeroacoustics}. Aeroacoustic instabilities can also occur in industrial machines, where they cause noise pollution and fatigue damage of components \cite{rockwell1979self}. \fc{A discussion of the aeroacoustic instability mechanism from an industrial perspective, with a focus on mitigation measures and design rules, is found in \cite{ziada2014flow}.}
 
The acoustic coupling is incorporated in the oscillator model by linear and \fc{Van der Pol-type} nonlinear resistive \fc{(damper-like)} terms. The linear terms are analogous to damper and resistor elements in mechanical and electrical oscillators, respectively. 

This linear resistive coupling describes the small-amplitude aeroacoustic response of the turbulent wake in the aperture to pressure differences between neighboring cans. Consistent with the low-order model of an aeroacoustic instability validated against experiments in Ref. \cite{bourquard_faure-beaulieu_noiray_2021}, we include the nonlinear saturation of the aeroacoustic response with increasing pressure amplitude by adding a quadratic term in the pressure difference to the linear coupling \fc{$\lambda$}. When the nonlinear coupling term is neglected, the governing equations of our model describe a network of nonlinear oscillators with linear coupling. Such systems are ubiquitous in classical mechanics \cite{gendelman2001energy,vakakis2001energy,coulombe2017computing,chakraborty2018entanglement}, chemistry \cite{neu1979coupled,bar1985stability,crowley1989experimental} and biology \cite{Turing52,Linkens76,collins1993coupled}. 
 
 \subsection{\fc{Literature review} \label{Relation to previous research}}
The study of synchronization and rotating waves in rings of limit cycle oscillators is not novel. We give below a review of related analyses and discuss \fc{how our work differs from these previous efforts.}
A pair of coupled oscillators is a limiting case of a ring (see, for instance,  \cite{OnCoupledOscillatorsRichie66,SeriesConnectedDeeley72,storti1982dynamics,wirkus2002dynamics,low2003investigation}). For compactness, we do not include these works in the review below, \fc{but we note that our results apply also to this special case.}

We find the first modern study of rings of coupled Van der Pol oscillators in the work of Endo and Mori from the 1970s \cite{ModeAnalysis78}, wherein the authors classify the limit cycle solutions in a symmetric ring of a large number of oscillators with resistive coupling. Deterministic averaging is performed to identify limit cycle amplitudes. All limit cycle solutions are found to be linearly stable, and this is confirmed by experiments on an electrical circuit with 4 elements. 
Their study is continued in \cite{MoriDelay81}, where the effect of delay in the coupling term is studied. It is found that the addition of such a delay can lead to instabilities of certain modes, which is confirmed experimentally. 

A group-theoretical approach to the classification of phase-locked solutions in rings of coupled oscillators is taken in \cite{collins1994group}. 

In Ref. \cite{Endo98DeviationOfElementValues}, averaging, linear stability analysis, simulations and experiments are used to study the effect of differing parameters in members of rings of 3 and 4 oscillators. In \cite{nana2006synchronization}, theory and experiments are combined to study the effect of external forcing and non-uniform parameter distribution on the synchronization of a 4-member ring of Van der Pol oscillators. Following up on their earlier work in Ref. \cite{Barron2013}, in \cite{BARRON201662}, the authors study the effect of a non-uniform distribution of the reactive coupling parameter along the ring on the system stability by transforming the system into a Hill equation. They show that a harmonic variation of the reactive coupling parameter along the ring can affect the linear stability of limit cycles. We mention also Ref. \cite{Takashi16}, where it is shown that particular asymmetric parameter distributions can favor the stability of the perfectly symmetric, synchronized solution in a ring of nonidentical limit cycle oscillators.

An approach based on Floquet analysis of Hill's equation is taken by \cite{enjieu2007spatiotemporal} to study the stability of synchronized states in a large ring of Van der Pol oscillators. Therein, the authors put emphasis on linking their results to intercellular dynamics in biology. Concepts from neuroinformatics are invoked in Ref. \cite{uwate2010complex}, focusing on synchronization phenomena in a ring of Van der Pol oscillators coupled by time-varying resistors.

In Ref. \cite{emenheiser2016patterns}, the authors begin by characterizing limit cycle solutions in a ring of (linearly) resistively coupled limit cycle oscillators. They perform a linear stability analysis and derive a quasi-potential (a candidate function for Lyapunov's second method of stability \cite{lyapunov1992general}) to determine the nonlinear stability of these solutions. After a change to amplitude-phase coordinates, deterministic averaging is applied to eliminate the fast oscillating terms. Subsequently, external noise forcing is added to the phase equation to numerically investigate stochastic effects on the synchronization between the oscillators. Using a small-noise approximation, the authors obtain analytical formulas for the mean time of a state to leave an attractor basin. Because the phase noise is added phenomenologically after deterministic averaging is carried out on the original\fc{, fast} system, it is unclear how Eqs. (11) and (12) relate to Eq. (1) in Ref. \cite{emenheiser2016patterns}. 

\fc{In this work, we go further than previous studies by accounting for nonlinearity in the (purely resistive) coupling, in combination with additive stochastic forcing of the fast variables, which is consistently included in the averaging procedure. To the authors' knowledge, this case, which leads to novel and unexpected results, has not been investigated so far.}

\fc{We} note that noise-induced phase synchronization of limit cycle oscillators without a ring-like topology has been studied extensively before (see, for instance, \cite{pikovskii1984synchronization,teramae2004robustness,goldobin2005synchronization,teramae2006noise,chik2006noise,nagai2010noise,nakada2012noise,kurebayashi2012colored,lai2013noise,kawamura2014collective,kawamura2016optimization,pimenova2016interplay,vaidya2021using}). Instabilities of azimuthal waves in a discrete, ring-shaped fluid-dynamical system are \fc{also} encountered in Refs. \cite{couchman_turton_bush_2019,couchman_bush_2020}, where theoretical and experimental methods are combined to study of the dynamics in a ring of bouncing droplets. Amplitude death and phase-flip bifurcations between in-phase and anti-phase synchronization in a system of coupled chemical oscillators are investigated in \cite{manoj2018experimental}.

Due to the symmetry of the system considered in this work, we find that \fc{different} emergent patterns \fc{(synchronized state, rotating waves  etc.)} occur at similar amplitudes and are mainly distinguished by different phase dynamics. \fc{In regard to this, we} mention Adler's equation for phase injection locking \cite{mirzaeisep2006analysis,atkinsondec2006seed,Munsberg2020} and Kuromoto's classic model \cite{kuramoto1975self}, which has become a paradigmatic example of phase synchronization in coupled oscillators \cite{yeung1999time,acebron2005kuramoto,rodrigues2016kuramoto}.

\subsection{Overview}
In Sec. \ref{Section 2: Model dynamics}, we \fc{present the model and derive the potential governing the amplitude and phase dynamics.} In Sec. \ref{Section 3}, numerical experiments on the deterministic, noise-free system are performed and analyzed. In Sec. \ref{Section 4}, the steady state statistics of the noise-driven system are derived and studied. Selected results and topics for future research are discussed in Sec. \ref{Section 5}. In Sec. \ref{Section 6}, we summarize our conclusions.

\section{\fc{Theoretical model} \label{Section 2: Model dynamics}}
\fc{In an isolated can, the interaction of a single, dominant acoustic mode with the flame can be represented by a self-excited (Van der Pol) oscillator \cite{Culick19711,Lieuwen20033JSV,Noiray16}. Such} low-order oscillator models of thermoacoustic instabilities in silo-type combustors are well understood in terms of their accuracy compared to more detailed \fc{modeling} approaches and experimental results (see, for instance, Fig. 17 in Ref. \cite{BONCIOLINI2021396}). \fc{As shown in the latter reference,} the linear growth rate $\nu$ and the nonlinear saturation $\kappa$ can be adjusted to obtain a qualitative approximation of the experimentally measured phase space spanned by the acoustic pressure $p$ and its time derivative $\dot{p}$. Our model combines $N$ of such oscillators to approximate the acoustic pressure dynamics in a can-annular combustor. 

Following Ref. \cite{bourquard_faure-beaulieu_noiray_2021}, we account for the aeroacoustic interaction between neighboring cans \fc{with the} linear resistive coupling $\lambda$ \fc{and its} nonlinear saturation $\vartheta$.

Stochastic forcing of the acoustic field by broadband noise production in the turbulent flow in the cans is modeled by additive, zero-mean white noise signals $\xi_j$ of equal intensity $\Gamma$ with the autocorrelation functions \fc{$\langle \xi_j \xi_{j,\tau}\rangle_\mathbb{R}=\int_\mathbb{R} \xi_j(t)\xi_j(t+\tau) dt=\Gamma\delta(\tau)$, where $\langle\cdot\rangle_\mathbb{R}$ denotes the time integral over the real line $[-\infty,\infty]$ and $\delta$ is the Dirac delta function.}

Based on the above discussion on low-order models of thermo- and aeroacoustic instabilities, we \fc{propose} the following model of $N$ coupled Van der Pol oscillators \fc{for the acoustic pressure dynamics in the can-annular combustor}:
\begin{eqnarray}
     \ddot{\eta}_{j}+\omega_{0,j} ^2{\eta}_j &=&2\nu\dot{\eta}_j-\kappa \eta_j^2\dot{\eta}_j+ \lambda(\dot{\eta}_{j+1}+\dot{\eta}_{j-1}-2\dot{\eta}_j)\nonumber\\
     &&+\vartheta\big[({\eta}_{j+1}-{\eta}_j)^2(\dot{\eta}_{j+1}-\dot{\eta}_j)\nonumber\\
     &&+({\eta}_{j-1}-{\eta}_j)^2(\dot{\eta}_{j-1}-\dot{\eta}_j)\big]+\xi_j, \label{Dynamics of dominant modal amplitude with swirl, nonlinear with noise}
\end{eqnarray}
\fc{for $j=1,\ldots,N$ and $j+k\equiv\mathrm{mod}(j+k,N)$, $k\in\mathbb{N}$.} The sign before $\vartheta$ is positive because the coupling is amplifying if $\lambda<0$ \fc{and we assume that, as in Ref. \cite{Pedergnana2021}, the constructive aeroacoustic coupling saturates at high amplitudes (monotonic decay of constructive feedback strength). For simplicity, this work focuses on situations where the nonlinear aeroacoustic coupling can be considered as a small perturbation of the thermoacoustic oscillations in the cans, so that the saturation associated with the thermoacoustic instability is significantly larger than the saturation of the aeroacoustic coupling: $\kappa\gg \vartheta$. We exclude the case of vanishing resistive coupling in our study, because a vanishing linear response $\lambda=0$ with nonzero saturation $\kappa\neq 0$ is not physical.} The parameter values used in the numerical examples throughout this work are listed in Table \ref{Table 1}. If multiple values are used, the default value is shown in bold. It is indicated below if values different from the default ones are used.

\begin{table*}
\caption{Parameter values used in the numerical examples throughout this work. If multiple values are used, the default value is shown in bold.}
\label{Table 1}       
\center
\begin{tabular}{lll}
\hline\noalign{\smallskip}
Parameter &Meaning & Value \\
\noalign{\smallskip}\hline\noalign{\smallskip}
$\omega_0/2\pi$ & Frequency of dominant acoustic mode & $255$ Hz\\
$\nu$ & Thermoacoustic growth rate& $\{\boldsymbol{25},50\}$ s$^{-1}$\\
$\kappa$ & Thermoacoustic saturation & $0.48$ s$^{-1}$ Pa$^{-2}$ \\
$G=\kappa \Gamma/8|\nu|^2 \omega_0 ^2 $ & Normalized white noise intensity & $2.88\times \{\boldsymbol{10^{-1}},10^1,10^3\}$\\
$\Lambda=2\lambda/\nu$ & Normalized linear resistive coupling & $\in [-1,2]$ \\
$K=16\vartheta/\kappa$ & Normalized coupling nonlinearity & $\{0, 1/6, \boldsymbol{1/3}\}$  \\
$N$ & Number of oscillators & $\{3,\boldsymbol{12}$\}\\
$b$ & Bloch wavenumber & $\{0,\pm 1,\ldots,\pm \mathrm{floor}(N/2)\}$\\
$\beta$ & Continuous Bloch wavenumber & $\in[-N/2,N/2]$\\
$B=\pi \beta/N$ & Normalized Bloch wavenumber & $\in[-\pi/2,\pi/2]$\\
\noalign{\smallskip}\hline
\end{tabular}
\end{table*}
 
 We define the coordinate change to amplitude-phase variables $\{A_j,\phi_j\}$, $j=1,\ldots,N$, as \fc{follows:}
\begin{eqnarray}
     A_j&=&\sqrt{\eta_j^2+\left(\dot{\eta}_j/\omega_0\right)^2}, \label{Amplitude}\\
     \varphi_j&=&-\arctan(\dot{\eta}_j/\omega_0,\eta_j)-\omega_0t, \label{Phase}
\end{eqnarray}
where $\eta_j=A_j\cos{\phi_j }$, $\dot{\eta}_j=-\omega_0A_j\sin{\phi_j }$, $\phi_j=\varphi_j+\omega_0 t$ and $\arctan(y,x)$ is the two-argument arctangent function. Assuming that \textbf{(A)} the variation of $A_j$ and $\varphi_j$ over an acoustic period is negligible and \textbf{(B)} the oscillators perform quasi-sinusoidal motion, deterministic and stochastic averaging (see Refs. \cite{krylov1950introduction,sanders2007averaging} and \cite{stratonovich1963topics,roberts1986stochastic}, respectively) can be performed on Eq. \eqref{Dynamics of dominant modal amplitude with swirl, nonlinear with noise} to eliminate the fast-oscillating terms. This is done in Appendix \ref{Appendix A} and yields the slow-flow dynamics in \fc{amplitude-phase} form:
\begin{eqnarray}
     &&\dot{A}_j=A_j\bigg(\nu-\frac{\kappa A_j^2}{8}\bigg)+\frac{\lambda}{2}\big[A_{j+1}\cos{(\Delta_{j+1})}+A_{j-1}\nonumber\\
     &&\cos{(\Delta_{j})}-2A_j\big]+\frac{\vartheta}{8}\bigg[(A_{j+1}^3+3A_j^2 A_{j+1})\cos(\Delta_{j+1})-\nonumber\\
     &&A_j A_{j+1}^2\cos(2\Delta_{j+1})+(A_{j-1}^3+3A_j^2 A_{j-1})\cos(\Delta_{j})-\nonumber\\
     &&A_j A_{j-1}^2\cos(2\Delta_{j})-2(A_j^3+A_j A_{j+1}^2+A_j A_{j-1}^2)\bigg]\nonumber\\
     &&+\frac{\Gamma}{4\omega_0^2 A_j}+\zeta_j \label{Amplitude dynamics, slow-flow}
\end{eqnarray}
and
\begin{eqnarray}
     &&\dot{\varphi}_j=\frac{\lambda}{2}\left[\frac{A_{j+1}}{A_j}\sin(\Delta_{j+1})-\frac{A_{j-1}}{A_j}\sin(\Delta_{j})\right]\nonumber\\
     &&+\frac{\vartheta}{8}\bigg[\Big(A_j A_{j+1}+\frac{A_{j+1}^3}{A_j}\Big)\sin(\Delta_{j+1})-A_{j+1}^2\nonumber\\
     &&\sin(2\Delta_{j+1})-\Big(A_j A_{j-1}+\frac{A_{j-1}^3}{A_j}\Big)\sin(\Delta_{j})\nonumber\\
     &&+A_{j-1}^2\sin(2\Delta_{j})\bigg]+\frac{\chi_j}{A_j}, \label{Phase dynamics, slow-flow}
\end{eqnarray}
for $j=1,\ldots,N$, where $\Delta_j=\varphi_j-\varphi_{j-1}$ \fc{is the phase difference} and $\zeta_j$ and $\chi_j$ are $2N$ uncorrelated white noise sources with the same noise intensity $\Gamma/2\omega_0^2$, respectively. \fc{The above definitions of $\varphi_j$ and $\Delta_j$ were used in the derivation of the slow-flow system, given by Eqs. \eqref{Amplitude dynamics, slow-flow} and \eqref{Phase dynamics, slow-flow}. For the presentation of our results in Secs. \ref{Section 3} and \ref{Section 4}, we set $\varphi_j\equiv \mathrm{mod}(\varphi_j,2\pi)$ and $\Delta_j\equiv \mathrm{mod}(\Delta_j+\pi,2\pi)-\pi$. These operations, which leave Eqs. \eqref{Amplitude dynamics, slow-flow} and \eqref{Phase dynamics, slow-flow} unchanged, are equivalent to adding integer multiples of $\pm 2\pi$ to $\varphi_j$ and $\Delta_j$ so that the sums lie in the domains $[0,2\pi]$ and $[-\pi,\pi]$, respectively.} 

The slow flow system \eqref{Amplitude dynamics, slow-flow} and \eqref{Phase dynamics, slow-flow} is the main focus of this work. It has the form of a \fc{$2N$-dimensional} Langevin equation \cite{risken1996fokker}:
\begin{eqnarray}
    &&\dot{A}_j=-\frac{\partial}{\partial A_j}\mathcal{V}(A_m,\varphi_m)+\zeta_j, \label{Langevin amplitude}\\
     &&A_j\dot{\varphi}_j=-\frac{\partial}{A_j\partial\varphi_j}\mathcal{V}(A_m,\varphi_m)+\chi_j, \label{Langevin phase}
\end{eqnarray}
where $m=1,\ldots,j,\ldots,N$, $j=1,\ldots,N$ and the potential $\mathcal{V}$ is defined as follows:
\begin{eqnarray}
&&\mathcal{V}(A_m,\varphi_m)=\sum_{l=1}^N \bigg(-\frac{\nu A_l^2}{2}+\frac{\kappa A_l^4}{32}-\frac{\lambda A_l}{2}\big[A_{l+1}\nonumber\\
&&\cos(\Delta_{l+1})-A_l\big]-\frac{\vartheta A_l}{8}\Big[(A_{l+1}^3+A_l^2 A_{l+1})\cos(\Delta_{l+1})\nonumber \\
&&-\frac{A_l A_{l+1}^2}{2}\cos(2\Delta_{l+1})-\frac{A_l}{2}(A_{l}^2+2A_{l+1}^2)\Big]\nonumber \\
&&-\frac{\Gamma}{4\omega_0^2}\ln(A_l)\,\bigg). \label{Potential for amplitudes and phases}
\end{eqnarray}
Equations \eqref{Langevin amplitude} and \eqref{Langevin phase} imply that trajectories $(A_j,\varphi_j)$, $j=1,\ldots,N$, are stochastically perturbed and attracted towards lower values of the potential $\mathcal{V}$ defined in Eq. \eqref{Potential for amplitudes and phases}.

\fc{In the deterministic case, trajectories of the fast and averaged systems are pointwise close to each other over time, and the degree of closeness is determined by how strongly the averaging assumptions \textbf{(A)} and \textbf{(B)} are satisfied \cite{krylov1950introduction,sanders2007averaging}. For a linear damped harmonic oscillator, the averaged system is equivalent to the fast system.} In contrast to averaging in deterministic systems, due to the lack of smoothness, we do not expect trajectories of the stochastically averaged system to be pointwise close to those of the fast system, but for the joint probability density function (PDF) $\mathcal{P}(\eta_m,\dot{\eta}_m/\omega_0,t)$, $m=1,\ldots,N$, \fc{computed from the fast and the averaged system, given by Eq. \eqref{Dynamics of dominant modal amplitude with swirl, nonlinear with noise} and Eqs. \eqref{Amplitude dynamics, slow-flow} and \eqref{Phase dynamics, slow-flow}, respectively,} to converge \fc{in the steady state, i.e., for} $t\rightarrow\infty$. \\
\indent Several \fc{theorems on stochastic averaging exist \cite{stratonovich1963topics,roberts1986stochastic}}. \fc{It is beyond the scope of this work to elaborate on these mathematically intricate results, which would require introducing additional notation and theory to understand under which conditions they apply to the fast system \eqref{Dynamics of dominant modal amplitude with swirl, nonlinear with noise}. Instead,} it is accepted that stochastic averaging is justified by a combination of physical and mathematical arguments, and careful numerical validation is performed to ensure that, \fc{for the parameter range considered,} the \fc{joint PDFs of the fast system \eqref{Dynamics of dominant modal amplitude with swirl, nonlinear with noise} and the slow amplitude-phase system \eqref{Amplitude dynamics, slow-flow} and \eqref{Phase dynamics, slow-flow} indeed coincide over long time spans $[0,t_\mathrm{end}]$, $t_\mathrm{end}\gg \mathrm{max}(1/\nu,\tau_{\xi_j},\tau_{A_j},\tau_{\varphi_j})$, where $1/\nu$ is the relaxation time ($\sim$ time is takes to reach the steady state) and $\tau_{a}$ is the correlation time of the signal ``$a$'', defined as $\tau_a= \int^{\infty}_{0^+} \langle \xi_j \xi_{j,\tau}\rangle_\mathbb{R} d\tau$ (see Ref. \cite{stratonovich1963topics}, p. 65, Eq. (3.61) with $\tau=t_2-t_1$), where $(\cdot)^+$ denotes the one-sided limit from above. This validation ensures that any qualitative statistical observations from long time series of the averaged system carry over to the fast system.} 

We \fc{note that, in real-world experiments and in numerical simulations, the signals $\xi_j$, $\zeta_j$ and $\chi_j$ will generally have nonzero correlation times. However, because we model $\xi_j$, $\zeta_j$ and $\chi_j$ as white noise signals with (theoretically) vanishing correlation times, we assume throughout this work that $1/\nu\gg\mathrm{max}(\tau_{\xi_j},\tau_{A_j},\tau_{\varphi_j})$, i.e., that the time to reach the steady state significantly exceeds the correlation time of the external stochastic forcing.}

\section{Potential landscape of the deterministic system \label{Section 3}}
In the absence of noise, $\Gamma=0$. Under the averaging assumptions \textbf{(A)} and \textbf{(B)}, the limit cycles of the deterministic part of the fast system (Eq. \eqref{Dynamics of dominant modal amplitude with swirl, nonlinear with noise} with $\xi_j\equiv 0$) are given by fixed points $\{\dot{A}_j=0,\dot{\varphi}_j=0\}$ of the slow-flow system \eqref{Amplitude dynamics, slow-flow} and \eqref{Phase dynamics, slow-flow} \fc{with $\zeta_j\equiv\chi_j\equiv 0$.} We find that these fixed points describe solutions with uniform amplitude along the ring and equal phase difference between neighboring oscillators:
\begin{eqnarray}
A_j&=&A_b^*,  \label{Uniform amplitude}\\
\Delta_j&=&\Delta_b^*\quad\forall j. \label{Phase-locked phase difference}
\end{eqnarray}
The periodicity condition $\varphi_j\equiv \varphi_{j+N}$ imposes the following additional restriction on the phase difference:
\begin{equation}
    \Delta_b^*=2\pi b/N,\quad b=0,\pm1,\ldots,\pm\mathrm{floor}(N/2). \label{Periodicity restriction on phase difference}
\end{equation} 
We call the solutions given by Eqs. \eqref{Uniform amplitude}-\eqref{Periodicity restriction on phase difference} Bloch modes and $b$ the Bloch wavenumber because the present result's similarity to Felix Bloch's classic theory of electron dynamics in perfectly periodic crystal lattices \cite{bloch1929quantenmechanik}. \fc{For compactness, we also introduce the continuous and normalized Bloch wavenumbers $\beta\in[-N/2,N/2]$ and $B=\pi \beta/N\in[-\pi/2,\pi/2]$, respectively.} 

Solutions with $|b|>0$ \fc{appear as} degenerate rotating waves along the ring \cite{emenheiser2016patterns}. \fc{At the limit cycle, $\varphi_j=\mathrm{const.}\hspace{0.1cm}\forall j$ and $\eta_j(t)=A^*_b \cos(\omega_0 t+\varphi_1 + (j-1)\Delta_b^*)$, so that, for $\Delta_b^*>0$, under a negative increment in $j$ (CCW shift along the ring) and a suitably chosen positive increment in $t$, the acoustic pressure $\eta$ remains constant. This implies that positive values of $b$ ($\Delta_b^*>0$) correspond to CCW rotating waves and negative values to CW rotating waves. In the absence of explicit asymmetries, both CW and CCW rotating waves have the same amplitude and frequency.} The case $b=0$ is the synchronized solution, \fc{for which all} oscillators are in phase. For even $N$, we call the solution with $b=N/2$ the ``push-pull mode'', which features a phase difference of $\pi$ between neighboring oscillators. 

\fc{We note that whereas the spinning direction the nodal lines of an apparent wave is determined by the sign of $b$, due to the finite number of members in the oscillator ring $N$, if $\mathrm{mod}(N,|b|)\neq 0$, there appear to be less nodal lines than expected (a Bloch mode with wavenumber $b$ has exactly $|b|$ nodal lines) and they appear to spin in the opposite direction than expected from the sign of $b$. This visual phenomenon is a consequence of a spatial analogue of the wagon-wheel effect \cite{Purves19963693} and is discussed more in Sec. \ref{Section 5}.}

\fc{For vanishing coupling $\lambda=\vartheta=0$, the system \eqref{Dynamics of dominant modal amplitude with swirl, nonlinear with noise} corresponds to $N$ decoupled Van der Pol oscillators, whose limit cycle solution $\{A_j\equiv\sqrt{8\nu/\kappa},\varphi_j=\mathrm{const.}\}$, $j=1,\ldots,N$, is asymptotically stable \cite{Noiray16}. By the hyperbolicity of this limit cycle and the implicit function theorem \cite{implicit}, linear stability persists for small enough coupling parameters $\{\lambda,\vartheta\}$, which act as small perturbations of the decoupled system. Indeed, for small enough $\lambda$, all fixed points are linearly stable. This can be seen by linearizing Eqs. \eqref{Amplitude dynamics, slow-flow} and \eqref{Phase dynamics, slow-flow} around the fixed points defined in Eqs. \eqref{Uniform amplitude}-\eqref{Periodicity restriction on phase difference} by writing $A_j\approx A_b^*+A'_j$ and $\Delta_j\approx\Delta_b^*+\Delta'_j$, where $A'_j$ and $\Delta'_j$ are small perturbations of the limit cycle amplitude \eqref{Uniform amplitude} and the phase difference \eqref{Periodicity restriction on phase difference}. Applying periodic boundary conditions to the linearization (see Refs. \cite{ghirardo18,VONSALDERN20216145,von2021analysis}) yields
\begin{eqnarray}
    \dot{A}_j'&=&-2 [\nu-2\lambda\sin(\pi b/N)^2 ]A'_j,\label{Linearization of amplitude eq. around limit cycles}\\
    \dot{\varphi}_j'&=&0.
\end{eqnarray}
For given $b$ and $\lambda<\nu/2 \sin(\pi b/N)^2$, the linear stability of the corresponding limit cycle demonstrated by Eq. \eqref{Linearization of amplitude eq. around limit cycles} shows that all Bloch modes are local minima of the potential $\mathcal{V}$ given by Eq. \eqref{Potential for amplitudes and phases}. Therefore, for small enough linear resistive coupling $\lambda$, trajectories may converge to any of these solutions, dependent on their initial conditions.}

\begin{figure}[h!]
\begin{psfrags}
\psfrag{a}{$0$}
\psfrag{c}{$2\pi$}
\psfrag{e}{$\varphi_j$}
\psfrag{l}{\hspace{-0.2cm}$\Delta_0^*=0$}
\psfrag{m}{\hspace{-0.17cm}$\Delta_{-1}^*=-\frac{2\pi}{3}$}
\psfrag{n}{\hspace{-0.2cm}$\Delta_{1}^*=\frac{2\pi}{3}$}
\psfrag{o}{$b=4$}
\psfrag{p}{$b=5$}
\psfrag{q}{$b=6$}

    \centering
    \includegraphics[width=0.45\textwidth,center]{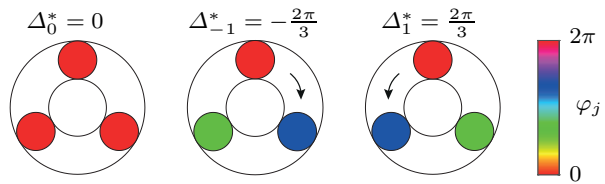}
    \end{psfrags}
    \caption{Illustration of the \fc{limit cycle} solutions corresponding to $b=0$ and $b=\pm1$ for $N=3$ oscillators. Shown are the \fc{values of the phase difference $\Delta^*_b$} and the \fc{phase distribution $\varphi_j$ along the ring.} A positive value of $b$ corresponds to a CCW rotating wave and a negative value to a CW rotating wave. The case $b=0$ is the synchronized solution, \fc{for which all} oscillators are in phase.}
    \label{Figure 2}
\end{figure}

For $N=3$ oscillators, the \fc{limit cycle} solutions corresponding to $b=0$ and $b=\pm 1$ are illustrated in Fig. \ref{Figure 2}. Shown are the \fc{values of the phase difference $\Delta^*_b$} and the \fc{phase distribution $\varphi_j$ along the ring.}

Substituting \fc{Eqs.} \eqref{Uniform amplitude}-\eqref{Periodicity restriction on phase difference} into the slow-flow system \eqref{Amplitude dynamics, slow-flow} and \eqref{Phase dynamics, slow-flow} yields the limit cycle amplitudes
\begin{eqnarray}
A^*_b=\sqrt{\frac{8\nu_b}{\kappa_b}}, \label{Amplitude fixed point}
\end{eqnarray}
where we have defined
\begin{eqnarray}
\nu_b&=&\nu-2\lambda \sin(\pi b/N)^2=\nu(1-\Lambda \sin(\fc{B})^2), \label{nu_b}\\
\kappa_b&=&\kappa+16\vartheta \sin(\pi b/N)^4=\kappa(1+K \sin(\fc{B})^4), \label{kappa_b}
\end{eqnarray}
\fc{$\Lambda=2\lambda/\nu$ is the normalized linear resistive coupling and $K=16\vartheta/\kappa$ the normalized coupling nonlinearity.} 

Limit cycles for a given value of the Bloch \fc{wavenumber} $b$ can only exist if 
\begin{equation}
    \lambda\leq\frac{\nu}{2\sin(\pi b/N)^2}, \label{Condition on coupling for limit cycles to exist}
\end{equation}
\fc{otherwise the amplitude $A^*_b$ defined in Eq. \eqref{Amplitude fixed point} does not exist. Indeed, as evidenced by Eq. \eqref{Linearization of amplitude eq. around limit cycles}, when} $\lambda$ exceeds this value for a given $b$, the corresponding rotating waves are linearly unstable. This occurs when the acoustic energy \fc{losses} from the response \fc{to small amplitude perturbations} of the turbulent wake in the apertures between the cans exceed the gain of acoustic energy from the coherent response of the flame. 

\fc{In the following, we frequently make use of the assumption that all oscillators converge to quasi-limit cycles so that the dynamics take place in the vicinity of a two-dimensional submanifold $\mathrm{P}(\Omega)$ of the $2N$-dimensional phase space $\Omega$, where
\begin{equation}
\Omega=\{A_1,\varphi_1,\ldots,A_N,\varphi_N\}, \label{Big omega phase space}
\end{equation}
defined by $\mathrm{P}(\Omega):\hspace{0.1cm}A_j\equiv A^*_{\beta},\hspace{0.1cm} \varphi_j \equiv \varphi_{j-1}+k(j)\pi\beta/N\hspace{0.1cm}\forall j$, where $k(j)\in\{-1,1\}$. These quasi-limit cycle solutions include the case of true limit cycles, which are fixed points of the slow-flow system \eqref{Amplitude dynamics, slow-flow} and \eqref{Phase dynamics, slow-flow}, for $k(j)\equiv 1$. However, we keep $k$ general because, as we show below, quasi-steady solutions with uniform amplitudes $A_j\equiv A^*_{\beta}\approx \mathrm{const.}$ and uniform \textit{absolute} phase differences $|\Delta_j|\equiv|\Delta_\beta^*|=|\pi \beta/N|\approx \mathrm{const.}$ play a crucial role in the nonlinear dynamics of the oscillator ring.} The projection of the potential $\mathcal{V}$ defined by Eq. \eqref{Potential for amplitudes and phases} onto the submanifold $\mathrm{P}(\Omega)$ reads
\begin{eqnarray}
\mathcal{V}_\mathrm{P}&=&\mathcal{V}(A_j\equiv A^*_\beta,\varphi_j\equiv\varphi_{j-1}+k(j)\Delta^*_\beta)\nonumber\\
&=&-2N\nu_b \left\lvert\frac{\nu_b}{\kappa_b}\right\rvert.\label{Potential at limit cycle}
\end{eqnarray}
Normalizing this result, and using the fact that \fc{for a thermoacoustic instability,} $\mathrm{sign}(\nu)=1$, yields
\begin{eqnarray}
\mathcal{U}_\mathrm{P}(B)&=&-(1-\Lambda\sin(B)^2) \left\lvert\frac{1-\Lambda\sin(B)^2}{1+K\sin(B)^4}\right\rvert,\label{Potential at limit cycle, normalized}
\end{eqnarray}
where $\mathcal{U}_\mathrm{P}=\kappa\mathcal{V}_\mathrm{P}/2N|\nu|^2$ is the normalized \fc{projection of the potential $\mathcal{V}$} onto the phase-locked quasi-limit cycle solutions. \fc{Note $\mathcal{U}_\mathrm{P}$ is independent of the signs in $k$ and is symmetric with respect to positive and negative $B$. We show below that $\mathcal{U}_\mathrm{P}$ provides a compact description of the (de-)synchronization transition in the parameter space which is independent of the number of oscillators. }

\begin{figure*}[h!]
\begin{psfrags}
\psfrag{a}{\hspace{0.15cm}$K=0$}
\psfrag{b}{\hspace{0.0088cm}$K=1/6$}
\psfrag{c}{\hspace{0.016cm}$K=1/3$}
\psfrag{d}{$\hspace{-0.05cm}\Lambda$}
\psfrag{e}{\hspace{-0.64cm}$2|B|=|\Delta_\beta^*|$}
\psfrag{f}{$-1$}
\psfrag{g}{$2$}
\psfrag{h}{$0$}
\psfrag{i}{$\hspace{0.12cm}\pi$}
\psfrag{j}{$\mathcal{U}_\mathrm{P}$}
\psfrag{k}{$0.95$}
\psfrag{l}{$-4$}
\psfrag{m}{$A_j$}
\psfrag{n}{$\sqrt{\frac{32\nu}{\kappa}}$}
\psfrag{o}{$\Delta_j$}
\psfrag{p}{$-\pi$}
\psfrag{q}{$\pi$}
\psfrag{r}{Cycles}
\psfrag{s}{$200$}
\psfrag{t}{$400$}
\psfrag{u}{$800$}
\psfrag{v}{$52/100$\hspace{0.4cm}\textbf{CCW}}
\psfrag{w}{$48/100$\hspace{0.4cm}\textbf{CW}}
\psfrag{x}{$97/200$}
\psfrag{y}{$52/97$}
\psfrag{z}{$43/97$}
\psfrag{A}{$2/97$}
\psfrag{B}{$\Delta_6^*=\pi$}
\psfrag{C}{$\Delta_5^*=\frac{5\pi}{6}$}
\psfrag{D}{$\Delta_4^*=\frac{2\pi}{3}$}
\psfrag{E}{$\hspace{-1.1cm}\Lambda_\mathrm{c}=-K$}
\psfrag{F}{\textbf{(a)}}
\psfrag{G}{\textbf{(b)}}
\psfrag{H}{\textbf{(c)}}
\psfrag{I}{\textbf{(d)}}
\psfrag{J}{\textbf{(e)}}
\psfrag{K}{\textbf{(f)}}
\psfrag{L}{\textbf{(g)}}
\psfrag{M}{\textbf{(h)}}
\psfrag{N}{\textbf{(i)}}
\psfrag{O}{\textbf{(j)}}
\psfrag{P}{\textbf{(k)}}
\psfrag{Q}{\textbf{(l)}}
\psfrag{R}{\textbf{(m)}}
\psfrag{S}{\textbf{(n)}}
\psfrag{T}{\hspace{-0.46cm}\bf Sync. state}
\psfrag{U}{\hspace{-0.32cm}\bf Push-pull}
\psfrag{V}{\hspace{0.05cm}$N=12$, $2\nu$}
\psfrag{W}{\hspace{0.2cm}$N=3$, $\nu$}
\psfrag{X}{$103/200$}
\psfrag{Y}{\hspace{-0.05cm}$\boldsymbol{\Lambda=0}$}
\psfrag{Z}{\hspace{-0.09cm}\bf LUW}
\psfrag{1}{$\sqrt{\frac{64\nu}{\kappa}}$}
\psfrag{2}{$600$}
\psfrag{3}{$1$}
\psfrag{4}{\textcolor{ablue}{$A_1$}}
\psfrag{5}{$5\pi/6$}
\psfrag{6}{$2\pi/3$}
\psfrag{7}{\bf \begin{tabular}{@{}c@{}}
Amplifying \\
coupling
\end{tabular}}
\psfrag{8}{\bf \begin{tabular}{@{}c@{}}
Dissipative \\
coupling
\end{tabular}}
\psfrag{9}{\hspace{-0.03cm}$\eta_1$}
\psfrag{0}{\hspace{-0.032cm}$2|B|$}

    \centering
    \includegraphics[width=0.75\textwidth,center]{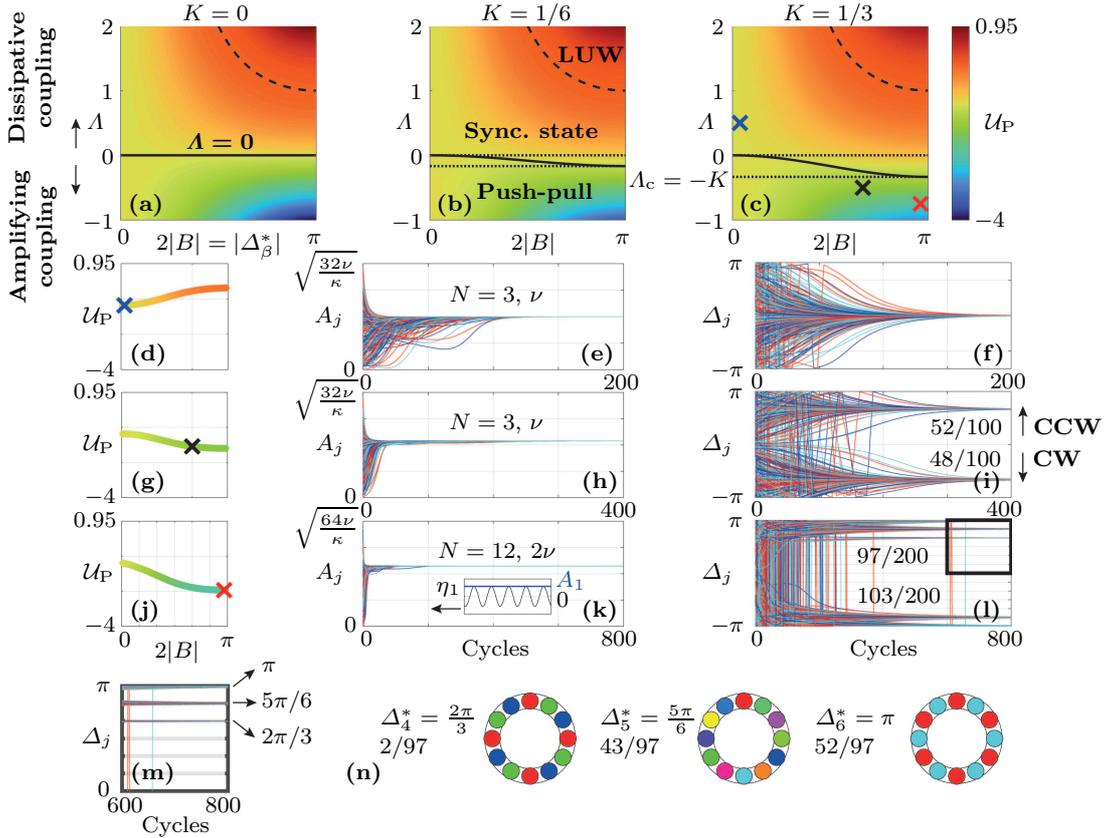}
    \end{psfrags}
    \caption{\fc{Numerical examples performed on the slow-flow system \eqref{Amplitude dynamics, slow-flow} and \eqref{Phase dynamics, slow-flow} in the deterministic limit with $\Gamma=0$.} The distribution of the projected potential $\mathcal{U}_\mathrm{P}$, given by Eq. \eqref{Potential at limit cycle, normalized}, as a function of $\Lambda$ and $2|B|=|\Delta^*_\beta|$ is shown in \textbf{(a)}-\textbf{(c)} for increasing values of the coupling nonlinearity $K$. The solid black curves parametrize the minima of $\mathcal{U}_\mathrm{P}$ with respect to $2B$ over $\Lambda$. The dotted lines bound the transition region, separating the domains where the synchronized state (\textbf{Sync. state}) and the push-pull mode (\textbf{Push-pull}) are the global minimum of the potential $\mathcal{V}$, given by Eq. \eqref{Potential for amplitudes and phases}, respectively. Beyond the dashed black line are \fc{linearly unstable rotating waves (\textbf{LUW})}. The blue, black and red crosses in \textbf{(c)} mark the location of the minimum of the projected potential $\mathcal{U}_\mathrm{P}$, given by Eq. \eqref{Potential at limit cycle, normalized}, for $\Lambda=0.5$, $-0.5$ and $-0.75$, respectively. The variation of $\mathcal{U}_\mathrm{P}$ over $2|B|$ is shown in \textbf{(d)}, \textbf{(g)} and \textbf{(j)}. Time traces from random initial conditions of the amplitudes $A_j$ and phase differences $\Delta_j$ are shown in \textbf{(e)}, \textbf{(f)}, \textbf{(h)} and \textbf{(i)} for $N=3$ (100 realizations) and in \textbf{(k)} and \textbf{(l)} for $N=12$ (200 realizations). The oscillators $j=1,2,3$ are indicated by blue, red and cyan color, respectively. In the small inset in \textbf{(k)}, the numerical solution $\eta_1$ (black) of the fast system \eqref{Dynamics of dominant modal amplitude with swirl, nonlinear with noise} is compared with the amplitude $A_1$ (blue) computed from the the slow-flow system \eqref{Amplitude dynamics, slow-flow} and \eqref{Phase dynamics, slow-flow} for $5$ cycles, starting at 200. \fc{The numbers above the trajectories in \textbf{(i)} and \textbf{(l)} indicate} how many realizations converge to the solution classes with $\Delta_j\lessgtr 0$ (CCW and CW rotating waves, respectively). The inset \textbf{(m)} shows an enlarged version of the time traces with $\Delta_j>0$ in \textbf{(l)} between 600 and 800 cycles. In \textbf{(n)}, it is indicated how the 97 realizations in \textbf{(l)} with $b>0$ are distributed among different Bloch modes.}
    \label{Figure 3}
\end{figure*}

\begin{figure*}[h!]
\begin{psfrags}
\psfrag{a}{\hspace{0.05cm}$K=1/3$}
\psfrag{b}{$2$}
\psfrag{c}{$-1$}
\psfrag{d}{$0$}
\psfrag{e}{\hspace{0.05cm}$\pi$}
\psfrag{f}{$\Delta_b^*$}
\psfrag{g}{$0.95$}
\psfrag{h}{$-4$}
\psfrag{i}{$-2$}
\psfrag{j}{$\mathcal{U}_\mathrm{P}$}
\psfrag{k}{$*$}
\psfrag{l}{\hspace{0.05cm}$\pi$}
\psfrag{m}{$\Delta_j$}
\psfrag{n}{$70\%$}
\psfrag{o}{\hspace{-0.07cm}$30$'$000$}
\psfrag{p}{\hspace{0.05cm}$-\pi$}
\psfrag{q}{$\pi$}
\psfrag{r}{Cycles}
\psfrag{s}{\hspace{-0.35cm}$R(\Delta_1)/R_\mathrm{tot}$}
\psfrag{t}{\hspace{-0.6cm}$\langle  |\Delta_1(t_\mathrm{end})| \rangle_R=0.195$}
\psfrag{u}{\hspace{-0.6cm}$\langle |\Delta_1(t_\mathrm{end})| \rangle_R=0.211$}
\psfrag{v}{\hspace{-0.6cm}$\langle |\Delta_1(t_\mathrm{end})| \rangle_R=0.244$}
\psfrag{w}{\hspace{-0.6cm}$\langle |\Delta_1(t_\mathrm{end})| \rangle_R=1.15$}
\psfrag{x}{\hspace{-0.6cm}$\langle |\Delta_1(t_\mathrm{end})| \rangle_R=2.50$}
\psfrag{y}{\hspace{-0.6cm}$\langle |\Delta_1(t_\mathrm{end})| \rangle_R=2.88$}
\psfrag{z}{\hspace{-0.1cm}$\Lambda$}
\psfrag{A}{{$\Delta^*_\beta$}}
\psfrag{B}{\hspace{-0.13cm}$-\pi$\hspace{0.5cm}$0$\hspace{0.5cm}$\pi$}
\psfrag{F}{\textbf{(a)}}
\psfrag{G}{\textbf{(b)}}
\psfrag{H}{\textbf{(c)}}
\psfrag{I}{\textbf{(d)}}
\psfrag{J}{\textbf{(e)}}
\psfrag{K}{\textbf{(f)}}
\psfrag{L}{\textbf{(g)}}
\psfrag{M}{\textbf{(h)}}
\psfrag{N}{\textbf{(i)}}
\psfrag{O}{\textbf{(j)}}
\psfrag{P}{\textbf{(k)}}
\psfrag{Q}{\textbf{(l)}}
\psfrag{R}{\textbf{(m)}}
\psfrag{S}{ \textcolor{darkblue2}{$\boldsymbol{\Lambda=0.5}$}}
\psfrag{T}{ \textcolor{mygreen1}{$\boldsymbol{\Lambda=0.3}$}}
\psfrag{U}{ \textcolor{cyan}{$\boldsymbol{\Lambda=0.1}$}}
\psfrag{V}{\textcolor{red}{$\boldsymbol{\Lambda=-0.1}$}}
\psfrag{W}{\textcolor{magenta}{$\boldsymbol{\Lambda=-0.3}$}}
\psfrag{X}{\textcolor{black}{$\boldsymbol{\Lambda=-0.5}$}}
\psfrag{Z}{\hspace{-0.05cm}$2|B|$}
\psfrag{1}{\hspace{-0.95cm}$\Lambda_\mathrm{c}=-K$}
\psfrag{2}{$0$}
\psfrag{3}{\hspace{-0.45cm}\bf Sync. state }
\psfrag{4}{\hspace{-0.85cm}\bf Push-pull mode}
\psfrag{5}{$1.16$}
\psfrag{6}{$2.50$}
\psfrag{7}{$1$}
\psfrag{8}{$\hspace{-0.55cm}2|B|=|\Delta_\beta^*|$}
\psfrag{9}{\hspace{-0.05cm}\bf LUW}
    \centering
    \includegraphics[width=0.75\textwidth,center]{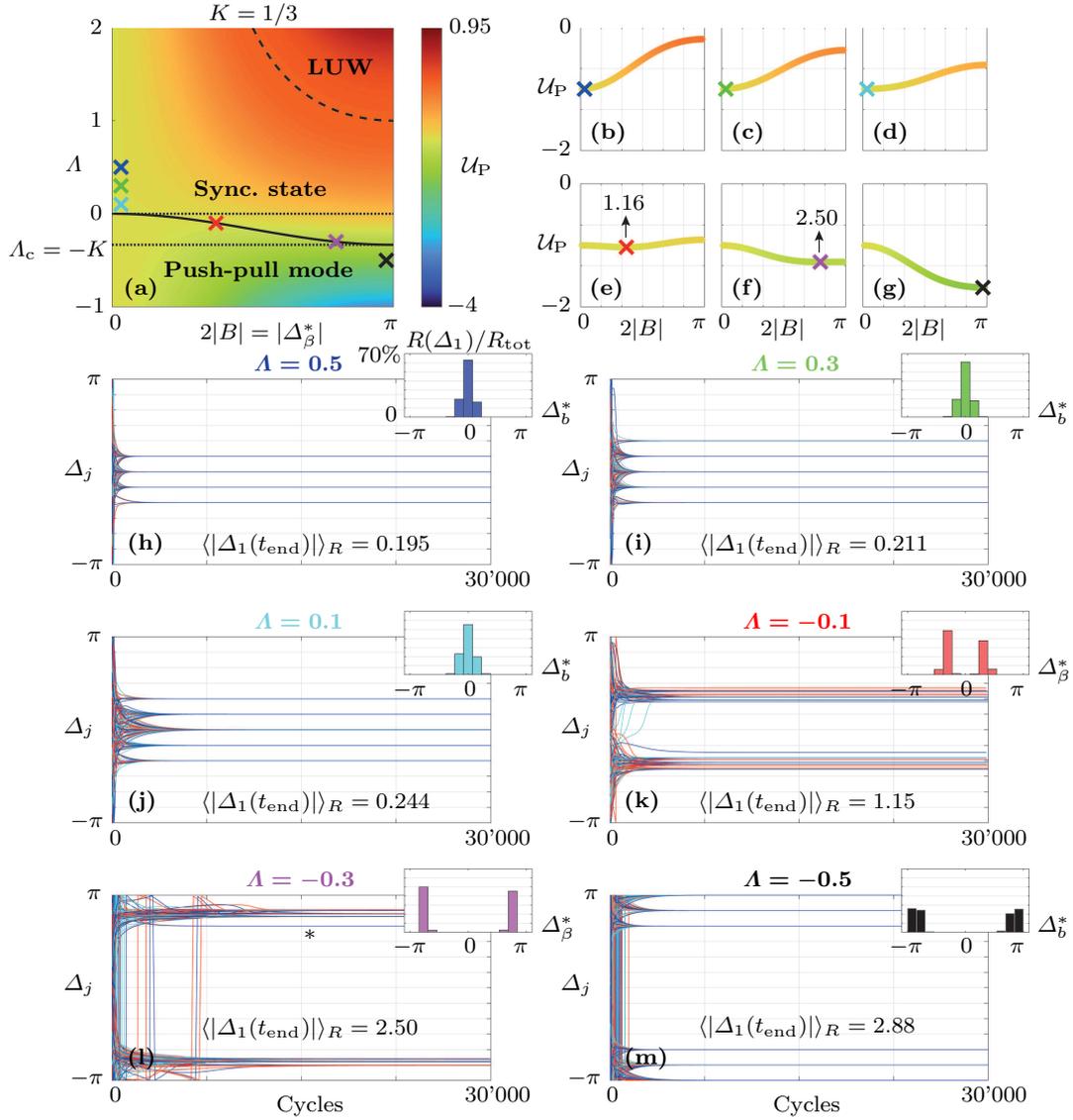}
    \end{psfrags}
    \caption{\fc{Numerical experiments near and in the transition region, which separates the two domains where the synchronized state (\textbf{Sync. state}) and the push-pull mode (\textbf{Push-pull mode}) are the global minimum of the potential $\mathcal{V}$, given by Eq. \eqref{Potential for amplitudes and phases}, respectively, for an oscillator ring with $N=12$ members.} \textbf{(a)} Distribution of the projected potential $\mathcal{U}_\mathrm{P}$, given by Eq. \eqref{Potential at limit cycle, normalized}, as a function of $\Lambda$ and $2|B|=|\Delta^*_\beta|$. The colored crosses indicate the location of the minimum of the projected potential $\mathcal{U}_\mathrm{P}$, given by Eq. \eqref{Potential at limit cycle, normalized}, over $B$ for six evenly spaced values of the linear resistive coupling $\Lambda$, respectively. The corresponding distributions of the projected potential $\mathcal{U}_\mathrm{P}$ over the phase difference $2|B|$ is shown in \textbf{(b)}-\textbf{(g)}. The positions of the minima of $\mathcal{U}_\mathrm{P}$ are indicated in \textbf{(e)} and \textbf{(f)}, where they do not correspond to discrete Bloch modes. For each value of $\Lambda$, $500$ realizations of the slow-flow system \eqref{Amplitude dynamics, slow-flow} and \eqref{Phase dynamics, slow-flow} were computed over $5000$ cycles \fc{from} random initial conditions. In \textbf{(h)}-\textbf{(m)}, the phase differences $\Delta_j$, $j=1,2,3$, are shown in blue, red and cyan\fc{, respectively.} Time traces of other oscillators appear similar. In \textbf{(h)}, \textbf{(i)}, \textbf{(j)} and \textbf{(m)}, the first 20 realizations converging to each Bloch mode are shown and in \textbf{(k)} and \textbf{(l)} the first 50 of all realizations are shown. The average of $|\Delta_1|$ over all realizations at 5000 cycles\fc{, $\langle  |\Delta_1(t_\mathrm{end})| \rangle_R$,} is displayed in the same insets. In \textbf{(h)}-\textbf{(m)}, the smaller insets in the upper right corners show the relative share $R(\Delta_1)/R_\mathrm{tot}$ of the total realizations $R_\mathrm{tot}$ of $\Delta_1$ that fall into the bins corresponding to the different Bloch modes. Note that while in \textbf{(h)}-\textbf{(j)} and \textbf{(m)}, all realizations converge to discrete Bloch modes, in \textbf{(k)} and \textbf{(l)} (transition region), trajectories generally converge to unsteady quasi-limit cycles which are superpositions of CW and CCW rotating waves, see Fig. \ref{Figure 5}. However, Bloch modes are also observed in the transition region, see for example the time trace marked with an asterisk ($*$) in \textbf{(l)}, which corresponds to $b=4$.}
    \label{Figure 4}
\end{figure*}
\begin{figure*}[h!]
\begin{psfrags}
\psfrag{a}{$-1.25$}
\psfrag{b}{$-1.28$}
\psfrag{c}{$\pi/2$}
\psfrag{d}{$\pi$}
\psfrag{e}{$\mathcal{U}_\mathrm{P}$}
\psfrag{f}{\hspace{-0.58cm}$2|B|=|\Delta_\beta^*|$}
\psfrag{g}{Cycles}
\psfrag{h}{$\Delta_j$}
\psfrag{i}{$2.498218$}
\psfrag{k}{$2.497948$}
\psfrag{j}{$-2.497956$}
\psfrag{l}{$-2.498225$}
\psfrag{m}{$29$'$700$}
\psfrag{n}{$30$'$000$}
\psfrag{o}{$2\pi$}
\psfrag{p}{$0$}
\psfrag{q}{$\varphi_j$}
\psfrag{r}{\bf (a)}
\psfrag{s}{\bf (c)}
\psfrag{t}{\hspace{0.1cm}\bf (c)}
\psfrag{u}{$2.4965$}
\psfrag{v}{$1.001 A_0^*$}
\psfrag{w}{$0.999 A_0^*$}
\psfrag{x}{$A_j$}
\psfrag{y}{\bf (b)}
\psfrag{z}{\bf (d)}
    \centering
    \includegraphics[width=0.6\textwidth,center]{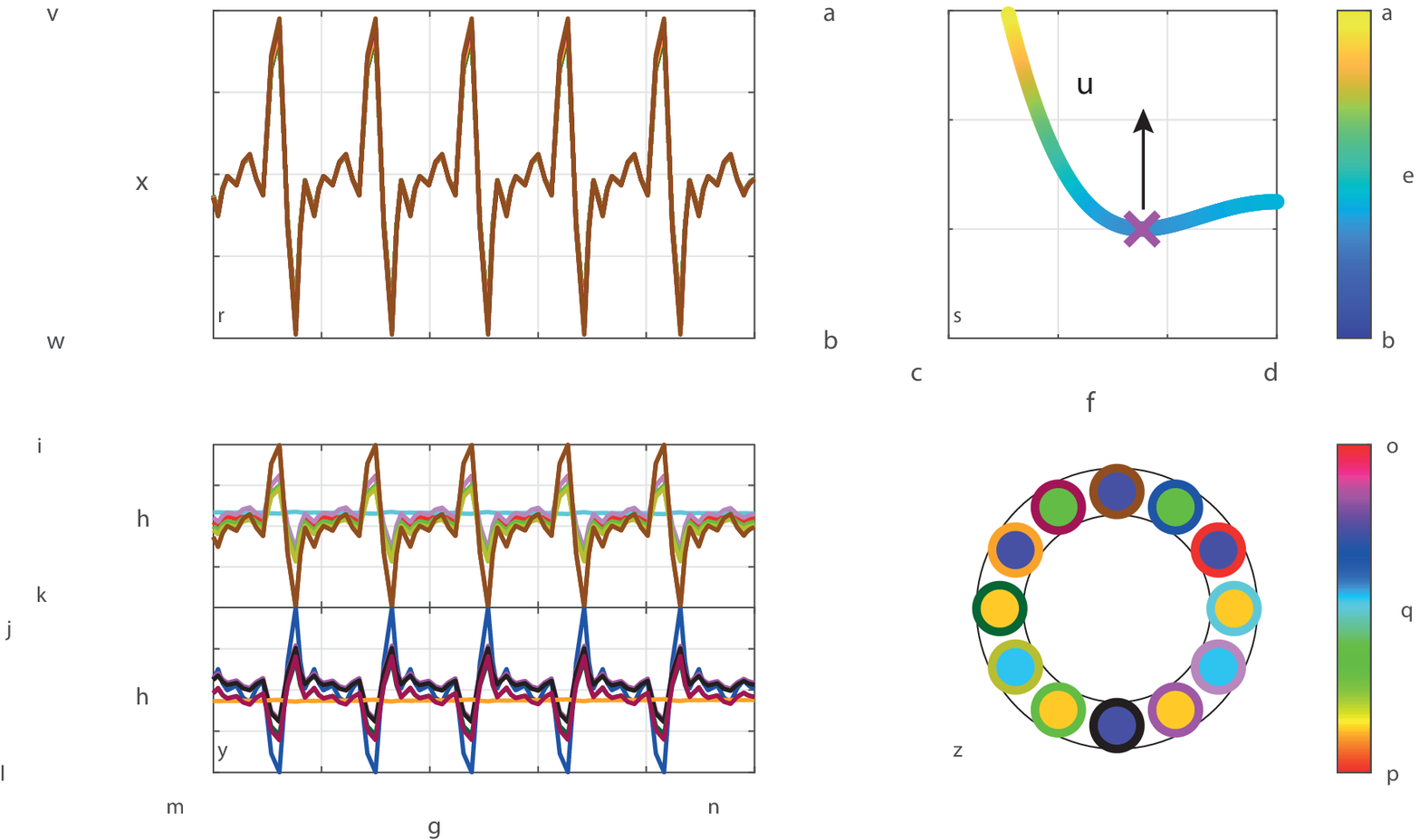}
    \end{psfrags}
    \caption{Example of a transition mode in the steady state, computed from the slow-flow system \eqref{Amplitude dynamics, slow-flow} and \eqref{Phase dynamics, slow-flow} for $\Lambda=-0.3$ ($\Lambda_\mathrm{c}=-0.\overline{3}$). Shown is \textbf{(a)} the variation of the amplitudes $A_j$ and \textbf{(b)} the phase differences $\Delta_j\equiv\mathrm{mod}(\varphi_j-\varphi_{j-1}+\pi,2\pi)-\pi$ for a single realization over $300$ cycles in the steady state, \textbf{(c)} the distribution of the projected potential $\mathcal{U}_\mathrm{P}$ over $2|B|=|\Delta_\beta^*|$ \fc{(magnified view of Fig. \ref{Figure 4}\textbf{(f)} with a different colormap)} and \textbf{(d)} the phases $\varphi_j$ at 30'000 cycles. The colors of the time traces in \textbf{(a)} and \textbf{(b)} correspond to those of the small rings in \textbf{(d)}. Time traces of other transition modes appear similar, with more or less pronounced asymmetry between different oscillators, but with absolute phase differences $|\Delta_j|$ near the minimum of $\mathcal{U}_\mathrm{P}$ over $2|B|$ (see Fig. \ref{Figure 4}\textbf{(l)}). The phase distribution $\varphi_j$ along the ring shown in \textbf{(c)} corresponds to a standing wave-type pattern of the acoustic pressure $\eta_j$ along the ring. Animated examples of transition mode time traces are contained in the \href{https://www.dropbox.com/sh/d4pvwzsns30izy9/AAAdW9aMZFzabITg71vU3YqEa?dl=0}{supplementary material}.}
    \label{Figure 5}
\end{figure*}

\fc{The Taylor expansion of the projected potential $\mathcal{U}_\mathrm{P}$, given by Eq. \eqref{Potential at limit cycle, normalized}, around $B=0$ (synchronized state) is} 
\fc{\begin{equation}
    \mathcal{U}_\mathrm{P}(B)\approx -1 + 2 \Lambda B^2 + O(B^4), \label{exp. around b=0}
\end{equation}
and the expansion around $B=\pm\pi/2$ is }
\fc{\begin{eqnarray}
    \mathcal{U}_\mathrm{P}(B)&\approx&-(\Lambda-1)^2/(K+1)\nonumber\\
    &&+2(\Lambda-1)(\Lambda+K)(B\mp\pi/2)^2/(K+1)^2\nonumber\\
    &&+O((B\mp\pi/2)^4). \label{exp. around b=N/2}
\end{eqnarray} 
For $\Lambda>0$ (dissipative coupling) and $B\approx 0$, we consider Eq. \eqref{exp. around b=0}, which shows that the global minimum of the potential $\mathcal{V}$ is at $B=0$, i.e., the synchronized state. Conversely, for $\Lambda<\Lambda_\mathrm{c}=-K$ (strong amplifying coupling) and $B\approx \pm \pi/2$, we consider Eq. \eqref{exp. around b=N/2} which shows the global minima of $\mathcal{V}$ are at $2B=\Delta^*_{\pm N/2}=\pm\pi$ (or $\Delta^*_{\pm\mathrm{floor}(N/2)}$ for odd $N$), corresponding to the push-pull mode.}

For \fc{$0>\Lambda>\Lambda_\mathrm{c}$, which we call the ``transition region'' in the following, the minima of the projected potential $\mathcal{U}_\mathrm{P}$, given by Eq. \eqref{Potential at limit cycle, normalized}, lie between $B=0$ and $B=\pm\pi/2$. This parameter range is further investigated below.} By computing the zero level set of the slope of $\mathcal{U}_\mathrm{P}$ with respect to $B$, we obtain the curve parametrizing the minima of $\mathcal{U}_\mathrm{P}$ \fc{in the transition region} as a function of $\Lambda$:
\begin{eqnarray}
    &&0\stackrel{!}{=}\frac{\partial \mathcal{U}_\mathrm{P}}{\partial B}=2\bigg(\frac{\Lambda\sin(B)\cos(B)}{1-\Lambda\sin(B)^2}+\frac{2K\sin(B)^3\cos(B)}{|1+K\sin(B)^4|^2}\nonumber \\
    &&\times(1+K\sin(B)^4)+\frac{\Lambda\sin(B)\cos(B)}{|1-\Lambda\sin(B)^2|}\bigg)(1-\Lambda\sin(B)^2)\nonumber\\
    &&\left\lvert\frac{1-\Lambda\sin(B)^2}{1+K\sin(B)^4}\right\rvert. \label{Domains of attraction boundary}
\end{eqnarray}

\fc{Numerical examples were performed and analyzed to explore the projected potential $\mathcal{U}_\mathrm{P}$, given by Eq. \eqref{Potential at limit cycle, normalized}, by launching a number of realizations of the slow-flow system \eqref{Amplitude dynamics, slow-flow} and \eqref{Phase dynamics, slow-flow} from random initial conditions. A part of the} results is reported in Fig. \ref{Figure 3}, \fc{where, for $N=3$ and $\nu=25$ s$^{-1}$, $R=100$ realizations were computed for $\Lambda=\pm 0.5$, respectively, and $R=200$ realizations were computed for $N=12$, $\nu=50$ s$^{-1}$ and $\Lambda=-0.75$. Trajectories of the oscillators $j=1,2$ and $3$ are shown in blue, red and cyan, respectively. For $N=12$, only the first three oscillators are shown, but trajectories of other oscillators appear similar.} In Fig. \ref{Figure 3}\textbf{(a)}-\textbf{(c)}, the distribution of the projected potential $\mathcal{U}_\mathrm{P}$ as a function of $\Lambda$ and $2|B|=|\Delta_b^*|$ is shown. These insets \fc{illustrate} how a nonzero coupling nonlinearity $K>0$ leads to a transition region \fc{for $0>\Lambda>\Lambda_\mathrm{c}=-K$}, bounded by the dotted lines, which separates the domains \fc{where the synchronized state (\textbf{Sync. state}) and the push-pull mode (\textbf{Push-pull mode}) are the global minimum of the potential $\mathcal{V}$, given by Eq. \eqref{Potential for amplitudes and phases}, respectively}. \fc{The solid black curve indicates the location of the minimum of $\mathcal{U}_\mathrm{P}$ with respect to $2B$ in the transition region. The blue, black and red crosses in \ref{Figure 3}\textbf{(c)} mark the location of the minimum of the projected potential $\mathcal{U}_\mathrm{P}$, given by Eq. \eqref{Potential at limit cycle, normalized}, for $\Lambda=0.5$, $-0.5$ and $-0.75$, respectively. The dashed black curve demarcates the domain within which Bloch modes are linearly unstable (\textbf{LUW}: linearly unstable waves).} Figure \ref{Figure 3}\textbf{(d)}, \textbf{(g)} and \textbf{(j)} show the variation of the projected potential $\mathcal{U}_\mathrm{P}$ as a function of $2|B|$. In \fc{Fig. \ref{Figure 3}\textbf{(e)}, \textbf{(f)}, \textbf{(h)} and \textbf{(i)}}, it is shown how, in a ring of $N=3$ oscillators with dissipative coupling, all trajectories converge to the synchronized state and for strong amplifying coupling outside the transition region, all trajectories converge to rotating wave solutions, with a symmetric distribution between CW ($\Delta_j<0$) and CCW ($\Delta_j>0$) rotating waves. \fc{The number of realizations that converge to either solution class is indicated above the $\Delta_j$-curves.} As shown in Fig. 3\textbf{(k)} and \textbf{(l)}, for $N=12$, \fc{out of all trajectories that converge to solutions with positive $b$ (97 out of 200 total realizations),} \fc{the majority (52 out of 97) converge} to \fc{the limit cycle} with $b=N/2=6$, but a significant amount of them do not, and instead \fc{converge to} the nearby \fc{solutions} with $b=5$ (43 out of 97) and $b=4$ (2 out of 97). The small inset in Fig. \ref{Figure 3}\textbf{(k)} shows a comparison between a single realization of $\eta_1$ (black) computed from the fast system \eqref{Dynamics of dominant modal amplitude with swirl, nonlinear with noise} and the amplitude $A_1$ (blue) computed from the slow-flow system \eqref{Amplitude dynamics, slow-flow} and \eqref{Phase dynamics, slow-flow} for $5$ cycles, starting at 200. \fc{Fig. \ref{Figure 3}\textbf{(m)} shows an enlarged version of Fig. \ref{Figure 3}\textbf{(l)} with $\Delta_j>0$ between 600 and 800 cycles. The corresponding Bloch modes are visualized in Fig. \textbf{(n)}. Similar aggregating behavior around the synchronized state with $b=0$ is observed for $\Lambda>0$, see Fig. \ref{Figure 4} below.}

We see in Fig. \ref{Figure 3} that, while a change in $\Lambda$ \fc{from $0.5$ to $-0.5$} leads to $O(1)$ changes in the phase differences \fc{$\Delta_j$}, the corresponding change in the amplitudes $A_j$ in the steady state is small compared to $A_0^*$. \fc{To better understand this, note that on one hand, for $\Lambda>0$, the synchronized state is globally attracting, and the system aggregates around low-order Bloch modes $b\approx 0$, which implies a small influence of $\Lambda$ and $K$ on the limit cycle amplitude $A_b^*$, see Eqs. \eqref{Amplitude fixed point}-\eqref{kappa_b}. On the other hand, for $\Lambda<0$ and $0<K\ll 1$, we have $(A_b^*)^2\approx (A_0^*)^2(1-\Lambda\sin(\pi b/N)^2-K\sin(\pi b/N)^4)$, i.e., $\Lambda$ and $K$ tend to cancel each other out. For this reason, and because the aeroacoustic coupling is considered as a small perturbation of the thermoacoustic limit cycles ($|\Lambda|\ll 1$, $K\ll 1$), the coupling-induced changes in the amplitude are in general assumed to be small. For $\Lambda<\Lambda_\mathrm{c}=-K$, solutions with $|b|>0$ reach limit cycle amplitudes $A_b^*$ exceeding that of the synchronized state: $A_b^*>A_0^*$. However, such} small variations of the oscillation amplitude are difficult to quantify in experiments on real aero- and thermoacoustic systems, which are typically plagued by significant noise levels. For this reason, this work mostly focuses on the phase differences $\Delta_j$, \fc{whose distribution} quantifies emergent patters of the acoustic pressure $\eta_j$ along the ring.

\fc{More} numerical experiments were performed on the slow-flow system \eqref{Amplitude dynamics, slow-flow} and \eqref{Phase dynamics, slow-flow} to explore the transition region. The \fc{corresponding} results are reported in Fig. \ref{Figure 4}. \fc{Fig. \ref{Figure 4}\textbf{(a)} shows the projected potential $\mathcal{U}_\mathrm{P}$, given by Eq. \eqref{Potential at limit cycle, normalized}, as a function of $\Lambda$ and $2|B|=|\Delta_\beta^*|$. The colored crosses in \ref{Figure 4}\textbf{(a)} mark the location of the minimum of the projected potential $\mathcal{U}_\mathrm{P}$, given by Eq. \eqref{Potential at limit cycle, normalized}, over $B$ for six evenly spaced values of the linear resistive coupling $\Lambda$, respectively. For the same values of $\Lambda$, the insets in Fig. \ref{Figure 4}\textbf{(b)}-\textbf{(g)} show the distributions of $\mathcal{U}_\mathrm{P}$ over the phase difference $2|B|$. The positions of the minima of $\mathcal{U}_\mathrm{P}$ are indicated in Fig. \ref{Figure 4}\textbf{(e)} and \textbf{(f)}, where they do not correspond to discrete Bloch modes. $500$ realizations of the slow-flow system \eqref{Amplitude dynamics, slow-flow} and \eqref{Phase dynamics, slow-flow} were computed over $5000$ cycles from random initial conditions at each value of $\Lambda$. The insets in Fig. \ref{Figure 4}\textbf{(h)}-\textbf{(m)} show the evolution of the phase differences $\Delta_j$ of the oscillators with $j=1,2$ and $3$ in blue, red and cyan, respectively. In Fig. \ref{Figure 4}\textbf{(h)}-\textbf{(j)} and \textbf{(m)}, the first 20 realizations converging to each Bloch mode are shown and in Fig. \ref{Figure 4}\textbf{(k)} and \textbf{(l)}, the first 50 of all realizations are shown. The average of $|\Delta_1|$ over all realizations at 5000 cycles\fc{, $\langle  |\Delta_1(t_\mathrm{end})| \rangle_R$,} is displayed in the same insets. In Fig. \ref{Figure 4}\textbf{(h)}-\textbf{(m)}, the smaller insets in the upper right corner show the relative share $R(\Delta_1)/R_\mathrm{tot}$ of the total realizations $R_\mathrm{tot}$ of $\Delta_1$ that fall into the bins corresponding to the different Bloch modes. Note that while in Fig. \ref{Figure 4}\textbf{(h)}-\textbf{(j)} and \textbf{(m)}, all realizations converge to discrete Bloch modes, in Fig. \ref{Figure 4}\textbf{(k)} and \textbf{(l)} (transition region), trajectories generally converge to unsteady quasi-limit cycles which we call ``transition modes''. These solutions are superpositions of CW and CCW rotating waves, see Fig. \ref{Figure 5}. Consistent with the linear stability analysis performed above, in the transition region, Bloch modes are also observed, see for example the time trace marked with an asterisk ($*$) in Fig. \ref{Figure 4}\textbf{(l)}, which corresponds to $b=4$.} 

We observe \fc{in Fig. \ref{Figure 4}} that outside the transition region, \fc{for $\Lambda>0$ or $\Lambda<\Lambda_\mathrm{c}$, when} the global minimum of the potential $\mathcal{V}$, given by Eq. \eqref{Potential for amplitudes and phases}, coincides with a fixed point of the \fc{slow-flow system \eqref{Amplitude dynamics, slow-flow} and \eqref{Phase dynamics, slow-flow}}, all trajectories converge to discrete Bloch modes. For $\Lambda>0$ and $\Lambda<0$, nearly symmetric unimodal and bimodal distributions of the phase difference $\Delta_1$ over $\Delta_b^*$ and $\Delta_\beta^*$ are observed in the numerical experiments, respectively.  

Inside the transition region, for $0>\Lambda>\Lambda_\mathrm{c}$, \fc{we observed in our numerical experiments that} trajectories generally do not converge to the phase-locked fixed points of the slow-flow system \eqref{Amplitude dynamics, slow-flow} and \eqref{Phase dynamics, slow-flow}\fc{, but to} unsteady \fc{quasi-limit cycles}, the transition modes, which feature small-scale, piecewise linear periodic variation of the amplitudes $A_j$ and phase differences $\Delta_j$'. In the transition region, the mean $\langle |\Delta_1(t_\mathrm{end})|\rangle_R$ over all realizations matches well with the minimum of the projected potential $\mathcal{U}_\mathrm{P}$ over $|\Delta^*_\beta|$, see Fig. \ref{Figure 4}\textbf{(e)}, \textbf{(f)}, \textbf{(k)} and \textbf{(l)}. 

An example of a transition mode in the steady state, computed from the slow-flow system \eqref{Amplitude dynamics, slow-flow} and \eqref{Phase dynamics, slow-flow} for $\Lambda=-0.3$ ($\Lambda_\mathrm{c}=-0.\overline{3}$), is illustrated in Fig. \ref{Figure 5}. Shown in Fig. \ref{Figure 5}\textbf{(a)} and \textbf{(b)} are \tr{the variation of the amplitudes $A_j$} and the phase differences $\Delta_j=\mathrm{mod}(\varphi_j-\varphi_{j-1}+\pi,2\pi)-\pi$ for $j=1,\ldots,N$, respectively, for a single realization over $300$ cycles in the steady state. Figure \ref{Figure 5}\textbf{(c)} and \textbf{(d)} show the distribution of the projected potential $\mathcal{U}_\mathrm{P}$ over $2|B|=|\Delta_\beta^*|$ \fc{(magnified view of Fig. \ref{Figure 4}\textbf{(f)} with a different colormap)} and the phases $\varphi_j$ along the ring at the end time, respectively. \fc{The colors of the time traces in Fig. \ref{Figure 5}\textbf{(a)} and \textbf{(b)} correspond to those of the small rings in Fig. \ref{Figure 5}\textbf{(d)}.} The phase differences $\Delta_j$ perform slow, small-scale periodic motion over time. Because the transition modes feature both positive and negative phase differences $\Delta_j$, these solutions are superpositions of CW and CCW rotating waves which appear as standing wave-type patterns of the acoustic pressure $\eta_j$ along the ring. After a transient, the amplitudes $A_j$ all remain quasi-steady, performing small-scale, periodic motion around $A_0^*$. Time traces of other transition modes appear similar, with more or less pronounced asymmetry between different oscillators, but with absolute phase differences $|\Delta_j|$ near the minimum of $\mathcal{U}_\mathrm{P}$ over $2|B|$, see Fig. \ref{Figure 4}\textbf{(k)} and \textbf{(l)}. 

\fc{Animated examples of transition mode time traces ($A_j$, $\varphi_j$, $\Delta_j$ and $\eta_j$), computed from the slow-flow system \eqref{Amplitude dynamics, slow-flow} and \eqref{Phase dynamics, slow-flow}, are contained in the \href{https://www.dropbox.com/sh/d4pvwzsns30izy9/AAAdW9aMZFzabITg71vU3YqEa?dl=0}{supplementary} \href{https://www.dropbox.com/sh/d4pvwzsns30izy9/AAAdW9aMZFzabITg71vU3YqEa?dl=0}{material}.}

\fc{The numerical results reported in Figs. \ref{Figure 4} and \ref{Figure 5} suggest that in the transition region, the occurrence of quasi-limit cycles near the minimum of the projected potential $\mathcal{U}_\mathrm{P}$, given by Eq. \eqref{Potential at limit cycle, normalized}, is generic. Despite these solutions not being fixed points of the slow-flow system \eqref{Amplitude dynamics, slow-flow} and \eqref{Phase dynamics, slow-flow}, through this unsteady motion, the system reaches lower values of the potential $\mathcal{V}$, given by Eq. \eqref{Potential for amplitudes and phases}, and hence the transition modes are typically favored over the discrete Bloch modes in the transition region. Further research is required to explore the dynamic nature of these transition modes. Open topics are discussed in Sec. \ref{Section 5}.}

To better understand the transition modes' unsteadiness, we consider the continuous Bloch wavenumber \fc{$\beta$} and assume the system reaches a quasi-limit cycle where $A_j\equiv A^*_\beta$, $\Delta_j \equiv {k(j)} \Delta^*_\beta$, $k(j)\in\{-1,1\}$ and $j=1,\ldots,N$. Of course this is an approximation which will never be exactly observed, because these quasi-limit cycles are not fixed points of the deterministic part of the slow-flow system given by Eqs. \eqref{Amplitude dynamics, slow-flow} and \eqref{Phase dynamics, slow-flow}. \fc{The normalized potential $\mathcal{U}_\mathrm{P}$ is independent of the signs in $k$, see Eq. \eqref{Potential at limit cycle, normalized}.} However, the signs \fc{do impact} the projection of the slow-flow equations \fc{\eqref{Amplitude dynamics, slow-flow} and \eqref{Phase dynamics, slow-flow}} onto the \fc{quasi-}limit cycle solutions \fc{$\{A_j\equiv A^*_\beta,\Delta_j\equiv k(j) \Delta^*_\beta\}$}:
\begin{eqnarray}
\dot{A}_\beta^*&=&\nu_\beta A_\beta^*-\frac{\kappa_\beta (A_\beta^*)^3}{8}\label{Amplitude, explanation transition modes}\\
     k(j)\dot{\Delta}^*_\beta&=&\frac{\lambda\sin(\Delta^*_\beta)}{2}\left[{k(j+1)}-{k(j)}\right]\nonumber\\
     &&+\frac{\vartheta (A_\beta^*)^2}{8}
     \bigg(2\sin(\Delta^*_\beta)\Big[{k(j+1)}-{k(j)}]\Big]\nonumber\\
     &&-\sin(2\Delta^*_\beta)\Big[k(j+1)-k(j)\Big]\bigg),\label{Phase, explanation transition modes}
\end{eqnarray}
where the fact that, in the steady state, $\dot{\varphi}_j\approx 0$ was used and $\nu_\beta$, $\kappa_\beta$ \fc{are obtained by replacing $b$ with $\beta$ in Eqs. \eqref{nu_b} and \eqref{kappa_b}, respectively.} For a distribution of $\Delta_j$ with varying signs in $k$ along the ring, Eq. \eqref{Phase, explanation transition modes} implies unsteadiness of the Bloch wavenumber $\beta$. This unsteadiness carries over to the equation for $\dot{A}_\beta^*$ through $\nu_\beta$ and $\kappa_\beta$. The ad-hoc argument made above to explain the transition modes' unsteadiness is not an exact description of the nonlinear dynamics in the phase space $\Omega$ (see Eq. \eqref{Big omega phase space}), which are responsible for the occurrence of transition modes and require further investigation. 

In this section, the potential landscape of the deterministic system was investigated. To \fc{this end}, the potential $\mathcal{V}$ of the amplitude-phase system \eqref{Amplitude dynamics, slow-flow} and \eqref{Phase dynamics, slow-flow} was \fc{projected onto} the phase-locked, uniform-amplitude quasi-limit cycle solutions to obtain the normalized projected potential $\mathcal{U}_\mathrm{P}$, which compactly describes the \\(de-)synchronization transition, dividing the parameter space into two domains \fc{where the synchronized state and the push-pull mode are the global minimum of the potential $\mathcal{V}$, respectively. These domains are separated by a transition region, whose size is proportional to the coupling nonlinearity $K$ and which vanishes for $K=0$.} Notably, $\mathcal{U}_\mathrm{P}$ is both non-dimensional and independent of the number of oscillators \fc{$N$}. 

\section{Steady state statistics of the noise-driven system \label{Section 4}}
\begin{figure*}[h!]
\begin{psfrags}
\psfrag{a}{\hspace{-0.05cm}$\Lambda$}
\psfrag{b}{\hspace{-0.1cm}$2|B|$}
\psfrag{c}{$-1$}
\psfrag{d}{$2$}
\psfrag{e}{$0$}
\psfrag{f}{$\pi$}
\psfrag{g}{$\mathcal{U}_\mathrm{P}$}
\psfrag{h}{$K$}
\psfrag{i}{$G$}
\psfrag{j}{$-1.28\times 10^4$}
\psfrag{k}{$-1.26\times 10^4$}
\psfrag{l}{$-107$}
\psfrag{m}{$-88.0$}
\psfrag{n}{$-4.95$}
\psfrag{o}{$-0.376$}
\psfrag{p}{$0$}
\psfrag{q}{$\dfrac{1}{6}$}
\psfrag{r}{$\dfrac{1}{3}$}
\psfrag{s}{$2.88\times 10^{-1}$}
\psfrag{t}{$2.88\times 10^{1}$}
\psfrag{u}{$2.88\times 10^{3}$}
\psfrag{v}{\footnotesize\textbf{Normalized coupling nonlinearity}}
\psfrag{w}{\footnotesize\textbf{Normalized noise intensity}}
\psfrag{x}{\hspace{-0.45cm}\bf Sync. state}
\psfrag{y}{\hspace{-0.85cm}\bf Push-pull mode}
\psfrag{1}{$1$}
\psfrag{0}{$0$}
\psfrag{A}{\hspace{0.18cm}$\boldsymbol{\Lambda=0}$}
\psfrag{B}{$\boldsymbol{\Lambda_\mathrm{c}\approx-K}$}
\psfrag{C}{{$\boldsymbol{\Lambda_\mathrm{c}<-K}$}}
\psfrag{D}{\hspace{-0.65cm}$2|B|=|\Delta_\beta^*|$}
    \centering
    \includegraphics[width=0.75\textwidth,center]{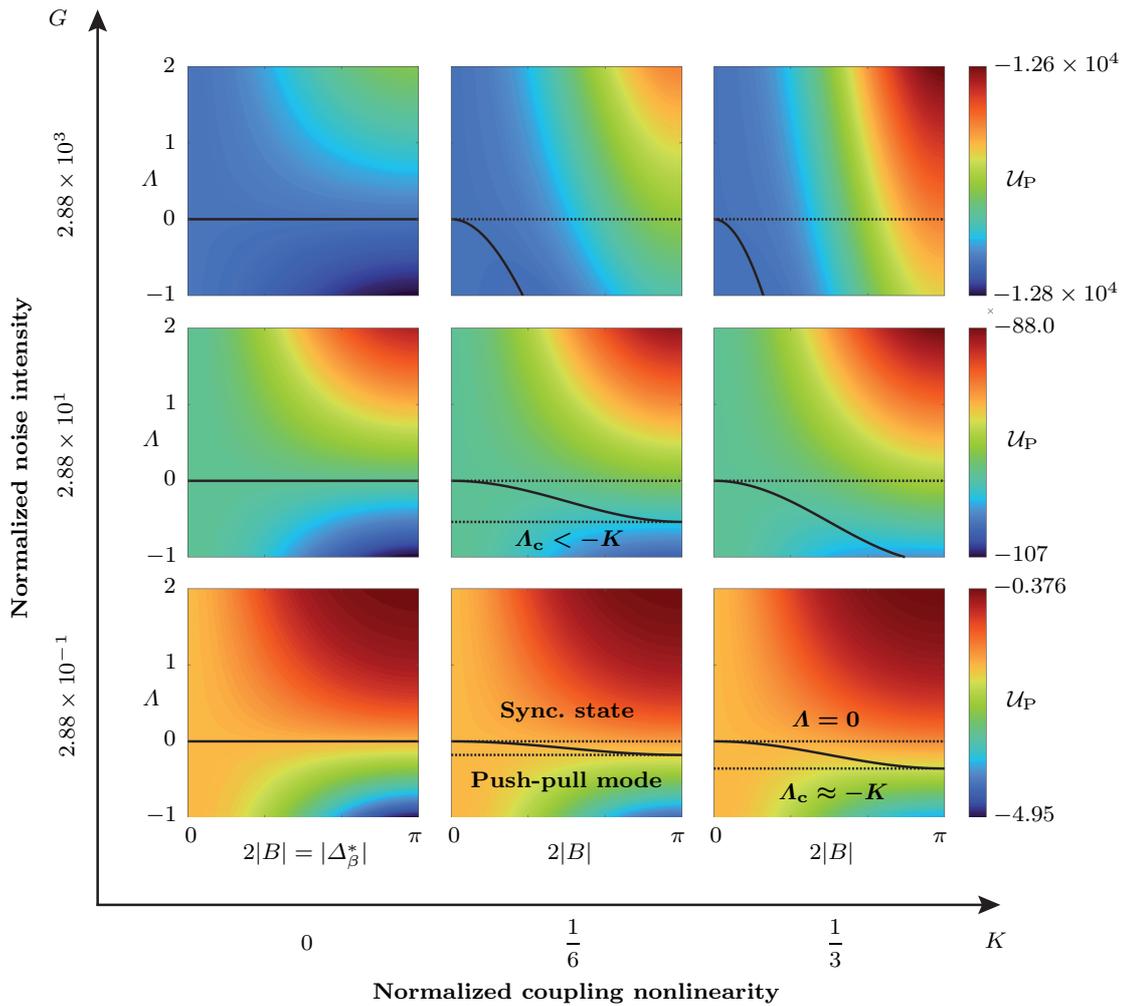}
    \end{psfrags}
    \caption{Distribution of the projected potential $\mathcal{U}_\mathrm{P}$, given by Eq. \eqref{Proj. potential stochastic case}, as a function of $\Lambda$ and $2|B|=|\Delta_\beta^*|$ for varying noise intensity $G$ and coupling nonlinearity $K$. \fc{The dashed black lines mark the transition region $0>\Lambda>\Lambda_\mathrm{c}$.  In the bottom right inset, $\Lambda_\mathrm{c}=-0.356$. Above and below the transition region, the synchronized state and the push-pull mode are the global minimum of the potential $\mathcal{V}$, given by Eq. \eqref{Potential for amplitudes and phases}, respectively. In the transition region, the solid black curve, defined by $\partial \mathcal{U}_\mathrm{P}/\partial B=0$, parametrizes the minima of $\mathcal{U}_\mathrm{P}$ with respect to $2B$ over $\Lambda$.}}
    \label{Figure 6}
\end{figure*}
\begin{figure}[h!]
\begin{psfrags}
\psfrag{a}{$A_j \cos\dfrac{\Delta_j}{2}$}
\psfrag{b}{$A_j \sin\dfrac{\Delta_j}{2}$}
\psfrag{c}{\textcolor{lightblue}{$\boldsymbol{\dfrac{\Delta_j}{2}}$}}
\psfrag{d}{\textcolor{lightblue}{$\boldsymbol{A_j}$}}
\psfrag{D}{$b=6$}
\psfrag{e}{$b=5$}
\psfrag{f}{$b=4$}
\psfrag{g}{$b=3$}
\psfrag{h}{$b=2$}
\psfrag{i}{$b=1$}
\psfrag{j}{$b=0$}
\psfrag{k}{\hspace{-0.06cm}\bf \footnotesize\begin{tabular}{@{}c@{}}
Push-pull \\
{   }mode
\end{tabular}}
\psfrag{l}{\hspace{-0.05cm}\bf \footnotesize\begin{tabular}{@{}c@{}}
Sync. \\
state
\end{tabular}}
\psfrag{m}{\hspace{0.2cm}\bf \footnotesize\begin{tabular}{@{}c@{}}
CCW\\
waves
\end{tabular}}

\psfrag{S}{ \textcolor{darkblue2}{$\boldsymbol{\Lambda=0.5}$}}
\psfrag{T}{ \textcolor{mygreen1}{$\boldsymbol{\Lambda=0.3}$}}
\psfrag{U}{ \textcolor{cyan}{$\boldsymbol{\Lambda=0.1}$}}
\psfrag{V}{\textcolor{red}{$\boldsymbol{\Lambda=-0.1}$}}
\psfrag{W}{\textcolor{magenta}{$\boldsymbol{\Lambda=-0.3}$}}
\psfrag{X}{\textcolor{black}{$\boldsymbol{\Lambda=-0.5}$}}

    \centering
    \includegraphics[width=0.30\textwidth,center]{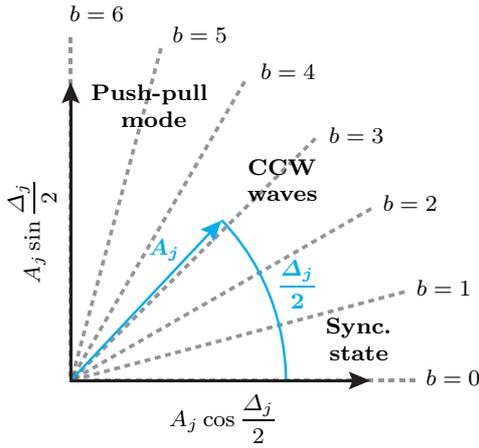}
    \end{psfrags}
    \caption{Geometric analogy between emergent patterns in the oscillator ring, quantified by the phase difference $\Delta_j$, and polar coordinates. Possible non-negative discrete Bloch wavenumbers $b\geq0$ are shown for $N=12$ oscillators. \fc{The $x$-axis ($b=0$) corresponds to the synchronized state and the $y$-axis ($b=N/2=6$) to the push-pull mode. Values of $b$ in between the two extremes correspond to CWW rotating waves.} The lower half-plane (not shown) is symmetric, with negative values of $b$ \fc{and CW waves}.}
    \label{Figure 7}
\end{figure}

\begin{figure*}[h!]
\begin{psfrags}
\psfrag{a}{$2A_0^*$}
\psfrag{b}{$0$}
\psfrag{c}{$5000$}
\psfrag{d}{Cycles}
\psfrag{e}{$2\pi$}
\psfrag{f}{$-2\pi$}
\psfrag{g}{$125$}
\psfrag{h}{$130$}
\psfrag{i}{$\varphi_j$}
\psfrag{j}{$A_j$}
\psfrag{k}{$\Delta_1$}
\psfrag{l}{$\Delta_1$}
\psfrag{m}{\textbf{(a)}}
\psfrag{n}{\textbf{(b)}}
\psfrag{o}{\textbf{(c)}}
\psfrag{p}{\textcolor{white}{\textbf{(d)}}}
\psfrag{q}{\textcolor{white}{\textbf{(e)}}}
\psfrag{r}{\textcolor{white}{\textbf{(f)}}}
\psfrag{s}{$1.1$}
\psfrag{t}{$0.9$}
\psfrag{u}{$t$}
\psfrag{v}{Power [dB]}
\psfrag{w}{$-35$}
\psfrag{x}{$35$}
\psfrag{y}{$0.6$}
\psfrag{z}{PDF}
\psfrag{A}{$-2 A_0^*$}
\psfrag{B}{$-\pi$}
\psfrag{C}{$\pi$}
\psfrag{D}{Norm. Freq.}
\psfrag{E}{\bf \textcolor{white}{CW}}
\psfrag{F}{\bf \textcolor{white}{CCW}}
\psfrag{G}{$A_1\cos{\dfrac{\Delta_1}{2}}$}
\psfrag{H}{$A_1\sin{\dfrac{\Delta_1}{2}}$}
\psfrag{I}{\bf \begin{tabular}{@{}c@{}}
\textcolor{myred}{Theory}\\
\textcolor{darkblue3}{Simulation}
\end{tabular}}
\psfrag{J}{$\pi$}
\psfrag{K}{$-\pi$}
\psfrag{L}{\hspace{-0.07cm}\textcolor{white}{$\langle \text{dB} \rangle_{f,t}=-21.5$}}
\psfrag{M}{\textcolor{gramp}{$A^*_6$}}

    \centering
    \includegraphics[width=0.9\textwidth,center]{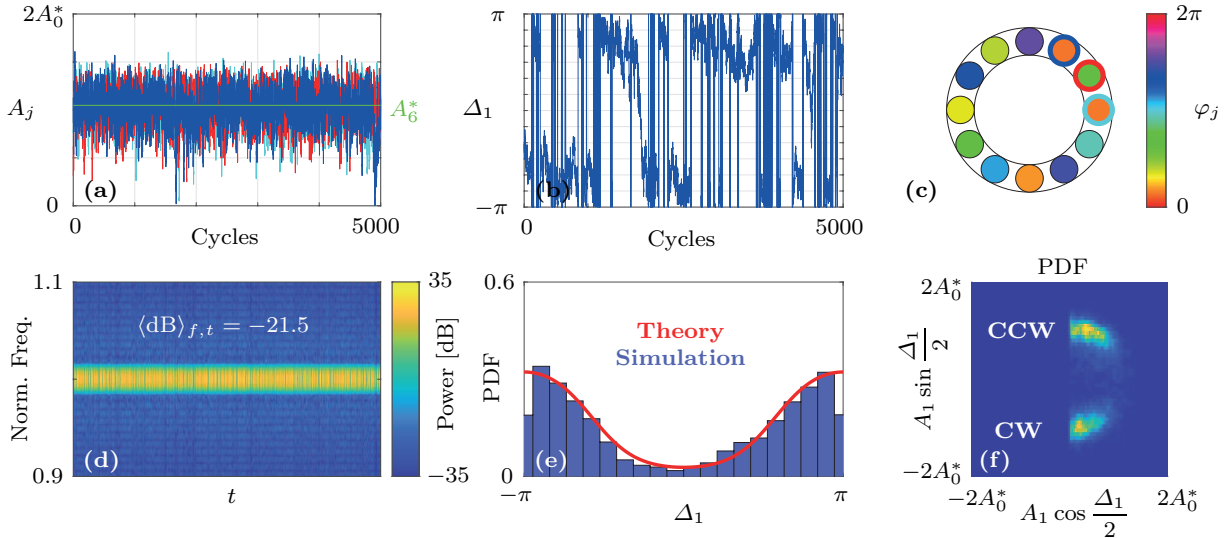}
    \end{psfrags}
    \caption{Analysis of a time trace of the slow-flow system given by Eqs. \eqref{Amplitude dynamics, slow-flow} and \eqref{Phase dynamics, slow-flow} with $\Lambda=-0.5$ ($\Lambda_\mathrm{c}=-0.356$) over $5000$ cycles. \textbf{(a)} Time traces of the amplitudes $A_j$. The oscillators $j=1,2,3$ are indicated by blue, red and cyan color, respectively. The limit cycle amplitude $A_6^*$, given by Eq. \eqref{Limit cycle amplitude with noise}, is shown in green. \textbf{(b)} Time traces of the phase difference $\Delta_1$. Time traces of other oscillators appear similar. \textbf{(c)} Phase distribution $\varphi_j$ along the ring at the end time. \textbf{(d)} Spectrogram of the reconstructed fast oscillating signal $\eta_1(t)$, showing the estimated signal power over time $t$ as a function of the normalized frequency $\omega/\omega_0$. The average of the signal power over the shown frequency and time domains, denoted by $\langle\mathrm{dB}\rangle_{f,t}$, is also indicated in the inset. \textbf{(e)} Comparison of the analytical PDF of the phase differences $\widetilde{P}^\infty_{\mathrm{P},\Delta}$ (red) with a histogram of $\Delta_1$ (blue) from the time trace shown in \textbf{(b)}. \textbf{(f)} Histograms of $A_1$ and $\Delta_1$, represented using the geometric analogy described in Fig. \ref{Figure 7}. Spectro-and histograms of other oscillators appear similar.}
    \label{Figure 8}
\end{figure*}
\begin{figure}[h!]
\begin{psfrags}
\psfrag{a}{$2A_0^*$}
\psfrag{b}{$4995$}
\psfrag{c}{$A_j$}
\psfrag{d}{\bf (a)}
\psfrag{e}{\bf (b)}
\psfrag{f}{Cyles}
\psfrag{g}{$A_j$}
\psfrag{h}{$5000$}
\psfrag{i}{$0$}

    \centering
    \includegraphics[width=0.45\textwidth,center]{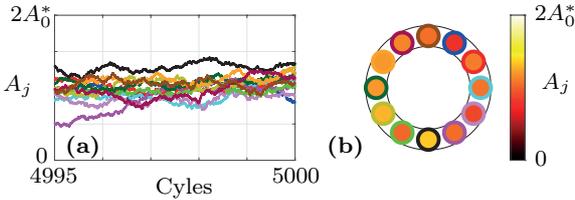}
    \end{psfrags}
    \caption{\textbf{(a)} Amplitudes $A_j$ of the time trace shown in Fig. \ref{Figure 8} over the last 5 cycles. \textbf{(b)} Distribution of the amplitudes $A_j$ along the ring after $5000$ cycles.}
    \label{Figure 9}
\end{figure}
\begin{figure}[h!]
\begin{psfrags}
\psfrag{a}{$2A_0^*$}
\psfrag{b}{$-2A_0^*$}
\psfrag{c}{$A_1 \sin\dfrac{\Delta_1}{2}$}
\psfrag{d}{$2A_0^*$}
\psfrag{e}{$-2A_0^*$}
\psfrag{f}{$\hspace{0.05cm}\pm A_1 \cos\dfrac{\Delta_1}{2}$}
\psfrag{g}{PDF}
\psfrag{h}{\bf Normalized Linear Coupling}
\psfrag{i}{\hspace{-0.2cm}\bf Normalized Noise Intensity}
\psfrag{j}{$2.88\times 10^{-1}$}
\psfrag{k}{$2.88\times 10^1$}
\psfrag{l}{$G$}
\psfrag{m}{$-0.5$}
\psfrag{n}{$0.5$}
\psfrag{o}{$\Lambda$}
\psfrag{p}{\textcolor{white}{\textbf{Fast}}}
\psfrag{q}{\textcolor{white}{\textbf{Av.}}}
\psfrag{1}{Low noise}
\psfrag{2}{High noise}
\psfrag{3}{\hspace{-0.02cm}Amplifying}
\psfrag{4}{\hspace{-0.1cm}Dissipative}

    \centering
    \includegraphics[width=0.45\textwidth,center]{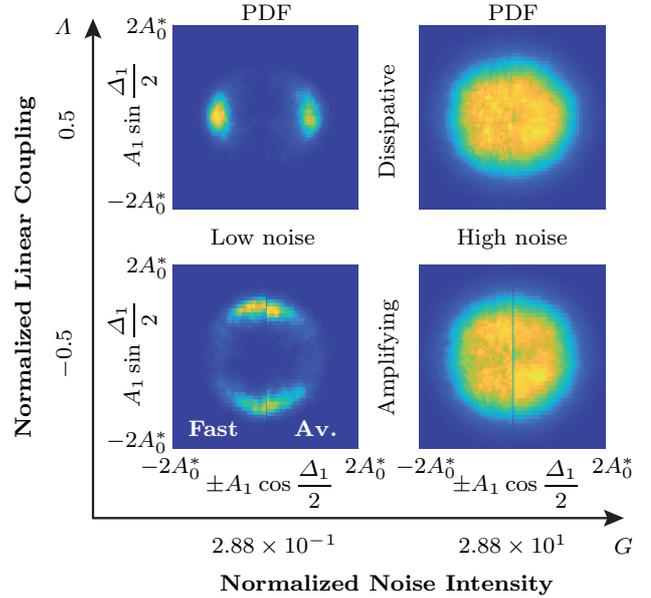}
    \end{psfrags}
    \caption{Validation of the stochastic averaging method against numerical simulations. Histograms of $A_1$ and $\Delta_1$, computed over $10$'$000$ cycles from the \fc{fast and the averaged system, given by Eq. \eqref{Dynamics of dominant modal amplitude with swirl, nonlinear with noise} and Eqs. \eqref{Amplitude dynamics, slow-flow} and \eqref{Phase dynamics, slow-flow},} respectively, are compared for varying linear resistive coupling $\Lambda$ and noise intensity $G$. Histograms of other oscillators appear similar. Note that the amplitude $A_0^*$ depends on the noise intensity $G$ and changes from the left to the right column.}
    \label{Figure 10}
\end{figure}

\begin{figure}[h!]
\begin{psfrags}
\psfrag{a}{$2A_0^*$}
\psfrag{b}{$-2A_0^*$}
\psfrag{c}{$A_1 \sin\dfrac{\Delta_1}{2}$}
\psfrag{d}{$2A_0^*$}
\psfrag{e}{$-2A_0^*$}
\psfrag{f}{$\hspace{0.05cm}\pm A_1 \cos\dfrac{\Delta_1}{2}$}
\psfrag{g}{PDF}
\psfrag{h}{\bf Normalized Linear Coupling}
\psfrag{i}{\hspace{-0.2cm}\bf Normalized Noise Intensity}
\psfrag{j}{$2.88\times 10^{-1}$}
\psfrag{k}{$2.88\times 10^1$}
\psfrag{l}{$G$}
\psfrag{m}{$-0.5$}
\psfrag{n}{$0.5$}
\psfrag{o}{$\Lambda$}
\psfrag{p}{\textcolor{white}{\textbf{Sim.}}}
\psfrag{q}{\hspace{-0.4cm}\textcolor{white}{\textbf{Theory}}}
\psfrag{1}{Low noise}
\psfrag{2}{High noise}
\psfrag{3}{\hspace{-0.02cm}Amplifying}
\psfrag{4}{\hspace{-0.1cm}Dissipative}

    \centering
    \includegraphics[width=0.45\textwidth,center]{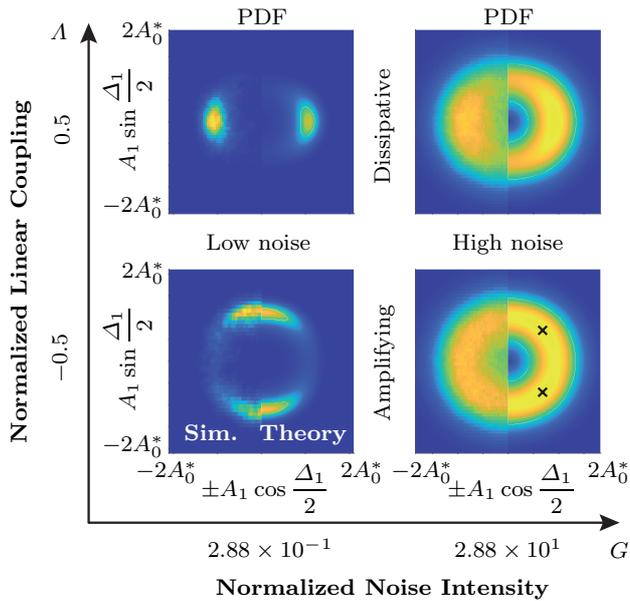}
    \end{psfrags}
    \caption{Validation of the stochastic averaging method against analytical results. Histograms of $A_1$ and $\Delta_1$, computed over $10'000$ cycles at low noise and $100'000$ cycles at high noise from the fast system \eqref{Dynamics of dominant modal amplitude with swirl, nonlinear with noise} (\textbf{Sim.}) are compared to the joint PDF $\widetilde{\mathcal{P}}_\mathrm{P}^\infty$ defined in Eq. \eqref{Projected Stationary FP solution PDF} (\textbf{Theory}) for varying linear resistive coupling $\Lambda$ and noise intensity $G$. For high noise and amplifying coupling, the maxima of the PDF are indicated by black crosses. Histograms of the other oscillators appear similar. Note that the amplitude $A_0^*$ changes from the left to the right column.}
    \label{Figure 11}
\end{figure}

\begin{figure*}[h!]
\begin{psfrags}
\psfrag{a}{$b=0$}
\psfrag{b}{$b=1$}
\psfrag{c}{$b=2$}
\psfrag{d}{$b=3$}
\psfrag{e}{$b=4$}
\psfrag{f}{$b=5$}
\psfrag{g}{$b=6$}
\psfrag{h}{$1.1$}
\psfrag{i}{$0.9$}
\psfrag{j}{Norm. Freq.}
\psfrag{k}{$t$}
\psfrag{l}{Power [dB]}
\psfrag{m}{$-35$}
\psfrag{n}{$35$}
\psfrag{A}{\hspace{-0.12cm}\textcolor{white}{$\langle \text{dB} \rangle_{f,t}=-34.5$}}
\psfrag{B}{\hspace{-0.12cm}\textcolor{white}{$\langle \text{dB} \rangle_{f,t}=-34.7$}}
\psfrag{C}{\hspace{-0.12cm}\textcolor{white}{$\langle \text{dB} \rangle_{f,t}=-34.5$}}
\psfrag{D}{\hspace{-0.12cm}\textcolor{white}{$\langle \text{dB} \rangle_{f,t}=-32.9$}}
\psfrag{E}{\hspace{-0.12cm}\textcolor{white}{$\langle \text{dB} \rangle_{f,t}=-31.7$}}
\psfrag{F}{\hspace{-0.12cm}\textcolor{white}{$\langle \text{dB} \rangle_{f,t}=-29.5$}}
\psfrag{G}{\hspace{-0.12cm}\textcolor{white}{$\langle \text{dB} \rangle_{f,t}=-27.9$}}

    \centering
    \includegraphics[width=0.9\textwidth,center]{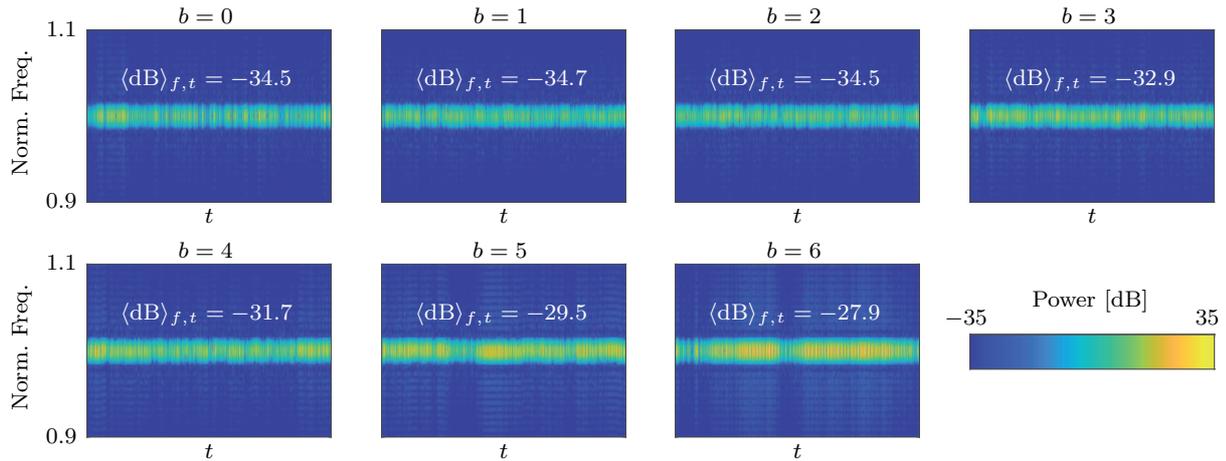}
    \end{psfrags}
    \caption{Spectrograms of the modal amplitudes $\eta_{b}$ for $b=0,\ldots,6$, showing the estimated signal power as a function of time $t$ and the normalized frequency $\omega/\omega_0$. The modal amplitudes $\eta_{b}$ are obtained by decomposing the reconstructed fast pressure signals $\eta_j$ from the example shown in Fig. \ref{Figure 8} (time trace of the slow-flow system \eqref{Amplitude dynamics, slow-flow} and \eqref{Phase dynamics, slow-flow} over $5000$ cycles for $\Lambda=-0.5$) into different Bloch mode components using the discrete Fourier transform, see Eq. \eqref{discrete fourier tf}. Spectrograms for positive and negative $b$ are identical. The average of the signal power over the shown frequency and time domains, denoted by $\langle\mathrm{dB}\rangle_{f,t}$, is also indicated in the insets. }
    \label{Figure 12}
\end{figure*}

\begin{figure}[h!]
\begin{psfrags}
\psfrag{a}{$0.6$}
\psfrag{b}{$0$}
\psfrag{c}{$\pi$}
\psfrag{d}{$\Delta_1$}
\psfrag{e}{PDF}
\psfrag{f}{\textbf{Normalized linear coupling}}
\psfrag{g}{\textbf{Normalized noise intensity}}
\psfrag{h}{$G$}
\psfrag{i}{$\Lambda$}
\psfrag{j}{$-0.5$}
\psfrag{k}{$0.5$}
\psfrag{l}{$2.88\times 10^{-1}$}
\psfrag{m}{$2.88\times 10^{1}$}
\psfrag{n}{\bf \textcolor{darkblue3}{Simulation}}
\psfrag{o}{\bf \begin{tabular}{@{}c@{}}
\textcolor{myred}{Theory}\\
\textcolor{darkblue3}{Simulation}
\end{tabular}}
\psfrag{p}{$A_1$}
\psfrag{q}{$\Delta_1$}
\psfrag{r}{$-\pi$}
\psfrag{1}{\hspace{0.05cm}Low noise}
\psfrag{2}{\hspace{0.05cm}High noise}
\psfrag{3}{Amplifying}
\psfrag{4}{\hspace{-0.04cm}Dissipative}

    \centering
    \includegraphics[width=0.45\textwidth,center]{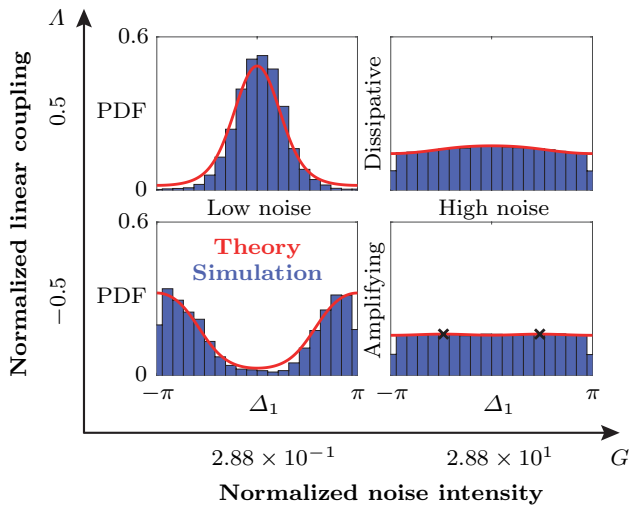}
    \end{psfrags}
    \caption{Comparison of the stationary PDF of the phase differences $\widetilde{\mathcal{P}}^\infty_{\mathrm{P},\Delta}(\Delta_b^*)$, given by Eq. \eqref{Stat. PDF of phase differences} (\textbf{Theory}), to histograms of $\Delta_1$ obtained from time series simulations of the fast system \eqref{Dynamics of dominant modal amplitude with swirl, nonlinear with noise} over $15$'$000$ cycles (\textbf{Simulation}). For high noise and amplifying coupling, the maxima of the PDF are indicated by black crosses. Histograms obtained from other oscillators or from the averaged system \eqref{Amplitude dynamics, slow-flow} and \eqref{Phase dynamics, slow-flow} appear similar.}
    \label{Figure 13}
\end{figure}

\begin{figure*}[h!]
\begin{psfrags}
\psfrag{a}{$2A_0^*$}
\psfrag{b}{$-2A_0^*$}
\psfrag{c}{$A_j \cos\dfrac{\Delta_j}{2}$}
\psfrag{d}{$A_j \sin\dfrac{\Delta_j}{2}$}
\psfrag{e}{$0$}
\psfrag{f}{\hspace{-1cm}\bf Decreasing linear resistive coupling $\boldsymbol{\Lambda}$}
\psfrag{g}{\bf \hspace{-2.3cm} Synchronized state}
\psfrag{h}{\hspace{-0.43cm}\bf Transition region}
\psfrag{i}{\hspace{-0.23cm}\bf Push-pull}
\psfrag{j}{$\boldsymbol{0}$}
\psfrag{k}{\hspace{0.27cm}\bf \textcolor{myyellow}{PDF}}
\psfrag{l}{$\boldsymbol{\Lambda}_\mathrm{c}$}

\psfrag{S}{\hspace{0.1cm} \textcolor{darkblue2}{$\boldsymbol{\Lambda=0.5}$}}
\psfrag{T}{\hspace{0.1cm} \textcolor{mygreen1}{$\boldsymbol{\Lambda=0.3}$}}
\psfrag{U}{\hspace{0.1cm} \textcolor{cyan}{$\boldsymbol{\Lambda=0.1}$}}
\psfrag{V}{\hspace{0.1cm}\textcolor{red}{$\boldsymbol{\Lambda=-0.1}$}}
\psfrag{W}{\hspace{0.1cm}\textcolor{magenta}{$\boldsymbol{\Lambda=-0.3}$}}
\psfrag{X}{\hspace{0.1cm}\textcolor{black}{$\boldsymbol{\Lambda=-0.5}$}}

    \centering
    \includegraphics[width=0.9\textwidth,center]{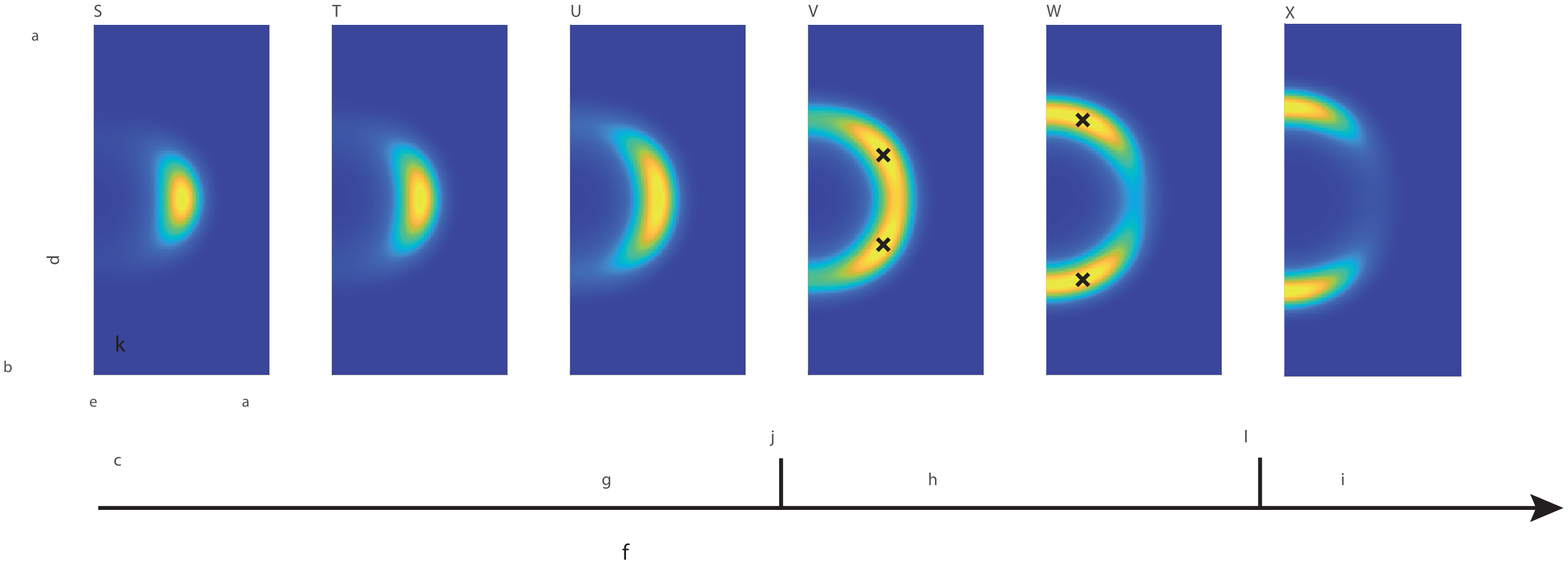}
    \end{psfrags}
    \caption{Visualization of the (de-)synchronization transition. Shown is the evolution of the joint PDF $\widetilde{\mathcal{P}}^\infty_\mathrm{P}$, given by Eq. \eqref{Projected Stationary FP solution PDF}, for varying $\Lambda$ with $G=2.2\times 10^{-1}$ (``Low noise'' in Fig. \ref{Figure 11}). The values of the normalized linear resistive coupling $\Lambda$ correspond to those used in the numerical experiments reported in Fig. \ref{Figure 4}. In the transition region, the maxima of the PDF are indicated by black crosses. For the given conditions, $\Lambda_\mathrm{c}=-0.356$.}
    \label{Figure 14}
\end{figure*}

Building on our analysis of the projected potential $\mathcal{U}_\mathrm{P}$ in the previous section, we begin our investigation of the noise-driven system by studying the behavior of $\mathcal{U}_\mathrm{P}$ under varying coupling nonlinearity $K$ and noise intensity $\Gamma$. \fc{For $\Gamma\neq 0$,} the \fc{fixed points} of the deterministic part of the slow-flow system \eqref{Amplitude dynamics, slow-flow} and \eqref{Phase dynamics, slow-flow} are still characterized by uniform amplitudes along the ring and constant phase differences between neighboring oscillators. The phase differences are still defined by Eq. \eqref{Periodicity restriction on phase difference}, while the limit cycle amplitudes are affected by the noise intensity $\Gamma$:
\begin{eqnarray}
A^*_b=2\sqrt{\frac{\nu_b+\sqrt{\nu_b^2+\kappa_b\Gamma/8\omega_0^2}}{\kappa_b}}. \label{Limit cycle amplitude with noise}
\end{eqnarray}

\noindent In the case $\Gamma\neq 0$, the trajectories of the slow-flow system are governed by a gradient system perturbed by white noise forcing, described by Eqs. \eqref{Langevin amplitude} and \eqref{Langevin phase}. \fc{Similar to the previous section, it is assumed that in the steady state, each oscillator performs stochastically fluctuating motion near the quasi-limit cycles of the deterministic system with uniform amplitude $A_j\equiv A^*_\beta$ and absolute phase differences $|\Delta_j|\equiv |\Delta^*_\beta|$.} The normalized projected potential $\mathcal{U}_\mathrm{P}$ depends now also on the normalized noise intensity $G=\kappa \Gamma/8|\nu|^2 \omega_0 ^2 $ and is computed for different values of $K$ and $G$ by substituting the limit cycle amplitude \eqref{Limit cycle amplitude with noise} and $\Delta_j\equiv \Delta_\beta^*$ into the potential $\mathcal{V}$, given by Eq. \eqref{Potential for amplitudes and phases}, and \fc{dividing the result} by the factor $2N|\nu|^2/\kappa$ for normalization. For compactness, the resulting expression is only listed implicitly here:
\begin{eqnarray}
    &&\mathcal{U}_\mathrm{P}(B)=\nonumber\\
    &&\frac{\kappa}{2N|\nu|^2}\mathcal{V}(A_j\equiv A^*_\beta,\varphi_j\equiv\varphi_{j-1}+k(j)\Delta^*_\beta), \label{Proj. potential stochastic case}
\end{eqnarray}
where $A^*_\beta$ and $\Delta^*_\beta$ are obtained by replacing $b$ with $\beta$ in Eqs. \eqref{Periodicity restriction on phase difference} and \eqref{Limit cycle amplitude with noise}, respectively.

The distribution of the projected potential $\mathcal{U}_\mathrm{P}$ as a function of $\Lambda$ and $2|B|=|\Delta^*_\beta|$ is studied in Fig. \ref{Figure 6}. \fc{The dashed black lines mark the transition region $0>\Lambda>\Lambda_\mathrm{c}$. Note that $\Lambda_\mathrm{c}$ depends now on both $K$ and $G$. Above and below the transition region, the synchronized state and the push-pull mode are the global minimum of the potential $\mathcal{V}$, given by Eq. \eqref{Potential for amplitudes and phases}, respectively. The solid black curves in the transition region parametrize the minima of $\mathcal{U}_\mathrm{P}$ with respect to $2B$, defined by $\partial \mathcal{U}_\mathrm{P}/\partial B=0$, over $\Lambda$. In the bottom right inset in Fig. \ref{Figure 6}, $\Lambda_\mathrm{c}=-0.356$. We observe that $\mathcal{U}_\mathcal{P}$ depends monotonously on both $K$ and $G$, and it is therefore justified for qualitative analyses to study the system for fixed values of $K$ and $G$ as we do in this work.} If $K$ or $G$ are increased, for $K>0$, the critical value of the linear resistive coupling $\Lambda_\mathrm{c}$, which is computed numerically in the noise-driven case, increases in magnitude, and hence the size the transition region increases. Furthermore, for increasing $K$ or $G$, the overall potential landscape becomes increasingly biased towards the synchronized solution. The top row of Fig. \ref{Figure 6} should be interpreted with care, because at high noise levels, the amplitudes $A_j$ are in general not all simultaneously close to some amplitude $A_\beta^*$ and the projection onto the limit cycle solutions loses its validity. Despite this, the figure illustrates correctly the monotonous trends of $\mathcal{U}_\mathrm{P}$, given by Eq. \eqref{Proj. potential stochastic case}, with increasing coupling nonlinearity $K$ and noise intensity $G$. The curve parametrizing the minima of $\mathcal{U}_\mathrm{P}$ with respect to $2B$ over $\Lambda$ is obtained by numerically computing the zero level set of $\partial\mathcal{U}_\mathrm{P}/\partial B$. 

For $t\rightarrow\infty$, in the steady state, the probability $P$ of the fast system \eqref{Dynamics of dominant modal amplitude with swirl, nonlinear with noise} being within a domain $\mathcal{D}$ in the $2N$-dimensional phase space \fc{ of the fast system \eqref{Dynamics of dominant modal amplitude with swirl, nonlinear with noise}, $\Sigma=\{\eta_1,\dot{\eta}_1,\ldots,\eta_N,\dot{\eta}_N\}$, is defined as}
\begin{eqnarray}
    &&P(\mathcal{D},t\rightarrow\infty)=\\
    &&\int_\mathcal{D} \mathcal{P}^\infty(\eta_1,\dot{\eta}_1,\ldots,\eta_N,\dot{\eta}_N) d\eta_1 d\dot{\eta}_1\ldots d\eta_N d\dot{\eta}_N,
\end{eqnarray}
\fc{where the stationary PDF $\mathcal{P}^\infty(\boldsymbol{z})$ of the fast variables 
\begin{equation}
    \boldsymbol{z}=(\eta_1,\dot{\eta}_1,\ldots,\eta_N,\dot{\eta}_N)
\end{equation}
can be deduced from the stationary Fokker--Planck equation (see Ref. \cite{risken1996fokker}) of the slow-flow system \eqref{Amplitude dynamics, slow-flow} and \eqref{Phase dynamics, slow-flow}, given by
\begin{eqnarray}
     0=\frac{\partial }{\partial \boldsymbol{x}}\bigg[\frac{\partial \mathcal{V}(\boldsymbol{x})}{\partial \boldsymbol{x}} \mathcal{P}^\infty(\boldsymbol{x})+\frac{\Gamma}{4\omega_0^2}\frac{\partial \mathcal{P}^\infty(\boldsymbol{x})}{\partial \boldsymbol{x}}\bigg]^T, \label{Stat. FP eq., main text}
\end{eqnarray}
where the newly introduced (slow) variables 
\begin{equation}
    \boldsymbol{x}=(U_1,V_1,\ldots,U_N,V_N)
\end{equation}
are related to ($\eta_j$, $\dot{\eta}_j$) by
$\eta_j=A_j\cos{\phi_j}$ and $\dot{\eta}_j=-\omega_0 A_j\sin{\phi_j}$, where
\begin{eqnarray}
     A_j&=&\sqrt{U_j^2+V_j^2} \label{Amp def main} \\
     \phi_j&=&\arctan(V_j,U_j)+\omega_0 t.  \label{Phase def main}
\end{eqnarray}
The derivation of $\mathcal{P}^\infty$ is detailed in Appendix \ref{Appendix B}. The final result is
\begin{eqnarray}
    &&\mathcal{P}^\infty(\eta_m, \dot{\eta}_m)= \nonumber\\
   &&\mathcal{N} \exp{(-4\omega_0^2 \mathcal{V}(A_m(\eta_m,\dot{\eta}_m), \varphi_m(\eta_m,\dot{\eta}_m))/\Gamma)}, \label{Stationary FP solution for Amplitude Phase system (in main text)}
\end{eqnarray}
$m=1,\ldots,N$, where $A_m=\sqrt{\eta_m^2+\left(\dot{\eta}_m/\omega_0\right)^2}$, $\varphi_m=-\arctan(\dot{\eta}_m/\omega_0,\eta_m)-\omega_0 t$, the potential $\mathcal{V}$ is given by Eq. \eqref{Potential for amplitudes and phases} and $\mathcal{N}$ is a normalization constant such that $P(\mathcal{D}\rightarrow\Sigma,t)=1$. In the figures below, the PDFs are normalized by numerical integration over the shown, finite domains.}

In line with our analysis of the deterministic system in the previous section, we now project $\mathcal{P}^\infty$ onto the phase-locked, uniform-amplitude quasi-limit cycle solutions $\{ A_j\equiv A^*_\beta,\Delta_j\equiv k(j)\Delta^*_{\beta} \}$ to obtain \fc{the projected stationary PDF} $\mathcal{P}^\infty_\mathrm{P}=\mathcal{N} (\widetilde{\mathcal{P}}^\infty_\mathrm{P})^N$,
where 
\begin{eqnarray}
    \widetilde{\mathcal{P}}^\infty_\mathrm{P}(A^*_\beta, \Delta^*_\beta)= \exp{(-4\omega_0^2 \mathcal{V}(A^*_\beta, \Delta^*_\beta)/\Gamma)}.\label{Projected Stationary FP solution PDF}
\end{eqnarray}
\fc{This result implies that on the submanifold of phase-locked, uniform-amplitude quasi-limit cycles, all oscillators perform uncorrelated stochastically fluctuating motion around the deterministic limit cycles and that in the steady state, the PDF of each of the oscillators, up to normalization, is given by $\widetilde{P}^\infty_\mathrm{P}$.} 

The PDF of the phase differences $\widetilde{\mathcal{P}}_{\mathrm{P},\Delta}$ is obtained by \fc{evaluating the limit cycle amplitude $A_\beta^*$ in Eq. \eqref{Limit cycle amplitude with noise} in terms of $\Delta^*_\beta$ using Eqs. \eqref{nu_b} and \eqref{kappa_b} with $b\rightarrow \beta$ and substituting the result into the potential $\mathcal{V}$ in Eq. \eqref{Projected Stationary FP solution PDF}. The result is, up to normalization,
\begin{eqnarray}
    \widetilde{\mathcal{P}}_{\mathrm{P},\Delta}^\infty(\Delta^*_\beta)= \exp{(-4\omega_0^2 \mathcal{V}(A^*_\beta(\Delta^*_\beta), \Delta^*_\beta)/\Gamma)}. \label{Stat. PDF of phase differences}
\end{eqnarray}}
\fc{To graphically compare} the joint PDF $\widetilde{\mathcal{P}}^\infty_\mathrm{P}(A^*_\beta, \Delta^*_\beta)$ to normalized histograms of $A_j$ and $\Delta_j$ obtained from time traces, we propose a geometric analogy between emergent patterns in the oscillator ring, quantified by the phase difference $\Delta_j$, and polar coordinates. The analogy is illustrated in Fig. \ref{Figure 7}. Possible non-negative discrete Bloch wavenumbers $b\geq0$ are shown for $N=12$ oscillators. \fc{The $x$-axis ($b=0$) corresponds to the synchronized state and the $y$-axis ($b=N/2=6$) to the push-pull mode. Values of $b$ in between the two extremes correspond to CWW rotating waves.} The lower half-plane (not shown) is symmetric, with negative values of $b$ \fc{and CW waves}.

In Fig. \ref{Figure 8}, we analyze a time trace of the slow-flow system \eqref{Amplitude dynamics, slow-flow} and \eqref{Phase dynamics, slow-flow} with $\Lambda=-0.5$ ($\Lambda_\mathrm{c}=-0.356$), so that the considered condition lies outside the transition region. The oscillators $j=1,2,3$ are indicated by blue, red and cyan colors, respectively. In Fig. \ref{Figure 8}\textbf{(a)}, we observe that the amplitudes $A_j$ fluctuate stochastically near the deterministic limit cycle amplitude $A^*_6$, which is indicated in green. The phase difference $\Delta_1$, shown in Fig. \ref{Figure 8}\textbf{(b)} performs similar fluctuations and jumps intermittently between $\pm \pi$. Time traces of other oscillators appear similar. The phase distribution $\varphi_j$ along the ring after $5000$ cycles is shown in Fig. \ref{Figure 8}\textbf{(c)}. Figure \ref{Figure 8}\textbf{(d)} shows the spectrogram of the reconstructed fast oscillating signal $\eta_1(t)$, i.e., the estimated signal power $\mathcal{L}_{\eta_1 \eta_1}(f)\mathrm{\Delta}f$, where $\mathcal{L}_{\eta_1 \eta_1}$ is the power spectral density of the signal $\eta_1$ and $\Delta f$ is the equivalent noise bandwith \cite{Kittel19771214} of a small frequency increment around $f$ (the shown frequency domain is discretized into 257 such increments), over time $t$ as a function of the normalized frequency $\omega/\omega_0$. We observe that the signal power is mainly concentrated near the natural eigenfrequency $\omega_0$. The average of the signal power over the shown frequency and time domains, denoted by $\langle\mathrm{dB}\rangle_{f,t}$, is also indicated in the inset. Figure \ref{Figure 8}\textbf{(e)} shows a comparison of the analytical PDF of the phase differences $\widetilde{P}^\infty_{\mathrm{P},\Delta}$ (red) with a histogram of $\Delta_1$ (blue) obtained from the time trace shown in Fig. \ref{Figure 8}\textbf{(b)}. Figure \ref{Figure 8}\textbf{(f)} shows histograms of $A_1$ and $\Delta_1$, visualized using the geometric analogy described in Fig. \ref{Figure 7}. As expected, for the considered conditions (strong amplifying coupling), we find symmetric bimodal distributions of the PDFs, with peaks around the push-pull mode with $\Delta_{\pm 6}^*=\pm \pi$. Spectro- and histograms of other oscillators appear similar.

Figure \ref{Figure 9}\textbf{(a)} depicts the amplitudes of the time trace shown in Fig. \ref{Figure 8}\textbf{(a)} over the last 5 cycles for $j=1,\ldots,N$. In Fig. \ref{Figure 9}\textbf{(b)}, the distribution of the amplitudes $A_j$ along the ring at the end time is shown. We observe that the assumption of slowly varying, uniform amplitudes is approximately satisfied.

The stochastic averaging method is validated against numerical simulations in Fig. \ref{Figure 10} by comparing histograms of $A_1$ and $\Delta_1$, computed over $10$'$000$ cycles from the \fc{fast and the averaged system, given by Eq. \eqref{Dynamics of dominant modal amplitude with swirl, nonlinear with noise} and Eqs. \eqref{Amplitude dynamics, slow-flow} and \eqref{Phase dynamics, slow-flow},} respectively, for varying linear resistive coupling $\Lambda$ and noise intensity $G$. Histograms of other oscillators appear similar. The results show good qualitative agreement between the two methods. Both methods describe a symmetric, bimodal distribution for $\Lambda<0$. At high noise intensity, the PDF becomes more spread out and less discernible patterns are visible. Note that the amplitude $A_0^*$ depends on the noise intensity $G$ and changes from the left to the right column. Sharper shapes of the PDFs emerge for longer times, but the computational cost of simulating the averaged system \eqref{Amplitude dynamics, slow-flow} and \eqref{Phase dynamics, slow-flow} became prohibitive for longer time traces than those shown in Fig. \ref{Figure 10}. Nevertheless, the analytical results derived above in this section enable further validation of the stochastic averaging method. 

This is done in Fig. \ref{Figure 11}, where the stochastic averaging method is validated against analytical results by comparing histograms of $A_1$ and $\Delta_1$ computed from the fast system \eqref{Dynamics of dominant modal amplitude with swirl, nonlinear with noise} (\textbf{Sim.}) to the joint PDF $\widetilde{\mathcal{P}}_\mathrm{P}^\infty$, given by Eq. \eqref{Projected Stationary FP solution PDF} (\textbf{Theory}), for varying linear resistive coupling $\Lambda$ and noise intensity $G$. Histograms of other oscillators appear similar. At low noise, time traces were computed for $10$'$000$ cycles and at high noise for $100$'$000$ cycles. For high noise and amplifying coupling, the maxima of the PDF are indicated by black crosses. The results show good qualitative agreement between the two methods. Note that the amplitude $A_0^*$ depends on the noise intensity $G$ and changes from the left to the right column.

The spectrograms of $\eta_j$, shown in Fig. \ref{Figure 8}\textbf{(d)} for $j=1$, can be decomposed into spectrograms of different rotating wave components by defining the modal amplitude $\eta_{b}$ of the Bloch mode with wavenumber $b$ using the discrete Fourier transform \cite{ghirardo18}:
\begin{equation}
    \eta_b(t)=\frac{1}{N}\sum^N_{k=1} \eta_k(t) e^{-i b 2\pi k /N}. \label{discrete fourier tf}
\end{equation}
In Fig. \ref{Figure 12}, we plot spectrograms of $\eta_b$ for $b=0,\ldots,6$. Spectrograms for positive and negative $b$ are identical. Random, intermittent energy transfer between different modes is observed. The average of the signal power over the shown frequency and time domains, denoted by $\langle\mathrm{dB}\rangle_{f,t}$, is also indicated in the insets. As expected, for the given conditions ($\Lambda=-0.5$, $\Lambda_\mathrm{c}=-0.356$), the push-pull mode with $b=\pm6$ dominates the power spectrum.

In Fig. \ref{Figure 13}, the stationary PDF of the phase differences $\widetilde{\mathcal{P}}^\infty_{\mathrm{P},\Delta}$, given by Eq. \eqref{Stat. PDF of phase differences}, is compared to histograms of $\Delta_1$ obtained from time series simulations of the fast system \eqref{Dynamics of dominant modal amplitude with swirl, nonlinear with noise} over $15$'$000$ cycles for varying noise intensity $G$ and coupling nonlinearity $K$. For high noise and amplifying coupling, the maxima of the PDF are indicated by black crosses. The plots show \fc{good} agreement between the two methods. Histograms obtained from other oscillators or from time traces of the averaged system \eqref{Amplitude dynamics, slow-flow} and \eqref{Phase dynamics, slow-flow} appear similar.

The evolution of the joint PDF $\widetilde{\mathcal{P}}^\infty_\mathrm{P}$, given by Eq. \eqref{Projected Stationary FP solution PDF}, is visualized in Fig. \ref{Figure 14} for varying $\Lambda$ ($\Lambda_\mathrm{c}=-0.356$), with $G=2.2\times 10^{-1}$ (``Low noise'' in Fig. \ref{Figure 11}). The values of the normalized linear resistive coupling $\Lambda$ correspond to those used in the numerical experiments reported in Fig. \ref{Figure 4}. In the transition region, the maxima of the PDF are indicated by black crosses. The figure demonstrates how the above analysis, combined with the geometric analogy introduced in Fig. \ref{Figure 7}, enables a compact description of the (de-)synchronization transition in dependence of $\Lambda$, independent of the number of oscillators $N$. The joint PDF $\widetilde{\mathcal{P}}^\infty_\mathrm{P}$ reproduces the transition from a unimodal to a bimodal distribution as $\Lambda$ is varied from $0.5$ to $-0.5$, which was also observed in the numerical experiments on the deterministic system reported in Figs. \ref{Figure 3} and \ref{Figure 4}.  

In this section, we analyzed the steady state statistics of the noise-driven ring of oscillators.  We introduced a geometric analogy between the emergent patterns in the ring, quantified by the phase difference $\Delta_j$, and polar coordinates to enable a graphical description of the (de-)synchronization transition as a function of the linear resistive coupling. The stationary joint PDF $\widetilde{\mathcal{P}}^\infty_\mathrm{P}$ and the PDF of the phase differences $\widetilde{\mathcal{P}}^\infty_{\mathrm{P},\Delta}$ were obtained by projecting the exact solution of the stationary Fokker--Planck equation onto the phase-locked, uniform-amplitude quasi-limit cycle solutions. The stochastic averaging method was validated by comparing histograms from time series of the fast to those of the averaged system and to the analytical PDFs, showing good agreement over the parameter range considered.

\section{Discussion \label{Section 5}}
\begin{figure*}[h!]
\begin{psfrags}
\psfrag{a}{$b=0$}
\psfrag{b}{$b=1$}
\psfrag{c}{$b=2$}
\psfrag{d}{$b=3$}
\psfrag{e}{$b=4$}
\psfrag{f}{$b=5$}
\psfrag{g}{$b=6$}
\psfrag{h}{$1.1$}
\psfrag{i}{$0.9$}
\psfrag{j}{Norm. Freq.}
\psfrag{k}{$t$}
\psfrag{l}{Power [dB]}
\psfrag{m}{$-40$}
\psfrag{n}{$30$}
\psfrag{A}{\hspace{-0.12cm}\textcolor{white}{$\langle \text{dB} \rangle_{f,t}=-36.0$}}
\psfrag{B}{\hspace{-0.12cm}\textcolor{white}{$\langle \text{dB} \rangle_{f,t}=-33.7$}}
\psfrag{C}{\hspace{-0.12cm}\textcolor{white}{$\langle \text{dB} \rangle_{f,t}=-33.6$}}
\psfrag{D}{\hspace{-0.12cm}\textcolor{white}{$\langle \text{dB} \rangle_{f,t}=-33.1$}}
\psfrag{E}{\hspace{-0.12cm}\textcolor{white}{$\langle \text{dB} \rangle_{f,t}=-31.6$}}
\psfrag{F}{\hspace{-0.12cm}\textcolor{white}{$\langle \text{dB} \rangle_{f,t}=-31.6$}}
\psfrag{G}{\hspace{-0.12cm}\textcolor{white}{$\langle \text{dB} \rangle_{f,t}=-32.3$}}

    \centering
    \includegraphics[width=0.9\textwidth,center]{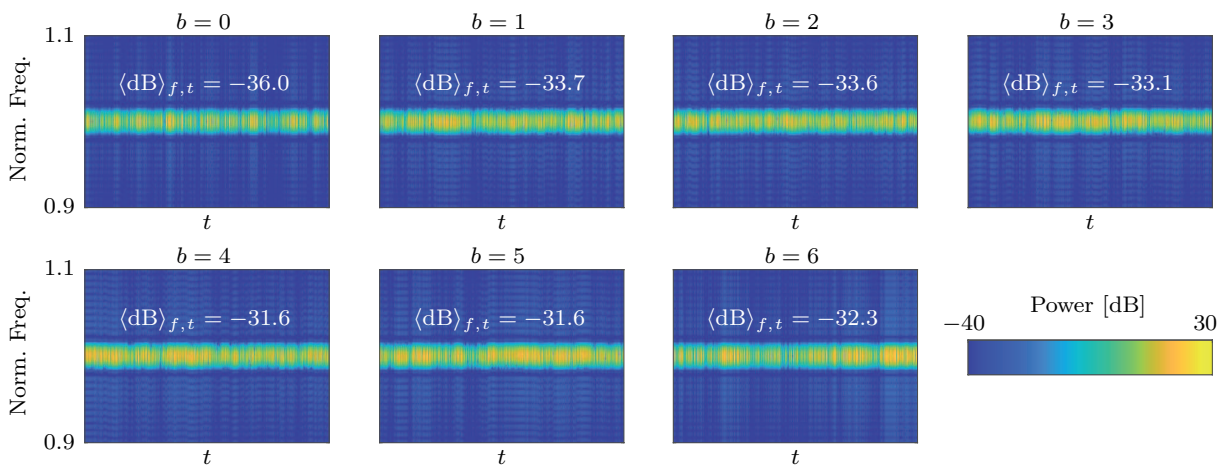}
    \end{psfrags}
    \caption{Spectrograms of the modal amplitudes $\eta_{b}$ for $b=0,\ldots,6$, showing the estimated signal power as a function of time $t$ and the normalized frequency $\omega/\omega_0$. The modal amplitudes $\eta_{b}$ are obtained by decomposing the pressure signals $\eta_j$ from time series of the fast system \eqref{Dynamics of dominant modal amplitude with swirl, nonlinear with noise} with $\Lambda=-0.3$ ($\Lambda_\mathrm{c}=-0.356$) over $5000$ cycles into different Bloch mode components using the discrete Fourier transform, see Eq. \eqref{discrete fourier tf}. Spectrograms for positive and negative $b$ are identical. The average of the signal power over the shown frequency and time domains, denoted by $\langle\mathrm{dB}\rangle_{f,t}$, is also indicated in the insets. }
    \label{Figure 15}
\end{figure*}

\begin{figure}[h!]
\begin{psfrags}
\psfrag{a}{$1.2 A_0^*$}
\psfrag{b}{$-1.2 A_0^*$}
\psfrag{c}{\hspace{-0.25cm}Ac. press. $\eta_j$}
\psfrag{e}{\bf \begin{tabular}{@{}c@{}}
\textcolor{darkblue3}{Nodal line spinning direction}\\
\textcolor{myred}{Apparent spinning direction}
\end{tabular}}
\psfrag{f}{$b=-1$}
\psfrag{g}{$b=5$}

    \centering
    \includegraphics[width=0.42\textwidth,left]{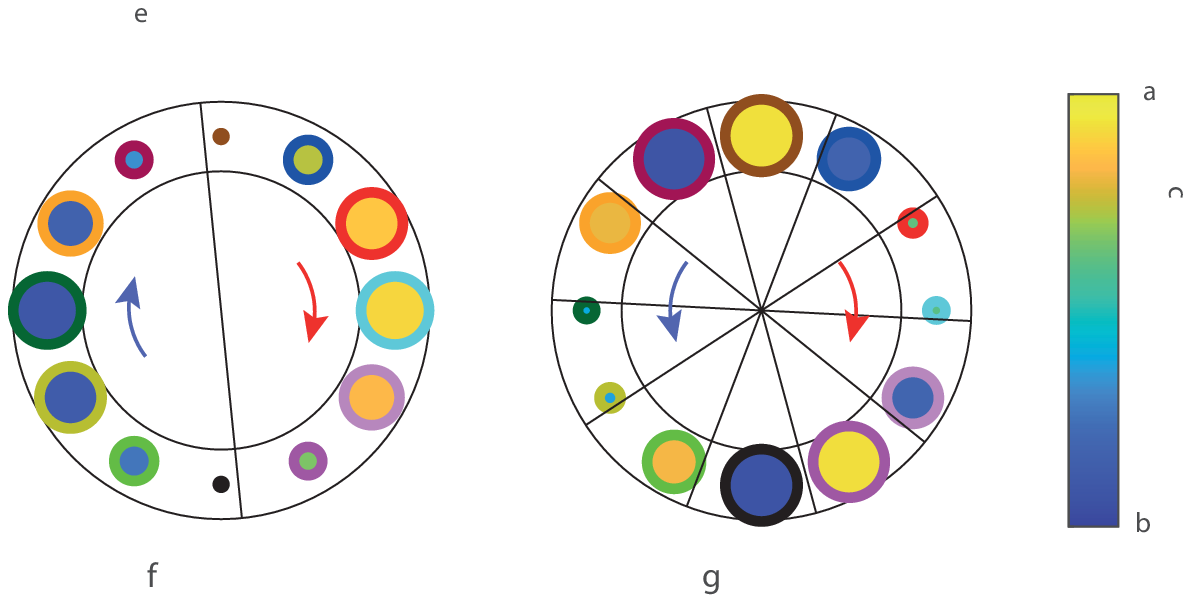}
    \end{psfrags}
    \caption{\fc{Spatial analogue of the wagon-wheel effect. The acoustic pressure $\eta_j$ along the ring is visualized at a fixed time $t$ for $b=-1$ and $b=5$, respectively. The diametric black lines indicate the nodal lines. The radii of the small circles scale with $|\eta_j|$. As the nodal lines of the $b=5$ mode spin in CCW direction, they first hit the oscillators with $j=3,9$ and then those with $j=4,10$ etc. Thus, there appears to be only a single nodal line which spins in CW direction, although the underlying Bloch mode is a CCW wave with $5$ nodal lines. Movies of the corresponding time traces are included in the \href{https://www.dropbox.com/sh/d4pvwzsns30izy9/AAAdW9aMZFzabITg71vU3YqEa?dl=0}{supplementary material}.}}
    \label{Figure 16}
\end{figure}
In Fig. \ref{Figure 12}, spectrograms of the modal amplitudes $\eta_b$ of Bloch modes with different wavenumber $b$ were shown. In Fig. 8 of Ref. \cite{ghirardo18}, similar spectrograms were obtained from acoustic pressure measurements at the can outlets ($\sim \eta_j$) of a real-world gas turbine with $N=12$ cans. The difference is that in the former, the Bloch modes with $b=5,6$ dominate the power spectrum while in the latter, the modes with $b=3,4$ are dominant. By decreasing $\Lambda$ to values within the transition region, this clustering of the signal power around non-maximal Bloch wavenumbers $b$ can be reproduced. The corresponding spectrograms are reported in the Fig. \ref{Figure 14}, showing the distribution of the estimated signal power over time $t$ as a function of the normalized frequency $\omega/\omega_0$. The time traces were computed for $\Lambda=-0.3$ ($\Lambda_\mathrm{c}=-0.356$) with the fast system \eqref{Dynamics of dominant modal amplitude with swirl, nonlinear with noise}. The average of the signal power over the shown frequency and time domains, denoted by $\langle\mathrm{dB}\rangle_{f,t}$, is also indicated in the insets. We observe that the signal power is now concentrated around the Bloch modes with $b=4,5$. If $\Lambda$ is further decreased, the signal power shifts towards lower values of $b$. Note that the inclusion of a coupling nonlinearity $K$ in the model is necessary to observe this effect, as without it, there is no transition region (see Fig. \ref{Figure 6}). The intermittent energy transfer between different Bloch waves observed in Ref. \cite{ghirardo18} is qualitatively reproduced by our model. It is an interesting insight that, despite the signal power appearing in the spectrograms of the Bloch modes with discrete wavenumber $b$, the underlying deterministic dynamics are mainly dominated by quasi-steady, non-classical transition modes which are superpositions of CW and CCW rotating waves. This fact cannot be understood from simple observation of the spectrograms, which may suggest that the Bloch modes exclusively dominate the underlying dynamics. The spectrograms reported in Ref. \cite{ghirardo18} show a shift of the frequency around which the signal power is concentrated with varying Bloch wavenumber $b$. This feature is not reproduced by our model, \tr{which may be} because the reactive coupling was neglected. Besides including reactive coupling effects, a topic for future research is to quantify in more detail the time-dependent energy transfer between different Bloch waves observed in both real-world experiments and our model. For this and other transient phenomena, which are not the focus of this study, differences between the fast system \eqref{Dynamics of dominant modal amplitude with swirl, nonlinear with noise} and the averaged system \eqref{Amplitude dynamics, slow-flow} and \eqref{Phase dynamics, slow-flow} are expected in the presence of noise, i.e., for $\Gamma\neq 0$.

We note that several questions remain open regarding the transition modes observed in Sec. \ref{Section 3} including, but not limited to, the following:

\begin{itemize}
    \item Are there different types of transition modes, distinguished by the distribution of $\Delta_j$ along the ring, and can they be systematically classified? 
    \item How is the ring size $N$ related to the complexity of the transition modes?
    \item What is their sensitivity with respect to initial conditions? Especially, is it possible to predict, from the initial conditions, if a trajectory will converge to a discrete Bloch mode or to a transition mode?
\end{itemize}

The plots in Fig. \ref{Figure 13} show that at low noise, the PDF of the phase differences is highly sensitive with respect to changes in $\Lambda$. Another topic for future research could be to investigate whether this sensitivity, described by the analytical expression for $\widetilde{P}^\infty_{\mathrm{P},\Delta}$ given in Eq. \eqref{Stat. PDF of phase differences}, can be exploited to perform parameter identification of the linear resistive coupling $\lambda$.

\fc{The following are also topics for future research: Asymmetries have been neglected in this study, but are inherent in real systems. To understand their effects, further investigation is required. Furthermore, in real gas turbines, ``true'' azimuthal waves may appear in the continuous, annular plenum concurrently with can-annular modes, i.e., apparent rotating waves that emerge due to communicating thermoacoustic modes in individual, but coupled control volumes. The classification of these combined modes is also a topic for future research.}

\fc{We conclude our discussion with the following remarks: As described in Sec. \ref{Section 3}, Bloch modes with positive or negative wavenumber $b$ appear as CW or CCW rotating waves, i.e., waves whose nodal lines spin in CW or CCW direction. A Bloch mode with wavenumber $b$ has exactly $|b|$ nodal lines. However, for finite $N$, if $\mathrm{mod}(N,|b|)\neq0$, there appears to be only a single nodal line which spins in the opposite direction. This can be understood as a spatial analogue of the wagon-wheel effect, where wheels appear to spin in the opposite direction as an effect of finite temporal sampling rates \cite{Purves19963693}. The effect is illustrated in Fig. \ref{Figure 16}, where the acoustic pressure $\eta_j$ along the ring is shown at a fixed time $t$. The diametric black lines indicate the nodal lines. The radii of the small circles scale with $|\eta_j|$. As the nodal lines of the $b=5$ mode spin in CCW direction, they first hit the oscillators with $j=3,9$ and then those with $j=4,10$ etc. Thus, there appears to be only a single nodal line which spins in CW direction, although the underlying Bloch mode is a CCW wave with $5$ nodal lines. Movies of the corresponding time traces are included in the \href{https://www.dropbox.com/sh/d4pvwzsns30izy9/AAAdW9aMZFzabITg71vU3YqEa?dl=0}{supplementary material}.}

\section{Conclusions \label{Section 6}}

In this work, a ring of noise-driven oscillators was analyzed. Deterministic and stochastic averaging was performed to eliminate the fast oscillating terms. Numerical experiments were performed on the noise-free system to motivate the direction of the study. By projecting the potential of the slow-flow variables onto the phase-locked, uniform-amplitude quasi-limit cycle solutions, a compact description of the (de-)synchronization transition in the ring was obtained. These results were adapted to the noise-driven ring of oscillators to derive, in analytical form, the steady state statistics of the fast system. Special focus was placed on the phase difference $\Delta_j$, which quantifies emergent patterns along the ring. The stochastic averaging procedure was validated against numerical simulations and analytical results. \tr{We have demonstrated that a simple oscillator model with symmetric, purely resistive coupling can reproduce the intermittent energy transfer between different Bloch mode components observed in real-world gas turbines.}

\section*{Declarations}
\textbf{Conflict of interest} The authors declare that they have no conflict of interest.\\ \\
\textbf{Data availability statement} The datasets used for generating the plots and
results in the present study can be directly obtained from the
numerical simulation of the related mathematical equations in
the manuscript.\\ \\
\textbf{Supplementary material \label{supp. mat}} Animated examples of time traces of $A_j$, $\Delta_j$, $\varphi_j$ and $\eta_j$, computed from the noise-free ($G=0$) and noise-driven ($G=2.2\times 10^{-1}$) slow-flow system \eqref{Amplitude dynamics, slow-flow} and \eqref{Phase dynamics, slow-flow} over $5$ acoustic cycles in the steady state for different values of $\Lambda$, are included in the \href{https://www.dropbox.com/sh/d4pvwzsns30izy9/AAAdW9aMZFzabITg71vU3YqEa?dl=0}{supplementary material}. For visualization purposes, the color bar of the time traces of $\eta_j$ in the noise-driven case is cut off at $\pm 1.2 A_0^*$. As in Fig. \ref{Figure 16}, in the plots showing the distribution of the acoustic pressure $\eta_j$ along the ring, the radii of the small circles scale with $|\eta_j|$.

\appendix
\section*{Appendix}
\renewcommand{\thesubsection}{\Alph{subsection}}
\subsection{Deterministic and stochastic averaging \label{Appendix A}}
Below, we perform deterministic and stochastic averaging on the fast system \eqref{Dynamics of dominant modal amplitude with swirl, nonlinear with noise} to obtain the slow-flow equations \eqref{Amplitude dynamics, slow-flow} and \eqref{Phase dynamics, slow-flow} for the amplitudes $A_j$ and the phases $\varphi_j$, respectively. These variables describe the dynamics of the envelope of the acoustic pressure oscillations in the cans. We follow closely the derivation presented in Ref. \cite{Noiray16} for a single Van der Pol oscillator. 

Considering the averaged system as an approximation for the amplitude dynamics in fast system is justified when \textbf{(A)} the fast time scale ($\sim 1/\omega_0$) is clearly separated from the slow time scale of the amplitude and phase variation ($\sim 1/\nu$), and \textbf{(B)} the fast system corresponds to weakly perturbed harmonic oscillators performing quasi-sinusoidal motion (see \cite{krylov1950introduction,sanders2007averaging} for deterministic and \cite{stratonovich1963topics,roberts1986stochastic} for stochastic averaging).  Thermoacoustic interactions in combustion chambers typically fall into the category of weakly damped/driven oscillations with $|\nu|\ll \omega_0$. Furthermore, the aeroacoustic coupling is assumed to be weak, i.e., $|\lambda|\ll\nu$, so that assumption \textbf{(A)} is generally satisfied. Furthermore, under those assumptions, the nonlinearity will also be small because at full saturation, when the limit cycle is reached, in the absence of coupling, $\eta_j\sim\sqrt{\nu/\kappa}$ so that $\kappa\eta_j^2\sim\nu$. As shown in Sec. \ref{Section 3}, under the present assumption of weak coupling ($|\Lambda|\ll 1$, $K\ll 1$), the limit cycle amplitude is not significantly affected by the resistive coupling terms. Therefore, the RHS of Eq. \eqref{Dynamics of dominant modal amplitude with swirl, nonlinear with noise} is generally small compared to the LHS, and assumption \textbf{(B)} is satisfied. 

\fc{We limit ourselves to the case of perfectly tuned oscillators with identical eigenfrequencies $\omega_0$.} To derive the slow-flow dynamics, we first make a coordinate change to amplitude-phase variables $\{A_j,\phi_j\}$, $j=1,\ldots,N$:
\begin{eqnarray}
     A_j&=&\sqrt{\eta_j^2+\left(\dot{\eta}_j/\omega_0\right)^2}, \label{Amplitude appendix}\\
     \varphi_j&=&-\arctan(\dot{\eta}_j/\omega_0,\eta_j)-\omega_0t, \label{Phase appendix}
\end{eqnarray}
where $\eta_j=A_j\cos{\phi_j }$, $\dot{\eta}_j=-\omega_0A_j\sin{\phi_j }$, $\phi_j=\omega_0t+\varphi_j$ and $\arctan(y,x)$ is the two-argument arctangent function. To transform Eq. \eqref{Dynamics of dominant modal amplitude with swirl, nonlinear with noise}, we rewrite it as
\begin{equation}
     \ddot{\eta}_{j}+\omega_0 ^2{\eta}_j =f(\eta_k,\dot{\eta}_k)+\xi_j, \label{Dynamics of dominant modal amplitude, implicit}
\end{equation}
for $k=j-1,j,j+1$, where 
\begin{eqnarray}
&&f(\eta_k,\dot{\eta}_k)=2\nu\dot{\eta}_j-\kappa \eta_j^2\dot{\eta}_j+ \lambda(\dot{\eta}_{j+1}+\dot{\eta}_{j-1}-2\dot{\eta}_j)\nonumber\\
&&+\vartheta\big[({\eta}_{j+1}-{\eta}_j)^2(\dot{\eta}_{j+1}-\dot{\eta}_j)+({\eta}_{j-1}-{\eta}_j)^2(\dot{\eta}_{j-1}\nonumber \\
&&-\dot{\eta}_j)\big]. \label{Function f}
\end{eqnarray}
Taking the time derivative of Eqs. \eqref{Amplitude appendix} and \eqref{Phase appendix} and substituting Eq. \eqref{Dynamics of dominant modal amplitude, implicit} into the resulting expressions yields
\begin{eqnarray}
     \dot{A}_j&=&f_1(A_k,\phi_k)+f_2(A_k,\phi_k), \label{Amplitude dynamics, implicit}\\
     \dot{\varphi}_j&=&f_3(A_k,\phi_k)+f_4(A_k,\phi_k), \label{Phase dynamics, implicit} 
\end{eqnarray}
where
\begin{eqnarray}
     f_1(A_k,\phi_k)&=&-\frac{\sin{\phi_j}}{\omega_0}f(A_k\cos{\phi_k },-\omega_{0}A_k\sin{\phi_k }),\label{refa1}\\
     f_2(A_k,\phi_k)&=&-\frac{\sin{\phi_j}}{\omega_0}\xi_j,\\
     f_3(A_k,\phi_k)&=&-\frac{\cos{\phi_j}}{\omega_0 A_j}f(A_k\cos{\phi_k },-\omega_{0}A_k\sin{\phi_k }), \label{refa2}\\
     f_4(A_k,\phi_k)&=&-\frac{\cos{\phi_j}}{\omega_0 A_j}\xi_j, \label{int ap 1}
\end{eqnarray}
for $k=j-1,j,j+1$. The expressions in Eqs. \eqref{refa1} and \eqref{refa2} are valid for any nonlinear function $f$. We now restrict ourselves the case where the function $f$ is given by Eq. \eqref{Function f}. Averaging is applied to keep only contributions that are relevant on the slow time scale associated with the modulation of the amplitudes $A_j$ and phases $\psi_j$. First, following the classic treatment of Krylov and Bogoliubov \cite{krylov1950introduction}, deterministic averaging is performed on $f_1$ and $f_3$. For this, we rewrite these functions by expanding the trigonometric terms. This yields
\begin{eqnarray}
     &&f_1(A_k,\phi_k)=A_j\bigg(\nu-\frac{\kappa A_j^2}{8}\bigg)+\frac{\lambda}{2}\big[A_{j+1}\cos{(\Delta_{j+1})}\nonumber\\
     &&+A_{j-1}\cos{(\Delta_{j})}-2A_j\big]+\frac{\vartheta}{8}\Big[(A_{j+1}^3+3A_j^2 A_{j+1})\nonumber\\
     &&\cos(\Delta_{j+1})-A_j A_{j+1}^2\cos(2\Delta_{j+1})+(A_{j-1}^3\nonumber\\
     &&+3A_j^2 A_{j-1})\cos(\Delta_{j-1})-A_j A_{j-1}^2\cos(2\Delta_{j-1})\nonumber\\
     &&-2(A_j^3+A_j A_{j+1}^2+A_j A_{j-1}^2)\Big]+\text{f.o.t.}\label{f1 before averaging}
\end{eqnarray}
and
\begin{eqnarray}
     &&f_3(A_k,\phi_k)=\frac{\lambda}{2}\left[\frac{A_{j+1}}{A_j}\sin(\Delta_{j+1})-\frac{A_{j-1}}{A_j}\sin(\Delta_{j})\right]\nonumber\\
     &&+\frac{\vartheta}{8}\Bigg[\bigg(A_j A_{j+1}+\frac{A_{j+1}^3}{A_j}\bigg)\sin(\Delta_{j+1})-A_{j+1}^2\nonumber\\
     &&\sin(2\Delta_{j+1})-\bigg(A_j A_{j-1}+\frac{A_{j-1}^3}{A_j}\bigg)\sin(\Delta_{j-1})\nonumber\\
     &&+A_{j-1}^2\sin(2\Delta_{j-1})\Bigg]+\text{f.o.t.}\label{f3 before averaging}
\end{eqnarray}
where $\Delta_j=\phi_j-\phi_{j-1}=\varphi_j-\varphi_{j-1}$ and ``f.o.t.'' stands for ``fast oscillating terms''. Note that the quantities $\phi_j+\phi_{j\pm1}=\varphi_j+\varphi_{j\pm1}+2\omega_0 t$ as well as $2\phi_j=2\varphi_j+2\omega_0 t$ define arguments of fast oscillating terms. We assume that the period of an acoustic cycle
$T=2\pi/\omega_0$ is small compared to the slow time scale $1/\nu$ of the modulation of the amplitude and phase signals. By taking the averages of Eqs. \eqref{f1 before averaging} and \eqref{f3 before averaging}, $\langle f_1(A_k,\varphi_k)\rangle_T$ and $\langle f_3(A_k,\varphi_k)\rangle_T$, respectively, over a time interval $\{t_0, t_0+T\}$ of length $T$ with arbitrary start time $t_0$, one removes the fast oscillating terms.

\fc{Stochastic averaging of the terms $f_2$ and $f_4$ can be performed in different ways, two of which are presented in Refs. \cite{stratonovich1963topics} (Vol. 2, pp. 105-113) and \cite{roberts1986stochastic}, respectively. The former method, due to Stratonovich, is adopted here. In our derivation, we follow the presentation of a similar result (stochastic averaging of a single thermoacoustic Van der Pol oscillator) given in Appendix A of Ref. \cite{Noiray16}.}

\fc{For compactness, we use the notation $f_\tau=f(t+\tau)$ and the dependence of $\phi_j$, $\varphi_j$ and $A_j$ on time is suppressed below. We introduce the random variables $\sigma_{j,1}=-\xi_j\sin{\phi_j}=f_2\omega_0$ and $\sigma_{j,2}=-\xi_j\cos{\phi_j}=f_4 A_j \omega_0$. We begin by estimating the expected values of $\sigma_{j,1}$ and $\sigma_{j,2}$. We consider a time interval of length $\theta$ satisfying $1/\nu \gg \theta \gg \mathrm{max}(\tau_{\xi_j},\tau_{A_j},\tau_{\varphi_j})\hspace{0.1cm}\forall j$, where $\tau_{a}$ denotes the correlation time of the signal ``$a$''. We note that although the signals $\xi_j$, $\zeta_j$ and $\chi_j$ are modeled as white noise sources with (theoretically) vanishing correlation times, in reality and in numerical simulations, these signal will generally have nonzero, albeit small correlation times. As a first step, we approximate $\sigma_{j,1}$ and $\sigma_{j,2}$ as follows}:
\fc{\begin{eqnarray}
    &&\sigma_{j,1}=-\xi_j(t)\sin{(\omega_0 t+\varphi_j)} \nonumber\\
    &&\approx -\xi_j \sin{(\omega_0 t+\varphi_{j,-\theta})}-\xi_j\cos{(\omega_0 t + \varphi_{j,-\theta})}\mathrm{\Delta} \varphi_j, \label{Int 3}\\
    &&\sigma_{j,2}=-\xi_j(t)\cos{(\omega_0 t+\varphi_j)}\nonumber\\
    &&\approx -\xi_j \cos{(\omega_0 t+\varphi_{j,-\theta})}+\xi_j\sin{(\omega_0 t + \varphi_{j,-\theta})}\mathrm{\Delta} \varphi_j,\label{Int 4}
\end{eqnarray}
where $\varphi_{j,-\theta}$=$\varphi_j(t-\theta)$ and $\mathrm{\Delta} \varphi_j=\varphi_j-\varphi_{j,-\theta}$. Using the fact that $\varphi_{j,-\theta}$ is not correlated with $\xi_j$ since $\theta\gg \mathrm{max}(\tau_{\xi_j},\tau_{\varphi_j})$, we write }
\fc{\begin{eqnarray}
    &&\langle \sigma_{j,1}\rangle_\mathbb{R} \approx -\cos{ \phi_j}\langle \xi_j \mathrm{\Delta} \varphi_j \rangle_\mathbb{R}\\
    &&\langle \sigma_{j,2}\rangle_\mathbb{R} \approx \sin{ \phi_j}\langle \xi_j \mathrm{\Delta} \varphi_j \rangle_\mathbb{R},
\end{eqnarray}
where $\langle\cdot\rangle_\mathbb{R}$ denotes the time integral over the real line $[-\infty,\infty]$. Next, we integrate Eq. \eqref{Phase dynamics, implicit} over the time interval $[t-\theta,t]$. Integrating Eq. \eqref{int ap 1} without the fast oscillation terms, which vanish after averaging over one period (see Eq. \eqref{f3 before averaging}), we can write }
\fc{\begin{eqnarray}
    &&\mathrm{\Delta}\varphi_j=\int^t_{t-\theta} -\frac{\xi_j}{ \omega_0 A_j} \cos{\varphi_j} dt\\
    &&\int^0_{-\theta}-\frac{\xi_j(\tau+t)}{\omega_0 A_j(\tau+t)}\cos\big(\omega_0(\tau+t)+\varphi_j(\tau+t)\big) d\tau.
\end{eqnarray}
Considering that $\xi_j$ is a stationary process and that $1/\nu\gg\theta$, i.e., the variation of the amplitudes $A_j$ and phases $\varphi_j$ over the considered time interval is negligible, we obtain}
\fc{\begin{eqnarray}
    &&\langle \xi_j \mathrm{\Delta} \varphi_j\rangle_\mathbb{R}=\nonumber\\
    &&-\frac{1}{\omega_0 A_j} \int^0_{-\theta} \langle \xi_j \xi_{j,\tau}\rangle_\mathbb{R} \cos{(\omega_0(\tau+t)+\varphi_j)} d\tau
\end{eqnarray}
and }
\fc{\begin{eqnarray}
    &&\langle \sigma_{j,1}\rangle_\mathbb{R} = \nonumber  \\
    &&\frac{\int^0_{-\theta} \langle \xi_j \xi_{j,\tau} \rangle_\mathbb{R} \cos{(\omega_0 t +\varphi_j)}\cos{(\omega_0 (\tau+t)+\varphi_j)} d\tau}{\omega_0 A_j}, \label{Int 1}\\
    &&\langle \sigma_{j,2}\rangle_\mathbb{R} = \nonumber\\
    &&-\frac{\int^0_{-\theta} \langle \xi_j \xi_{j,\tau} \rangle_\mathbb{R} \sin{(\omega_0 t +\varphi_j)}\cos{(\omega_0 (\tau+t)+\varphi_j)} d\tau}{\omega_0 A_j}. \label{Int 2} 
\end{eqnarray}
Expanding the products of cosines in Eqs. \eqref{Int 1} and \eqref{Int 2} and neglecting the fast oscillating terms gives}
\fc{\begin{eqnarray}
    \langle \sigma_{j,1}\rangle_\mathbb{R} &=& \frac{\int^0_{-\infty} \langle \xi_j \xi_{j,\tau} \rangle_\mathbb{R} \cos{(\omega_0 \tau +\varphi_j)} d\tau}{2\omega_0 A_j}\nonumber\\
    &=&\frac{\pi \mathcal{L}_{\xi_j \xi_j}(\omega_0)}{2\omega_0 A_j}=\frac{\Gamma}{4\omega_0 A_j} \\
    \langle \sigma_{j,2}\rangle_\mathbb{R} &=& -\frac{\int^0_{-\infty} \langle \xi_j \xi_{j,\tau} \rangle_\mathbb{R} \sin{(\omega_0 t)} d\tau}{\omega_0 A_j}\approx 0, 
\end{eqnarray}
where $\mathcal{L}_{\xi_j \xi_j}$ is the power spectral density of the noise signal $\xi_j$. }

\fc{We proceed by estimating the correlation function of the zero mean processes $\sigma_{j,1}'=\sigma_{j,1}-\langle\sigma_{j,1}\rangle_\mathbb{R}$ and $\sigma_{j,2}'=\sigma_{j,2}-\langle\sigma_{j,2}\rangle_\mathbb{R}=\sigma_{j,2}$. In Eqs. \eqref{Int 3} and \eqref{Int 4}, the mean values are determined by the second terms on the RHS and the fluctuating components result from the first terms:}
\fc{\begin{eqnarray}
    \sigma_{j,1}'&\approx& -\xi_j \sin{(\omega_0 t+\varphi_{j,-\theta})}\\
    \sigma_{j,2}'&\approx& -\xi_j \cos{(\omega_0 t+\varphi_{j,-\theta})}.
\end{eqnarray}
Since $\theta\gg\mathrm{max}(\tau_{\xi_j},\tau_\varphi)$, it is assumed that these processes are delta-correlated. Considering first $\sigma_{j,1}'$, one can write }
\fc{\begin{equation}
    \langle\sigma_{j,1}' \sigma_{j,1,\tau}'\rangle_\mathbb{R}=\delta(\tau)\int^\infty_{-\infty} \langle\sigma_{j,1}' \sigma_{j,1,\Theta}'\rangle d\Theta, \label{refthis}
\end{equation}
where the integral in Eq. \eqref{refthis} is expressed as}
\fc{\begin{equation}
    \int^\infty_{-\infty}\langle \xi_j \sin{(\omega_0 t+\varphi_{j,-\theta})}\xi_{j,\Theta}\sin{(\omega_0 (t+\Theta)+\varphi_{j,-\theta})}\rangle_\mathbb{R} d\Theta. \nonumber
\end{equation}
Because the correlation time $\tau_{\xi_j}$ is much shorter (theoretically zero) than the oscillation period $1/\omega_0$, we can neglect the fast oscillating terms and write}
\fc{\begin{eqnarray}
    \langle \sigma_{j,1}' \sigma_{j,1,\tau}'\rangle_\mathbb{R} &=&\delta(\tau) \int^\infty_{-\infty} \langle \xi_j \xi_{j,\Theta} \rangle_\mathbb{R} \frac{\cos(\omega_0 \Theta)}{2}d\Theta\\
   &=&\pi \mathcal{L}_{\xi_j \xi_j}(\omega_0)\delta(\tau)\\
   &=&\frac{\Gamma}{2}\delta(\tau).
\end{eqnarray}
Following the same derivation for $\sigma_{j,2}'$ yields the correlation function $\langle \sigma_{j,2}' \sigma_{j,2,\tau}'\rangle_\mathbb{R}=(\Gamma/2)\delta(\tau)$.
With the above results, we can write}
\fc{\begin{eqnarray}
     \langle f_2(A_k,\varphi_k)\rangle_\mathbb{R}&=&\frac{\Gamma}{4\omega_0^2 A_j}+\zeta_j, \label{stoch. av 1}\\
     \langle f_4(A_k,\varphi_k)\rangle_\mathbb{R}&=&\frac{\chi_j}{A_j}, \label{stoch. av 2}
\end{eqnarray}
for $j=1,\ldots,N$, where $\zeta_j$ and $\chi_j$ are $2N$ uncorrelated white noise sources with the noise intensity $\Gamma/2\omega_0^2$. Note that the mean of the averaged process $\langle f_2(A,\phi)\rangle_\mathbb{R}$ is not zero and that the stochastic averaging procedure gives rise to a deterministic term which is inversely proportional to $A_j$.}

Combining Eqs. \eqref{f1 before averaging}, \eqref{f3 before averaging}, \eqref{stoch. av 1} and \eqref{stoch. av 2}, together with the assumption that the variation of $A_j$ and $\varphi_j$ over an acoustic period is negligible, yields the slow-flow amplitude and phase dynamics of the fast system \eqref{Dynamics of dominant modal amplitude with swirl, nonlinear with noise}, given by the following set of equations:
\begin{eqnarray}
     &&\dot{A}_j=A_j\bigg(\nu-\frac{\kappa A_j^2}{8}\bigg)+\frac{\lambda}{2}\big[A_{j+1}\cos{(\Delta_{j+1})}+A_{j-1}\nonumber\\
     &&\cos{(\Delta_{j})}-2A_j\big]+\frac{\vartheta}{8}\bigg[(A_{j+1}^3+3A_j^2 A_{j+1})\cos(\Delta_{j+1})-\nonumber\\
     &&A_j A_{j+1}^2\cos(2\Delta_{j+1})+(A_{j-1}^3+3A_j^2 A_{j-1})\cos(\Delta_{j})-\nonumber\\
     &&A_j A_{j-1}^2\cos(2\Delta_{j})-2(A_j^3+A_j A_{j+1}^2+A_j A_{j-1}^2)\bigg]\nonumber\\
     &&+\frac{\Gamma}{4\omega_0^2 A_j}+\zeta_j \label{Amplitude dynamics, slow-flow appendix}
\end{eqnarray}
and
\begin{eqnarray}
     &&\dot{\varphi}_j=\frac{\lambda}{2}\left[\frac{A_{j+1}}{A_j}\sin(\Delta_{j+1})-\frac{A_{j-1}}{A_j}\sin(\Delta_{j})\right]\nonumber\\
     &&+\frac{\vartheta}{8}\bigg[\Big(A_j A_{j+1}+\frac{A_{j+1}^3}{A_j}\Big)\sin(\Delta_{j+1})-A_{j+1}^2\nonumber\\
     &&\sin(2\Delta_{j+1})-\Big(A_j A_{j-1}+\frac{A_{j-1}^3}{A_j}\Big)\sin(\Delta_{j})\nonumber\\
     &&+A_{j-1}^2\sin(2\Delta_{j})\bigg]+\frac{\chi_j}{A_j}, \label{Phase dynamics, slow-flow appendix}
\end{eqnarray}
for $j=1,\ldots,N$.  

\subsection{Derivation of the solution of the stationary Fokker-Planck equation \label{Appendix B}}
Below, we derive the solution $\mathcal{P}^\infty(\boldsymbol{z})=\mathcal{P}(\boldsymbol{z},t\rightarrow \infty)$, where $\boldsymbol{z}=(\eta_1,\dot{\eta}_1,\ldots,\eta_N,\dot{\eta}_N)$, of the stationary Fokker--Planck equation \eqref{Stat. FP eq., main text} associated with the slow-flow system \eqref{Amplitude dynamics, slow-flow} and \eqref{Phase dynamics, slow-flow}. Our derivation is based on an understanding of the transformation properties of the stationary Fokker--Planck equation under an invertible mapping $\boldsymbol{f}$ from rectilinear coordinates $\boldsymbol{x}$ to curvilinear coordinates $\boldsymbol{y}$:
\begin{equation}
    \boldsymbol{x}=\boldsymbol{f}(\boldsymbol{y}). \label{Coordinate change FP eq.}
\end{equation} 
We assume that the Jacobian $\boldsymbol{J_f}=\partial \boldsymbol{f}/\partial \boldsymbol{y}$ is invertible and can be written as 
\begin{eqnarray}
    \boldsymbol{J_f}(\boldsymbol{y})=\boldsymbol{Q}(\boldsymbol{y})\boldsymbol{h}(\boldsymbol{y}), \label{Assumption on Jacobian}
\end{eqnarray}
where $\boldsymbol{Q}=\boldsymbol{Q}^{-T}$ is an orthogonal rotation matrix with \\$|\mathrm{det}(\boldsymbol{Q})|=1$ and $\boldsymbol{h}$ is a diagonal matrix. This implies $\boldsymbol{h}$ is related to the metric tensor $\boldsymbol{g}=\boldsymbol{J_f}^T\boldsymbol{J_f}$ by 
\begin{eqnarray}
    \boldsymbol{h}^T(\boldsymbol{y}) \boldsymbol{h}(\boldsymbol{y}) =\boldsymbol{g}(\boldsymbol{y}).
\end{eqnarray} 
Now consider a stochastically perturbed gradient system, given by the following Langevin equation in rectilinear coordinates $\boldsymbol{x}$:
\begin{equation}
    \dot{\boldsymbol{x}}=-\frac{\partial \mathcal{V}}{\partial \boldsymbol{x}}+\boldsymbol{n}, \label{Langevin equation}
\end{equation} 
where $\boldsymbol{n}$ contains additive white noise components in the directions of the state variables $\boldsymbol{x}$, which are all assumed to have the same noise intensity $\Gamma/2\omega_0^2$. For this type of system, the stationary Fokker--Planck equation \cite{risken1996fokker} is given by Eq. \eqref{Stat. FP eq., main text}, which, for $\mathcal{P}^\infty\neq0$, has the following normalizable solution:
\begin{eqnarray}
    \mathcal{P}^\infty(\boldsymbol{x})=\mathcal{N} \exp{(-4\omega_0^2 \mathcal{V}(\boldsymbol{x})/\Gamma)},
\end{eqnarray}
where $\mathcal{N}\in\mathbb{R}$ is a normalization constant. In $\boldsymbol{y}$-coordinates, the probability of the stationary state of the system being within some domain $\boldsymbol{f}^{-1}(\mathcal{D})$ is given by 
\begin{eqnarray}
    P=\int_{\boldsymbol{f}^{-1}(\mathcal{D})} \fc{\mathcal{P}^\infty}(\boldsymbol{f}(\boldsymbol{y})) |\mathrm{det}(\boldsymbol{h})| d\widetilde{V},
\end{eqnarray}
where the fact that $|\mathrm{det}(\boldsymbol{J_f})|=|\mathrm{det}(\boldsymbol{h})|$ was used. With this, the stationary PDF reads, in $\boldsymbol{y}$-coordinates,
\begin{eqnarray}
    \mathcal{P}^\infty(\boldsymbol{y})&=&|\mathrm{det}(\boldsymbol{h})|\mathcal{N} \exp{(-4\omega_0^2 \mathcal{V}(\boldsymbol{f}(\boldsymbol{y}))/\Gamma)} \nonumber\\
    &=&|\mathrm{det}(\boldsymbol{h})|\mathcal{N} \exp{(-4\omega_0^2 \mathcal{V}(\boldsymbol{x})/\Gamma)}\nonumber\\
    &=&|\mathrm{det}(\boldsymbol{h})|\mathcal{P}^\infty(\boldsymbol{x}). \label{Transformed stationary FP solution}
\end{eqnarray}
Under the mapping \eqref{Coordinate change FP eq.}, the Langevin equation \eqref{Langevin equation} becomes
\begin{equation}
    \dot{\boldsymbol{y}}=-\boldsymbol{g}^{-1}(\boldsymbol{y})\frac{\mathrm{d}\mathcal{V}(\boldsymbol{y})}{\mathrm{d} \boldsymbol{y}}+\boldsymbol{h}^{-1}(\boldsymbol{y})\tilde{\boldsymbol{n}}, \label{Transformed Langevin equation}
\end{equation}
where $\tilde{\boldsymbol{n}}=\boldsymbol{Q}^T \boldsymbol{n}$ contains the noise components in the direction of the $\boldsymbol{y}$-coordinates. The solution of the stationary Fokker--Planck equation $\mathcal{P}^\infty(\boldsymbol{y})$ associated with the transformed Langevin equation \eqref{Transformed Langevin equation} is given by Eq. \eqref{Transformed stationary FP solution}. 

In the present case, the rectilinear coordinates $\boldsymbol{x}$, the curvilinear coordinates $\boldsymbol{y}$ and the mapping $\boldsymbol{f}(\boldsymbol{y})$ are given by
\begin{eqnarray}
\boldsymbol{x}&=&(U_1,V_1,\ldots,U_N,V_N), \label{UVcoords}\\
\boldsymbol{y}&=&(A_1,\varphi_1,\ldots,A_N,\varphi_N),\\
\boldsymbol{f}(\boldsymbol{y})&=&[A_1 \cos(\varphi_1), A_1 \sin(\varphi_1),\ldots,\nonumber\\
&&\hspace{1.25cm} A_N \cos(\varphi_N), A_N\sin(\varphi_N)].
\end{eqnarray}
The Jacobian of the mapping $\boldsymbol{f}$ reads
\begin{eqnarray}
   \boldsymbol{J_f}=  \begin{pmatrix}
\boldsymbol{M}_1 & \boldsymbol{0} &\ldots&\boldsymbol{0}&\boldsymbol{0} \\
0 & \boldsymbol{M}_2 & \ldots &\boldsymbol{0}&  \boldsymbol{0} \\
\ldots &  \ldots &\ldots  &\ldots &\ldots \\
0 & 0 &  \ldots & \boldsymbol{M}_{N-1}  & \boldsymbol{0} \\
0 & 0 &  \ldots &\boldsymbol{0} & \boldsymbol{M}_{N}
\end{pmatrix}\boldsymbol{h}, \label{Inverse of h 1}
\end{eqnarray}
where $\boldsymbol{h}=\mathrm{diag}(1,A_1,\ldots,1,A_N)$ and
\begin{eqnarray}
   \boldsymbol{M}_j= \begin{pmatrix}
\cos(\varphi_j) & -\sin(\varphi_j)  \\
\sin(\varphi_j) &  \cos(\varphi_j)\\
\end{pmatrix}. \label{Block matrix}
\end{eqnarray}
The matrix $\boldsymbol{Q}=\boldsymbol{J_f}\,\boldsymbol{h}^{-1}$ is a block-diagonal matrix of orthogonal matrices, and is thus itself orthogonal. Hence Eq. \eqref{Assumption on Jacobian} is satisfied for the present system.

To derive the solution of the stationary Fokker--Planck equation \eqref{Stat. FP eq., main text}, we note that the slow-flow system \eqref{Amplitude dynamics, slow-flow} and \eqref{Phase dynamics, slow-flow} is equivalent to a Langevin equation of the form given in Eq. \eqref{Transformed Langevin equation}, where $\boldsymbol{y}=(A_1,\varphi_1,\ldots,A_N,\varphi_N)$,
\begin{eqnarray}
   \boldsymbol{h}^{-1}&=&\mathrm{diag}(1,A_1^{-1},\ldots,1,A_N^{-1}) \in \mathbb{R}^{{2N}\times{2N}},\label{Inverse of h}\\
   \boldsymbol{g}^{-1}&=&\boldsymbol{h}^{-1} \boldsymbol{h}^{-T}\nonumber\\
   &=&\mathrm{diag}(1, A_1^{-2},\ldots,1,A_N^{-2})\in \mathbb{R}^{{2N}\times{2N}},
\end{eqnarray}
and the potential $\mathcal{V}$ is defined by Eq. \eqref{Potential for amplitudes and phases}. We find from Eq. \eqref{Inverse of h} that
\begin{equation}
    \mathrm{det}(\boldsymbol{h})=\prod^N_{j=1} A_j.
\end{equation}
With this result, the stationary solution of the Fokker--Planck equation associated with the slow-flow system \eqref{Amplitude dynamics, slow-flow} and \eqref{Phase dynamics, slow-flow} can be directly computed from Eq. \eqref{Transformed stationary FP solution}:
\begin{eqnarray}
    &&\mathcal{P}^\infty(A_m, \varphi_m)=\nonumber\\
    &&\mathcal{N}\Big[\prod^N_{j=1} A_j\Big] \exp{(-4\omega_0^2 \mathcal{V}(A_m, \varphi_m)/\Gamma)}, \label{Stationary FP solution for Amplitude Phase system}
\end{eqnarray}
$m=1,\ldots,N$. Combining Eqs. \eqref{Transformed stationary FP solution} and \eqref{Stationary FP solution for Amplitude Phase system} yields the stationary PDF in $\boldsymbol{x}$-coordinates (see Eq. \eqref{UVcoords}):
\begin{eqnarray}
     &&\mathcal{P}^\infty(U_m, V_m)= \nonumber\\
     &&\mathcal{N} \exp{(-4\omega_0^2 \mathcal{V}(A_m(U_m,V_m), \varphi_m(U_m,V_m))/\Gamma)}, \label{Stationary FP solution for fast system}
\end{eqnarray}
$m=1,\ldots,N$, where 
\begin{eqnarray}
     A_m(U_m,V_m)&=&\sqrt{U_m^2+V_m^2}, \label{Amp def} \\
     \varphi_m(U_m,V_m)&=&\arctan(V_m,U_m).  \label{Phase def}
\end{eqnarray}
We now consider the mapping $\boldsymbol{z}=\boldsymbol{d}(\boldsymbol{x})$ from $\boldsymbol{x}$ to the fast $\boldsymbol{z}$ coordinates given by
\begin{eqnarray}
\boldsymbol{z}&=&(\eta_1,\dot{\eta}_1,\ldots,\eta_N,\dot{\eta}_N),\\
\boldsymbol{x}&=&(U_1,V_1,\ldots,U_N,V_N),\\
\boldsymbol{d}(\boldsymbol{x})&=&[A_1 \cos(\phi_1),-\omega_0 A_1 \sin(\phi_1),\ldots,\nonumber\\
&&\hspace{1.25cm} A_N \cos(\phi_N),-\omega_0 A_N\sin(\phi_N)],
\end{eqnarray}
where $\phi_j=\varphi_j+\omega_0 t$ and the amplitude-phase coordinates ($A_j$, $\varphi_j$) are expressed in terms of $U_j$ and $V_j$ using Eqs. \eqref{Amp def} and \eqref{Phase def}, respectively. The Jacobian of the mapping $\boldsymbol{d}$ reads
\begin{eqnarray}
   &&\boldsymbol{J_d}=  \begin{pmatrix}
\boldsymbol{H}_1 & \boldsymbol{0} &\ldots&\boldsymbol{0}&\boldsymbol{0} \\
0 & \boldsymbol{H}_2 & \ldots &\boldsymbol{0}&  \boldsymbol{0} \\
\ldots &  \ldots &\ldots  &\ldots &\ldots \\
0 & 0 &  \ldots & \boldsymbol{H}_{N-1}  & \boldsymbol{0} \\
0 & 0 &  \ldots &\boldsymbol{0} & \boldsymbol{H}_{N}
\end{pmatrix}, \label{Inverse of h 1 2}
\end{eqnarray}
where 
\begin{eqnarray}
   \boldsymbol{H}_j&=& \frac{1}{\sqrt{U_j^2+V_j^2}}\begin{pmatrix}
H_{11}(U_j,V_j)& H_{12}(V_j,U_j)   \\
H_{21}(U_j,V_j) & H_{22}(V_j,U_j)\\
\end{pmatrix},\label{Block matrix 2}\\
H_{11}(U_j,V_j)&=&U_j\cos(\phi_j) + V_j\sin(\phi_j),\\
H_{12}(U_j,V_j)&=&V_j\cos(\phi_j) - U_j\sin(\phi_j),\\
H_{21}(U_j,V_j)&=&-\omega_0\big[U_j\sin(\phi_j) - V_j\cos(\phi_j)\big],\\
H_{22}(U_j,V_j)&=&-\omega_0\big[V_j\sin(\phi_j) + U_j\cos(\phi_j)\big],
\end{eqnarray}
 and the dependence of $\phi_j$ on $U_j$ and $V_j$ is suppressed for compactness. The modulus of the determinant of $\boldsymbol{J_d}$ is the positive real constant $|\det(\boldsymbol{J_d})|=\omega_0^{N}$, which is absorbed into the normalization factor $\mathcal{N}$. Since
\begin{eqnarray}
    \mathcal{P}^\infty(\boldsymbol{z})=|\mathrm{det}(\boldsymbol{J_d}) |\mathcal{P}^\infty(\boldsymbol{x}) \label{x to z det}
\end{eqnarray} 
and by the definition of $\mathcal{P}^\infty(\boldsymbol{x})$ in Eq. \eqref{Stationary FP solution for fast system}, the stationary PDF of the fast system \eqref{Dynamics of dominant modal amplitude with swirl, nonlinear with noise} is given by
\begin{eqnarray}
    &&\mathcal{P}^\infty(\eta_m, \dot{\eta}_m)= \nonumber\\
   &&\mathcal{N} \exp{(-4\omega_0^2 \mathcal{V}(A_m(\eta_m,\dot{\eta}_m), \varphi_m(\eta_m,\dot{\eta}_m))/\Gamma)}, \label{Stationary FP solution for fast system 2}
\end{eqnarray}
$m=1,\ldots,N$, where 
\begin{eqnarray}
    A_m(\eta_m,\dot{\eta}_m)&=&\sqrt{\eta_m^2+\left(\dot{\eta}_m/\omega_0\right)^2},\\
    \varphi_m(\eta_m,\dot{\eta}_m)&=&-\arctan(\dot{\eta}_m/\omega_0,\eta_m)-\omega_0 t,
\end{eqnarray}
the potential $\mathcal{V}$ is given by Eq. \eqref{Potential for amplitudes and phases} and $\mathcal{N}$ is a normalization constant.

\bibliographystyle{spmpsci_unsort_doi}      
\bibliography{article.bib}   

%
%

\end{document}